\documentclass[prd,twocolumn,amsmath,amssymb,floatfix,superscriptaddress]{revtex4-1}

\usepackage{graphicx}
\usepackage{amssymb}
\usepackage{amsmath}
\usepackage{color}
\usepackage{ wasysym }
\usepackage{hyperref}
\usepackage{tabu}
\usepackage{bm}
\def\barray{\begin{array}}
\def\earray{\end{array}}
\def\be{\begin{equation}}
\def\ee{\end{equation}}
\def\ben{\begin{equation} \nonumber}
\def\een{\end{equation}}
\def\ban{\begin{eqnarray*}}
\def\ean{\end{eqnarray*}}
\def\ba{\begin{eqnarray}}
\def\ea{\end{eqnarray}}

\def\({\left(}
\def\){\right)}

\def\mnras{MNRAS}

\def\apjs{ApJS}

\graphicspath{{./fig/}}

\begin{document}

\title{Cosmological constraints on dark energy parametrizations after DESI 2024: Persistent deviation from standard $\Lambda$CDM cosmology}
\author{Mohammad Malekjani}
\email{malekjani@basu.ac.ir}
\affiliation{Department of Physics, Bu-Ali Sina University, Hamedan
	65178, Iran}
\author{Zahra Davari}
\affiliation{School of Physics, Korea Institute for Advanced Study (KIAS), 85 Hoegiro, Dongdaemun-gu, Seoul, 02455, Korea}
\author{Saeed Pourojaghi}
\affiliation{Department of Physics, Bu-Ali Sina University, Hamedan
65178, Iran}
\affiliation{School of Physics, Institute for Research in Fundamental Sciences (IPM), P.O.Box 19395-5531, Tehran, Iran}
\begin{abstract}
In this work, we present a study on cosmological constraints of dark energy (DE) parametrizations after the observations of Dark Energy Spectroscopic Instruments (DESI) for Baryon Acoustic Oscillations (BAO) in 2024, suggesting potential deviations from the standard flat-$\Lambda$CDM cosmology. This study aims to put observational constraints on equation of state (EoS) parametrizations beyond the standard $\Lambda$CDM model using DESI BAO 2024 data, Cosmic Microwave Background (CMB) anisotropy observations 2018, and various Supernovae type Ia (SNIa) compilations: PantheonPlus, Union3, and Dark Energy Survey 5-year photometrically classified SNIa (DES-SN5YR). Our main goal is to validate the result of DESI collaborations \cite{DESI:2024mwx} in the context of some parametrizations beyond the Chevallier–Polarski–Linder (CPL) approximation known as Barboza-Alcaniz (BA) and Padé parametrizations. These investigations allow us to examine the potential biases in the results presented by DESI collaborations with respect to CPL parametrization. By comparing our findings using BA and Padé parametrizations to the CPL case, we indicate that the deviation from $w_{\Lambda} = -1.0$, observed in the CPL parametrization, is also evident in the BA and Padé parametrizations.

\end{abstract}
\maketitle

\section{Introduction} \label{sec:intro}
A multitude of independent astronomical observations have corroborated the accelerated expansion of the Universe. This expansion is evidenced by data from Type Ia supernovae (SNIa), which serve as 'standard candles' for measuring cosmic distances \citep{Riess:1998cb,Perlmutter:1998np,Kowalski:2008ez}. Complementary to this, the cosmic microwave background (CMB)  provides a snapshot of the early Universe, offering insights into its subsequent evolution \citep{Komatsu2009,Jarosik:2010iu,Ade:2015rim}. Weak gravitational lensing \citep{Benjamin:2007ys,Amendola:2007rr,Fu:2007qq}, baryon acoustic oscillations (BAO), and the large-scale structure of the Universe\citep{Tegmark:2003ud,Cole:2005sx,Eisenstein:2005su,Percival2010,Blake:2011en,Reid:2012sw} are further substantiation of this phenomenon.  Additionally, distributions of High redshift galaxies and galaxy clusters  add another layer of evidence of accelerated phase of expansion \citep{Bunker_2010}.\\
Within the general relativity (GR) framework, the observed cosmic acceleration is attributed to a mysterious component with negative pressure,  widely referred to as dark energy (DE). A prevalent theory posits that  DE consists of vacuum energy, also known as the Cosmological Constant $\Lambda$,  which is denoted by a constant EoS parameter $w_{\Lambda}=-1$ 
\citep{Peebles:2002gy}. This Cosmological Constant, in conjunction with Cold Dark Matter (CDM), constitutes the cornerstone of the standard cosmological model, known as the $\Lambda$CDM model. Theoretically, the $\Lambda$CDM framework grapples with issues such as the 'fine-tuning' and 'cosmic coincidence' conundrums  
\citep{Weinberg:1988cp,Sahni:1999gb,Carroll:2000fy,Padmanabhan:2002ji}. Observationally, it faces puzzles like the 'Hubble tension' and the '$S_8$ tension', which question the model's completeness within the standard $\ Lambda$-cosmology. To see a recent review addressing the observational tensions within the $\Lambda$CDM cosmology, we refer to \citep{Perivolaropoulos:2021jda}.
The persistent puzzles within the standard cosmological model have propelled cosmologists to explore alternatives that extend beyond the standard $\Lambda$-cosmology. Two principal suggestions are being pursued: one posits a dynamic energy density for DE with pronounced negative pressure, consistent with the principles of general relativity (GR); the other proposes a fundamental revision of GR itself, through the window of modified gravity (MG) theories (For a review, see \citep{Tsujikawa:2010zza}). For historical context and initial explorations into time-varying energy densities, see seminal works by \citep{Copeland:2006wr,Caldwell:1997ii,Armendariz-Picon:2000ulo,Caldwell:1999ew,Padmanabhan:2002cp}.
Accurate measurements of EoS parameter and its evolution over cosmic time are crucial for unraveling the dynamic behavior and intrinsic nature of DE \cite{Copeland:2006wr,Frieman:2008sn,Weinberg_2013}. Such insights are often introduced through parametrization methods. The literature presents many of EoS parametrizations, among which the CPL parametrization stands out: $ w_{de} = w_0 + w_a (1 - a) = w_0 + w_a \frac{z}{1+z} $, originally proposed by \citep{Chevallier:2000qy,Linder:2002et}. The CPL model is a Taylor series expansion around the present scale factor and has been generalized to include a second-order term: $ w_{de} = w_0 + w_a (1 - a) + w_b (1 - a)^2 $ \citep{SDSS:2004kqt}.

Beyond the CPL, various phenomenological parametrizations have emerged, such as power-law models like $ w_{de}(z) = w_0 + w_a \frac{z}{(1+z)^2}$ \citep{Jassal:2004ej}, $w_{de} (a) = w_0 + w_a (1-a^\beta)/\beta$ \citep{Barboza:2009ks}, and logarithmic forms such as $ w_{de}(a) = w_0 + w_a \ln{a}$ \cite{Efstathiou:1999tm}, $w_{de}(z) = w_0 [1 + b \ln{1+z}]^\alpha$ \citep{Wetterich:2004pv}, each with their own merits and limitations.  An alternative to the CPL is the Padé parametrization \citep{Rezaei:2017yyj}, which is mathematically robust, remaining finite across all future redshifts (see Sect. \ref{sect:EoSparam}). 
Recent studies have underscored the limitations of dynamical DE parametrizations. For instance, in general parametrization form $w_{de}(z) = w_0 + w_1 f(z)$, the authors of \citep{Colgain:2021pmf} demonstrated that the precision of the measurement of $w_1$ parameter from observational constraints, which dictates the EoS variation, depends on the behavior of the function $f(z)$ that modulates the EoS evolution. A rapid increase in $f(z)$ with redshift leads to smaller uncertainties in $ w_1$, and conversely. This negative correlation between $w_1$ and $f(z)$ could potentially challenge our interpretations of the dynamical DE parametrization. So, in this concern, it would be beneficial to examine any conclusion in a given DE parametrization by other parametrizations.

The concerns raised about the limitations of dynamical DE parametrizations are indeed reflected in recent studies \citep{Colgain:2021pmf}. The general form $ w_{de}(z) = w_0 + w_1 f(z)$ highlights the importance of the function $ f(z) $ in determining the precision of the EoS parameter $w_1$ from observational data. As noted, a rapid increase in $f(z)$ with redshift can lead to smaller uncertainties in $w_1$, suggesting a negative correlation between them.

This relationship implies that the choice of function $f(z)$ is crucial and can significantly affect the interpretation of the dynamical DE parametrization. Therefore, it's essential to cross-check conclusions drawn from one DE parametrization against other parametrizations to ensure robustness. This approach can help mitigate the risk of misinterpretation due to the dependency of measurement precision on the behavior of $ f(z)$. In summary, the dynamical nature of DE and its parametrization present challenges that require careful consideration of the functions used to model its behavior, as well as validation across different parametrizations to ensure accurate interpretations of cosmological data.

The recent cosmological results from the Dark Energy Spectroscopic Instrument (DESI) \citep{DESI:2024mwx} have indeed been groundbreaking. DESI's Stage IV survey is designed to enhance cosmological constraints significantly by measuring the clustering of galaxies, quasars, and the Lyman-$\alpha$ forest across a vast area of the sky and a wide redshift range. The first year of observations has provided robust measurements of the transverse comoving distance and Hubble rate in seven redshift bins, utilizing data from over 6 million extragalactic objects. The DESI BAO Data Release 1 (DR1) alone is consistent with the standard flat-$\Lambda$CDM cosmological model, indicating a matter density of $\Omega_{m0} = 0.295 \pm 0.015$.
 When combined with a BBN prior and the acoustic angular scale from the CMB, DESI constrains the Hubble constant to $H_0= 68.52\pm0.62$ km/s/Mpc. In conjunction with CMB anisotropies from Planck and CMB lensing data from Planck and ACT, the results are $\Omega_m=0.307\pm 0.005$ and $H_0= 67.97\pm0.38$ km/s/Mpc \cite{DESI:2024mwx}.

Furthermore, when extending the baseline model to include a constant EoS parameter $w$, DESI BAO data alone require $w=-0.99^{+0.15}_{-0.13}$. In CPL parametrization $w_{de} = w_0 + w_a \frac{z}{1+z}$, the combinations of DESI with CMB and SNIa (DESI BAO + CMB + SNIa) individually prefer $w_0>-1$ and $w_a<0$. This preference is statistically significant and persists for different PantheonPlus \cite{Scolnic:2021amr}, Union3 \cite{Rubin:2023ovl} and DES-SN5YR  \cite{DES:2024tys} samples \cite{DESI:2024mwx}.

Building upon the work of \cite{DESI:2024mwx}, our study in this work, aims to investigate different DE parametrizations to study the potential deviation from flat-$\Lambda$CDM cosmology beyond the CPL parametrization, by incorporating various sets of observational data, including DESI BAO 2024 \cite{DESI:2024mwx}, CMB anisotropy observations from \cite{Planck:2018vyg}, and SNIa compilations from PantheonPlus\cite{Scolnic:2021amr}, Union3 \cite{Rubin:2023ovl}, and DES-SN5YR \cite{DES:2024tys}. The use of rational Padé and BA parametrizations allows for a more general approach to parametrize the EoS parameter of DE. Our study can validate the potential deviation from the standard flat-$\Lambda$CDM cosmology reported in  \cite{DESI:2024mwx}, if there is significant evidence to rule out the $\Lambda$CDM cosmology in different DE parametrizations considered in this work.
Generally, our research is crucial as it could provide clarity on whether the $\Lambda$CDM model remains the preferred description of our Universe or if alternative DE parametrizations are more consistent with the observational data.

The structure of this paper is organized as follows: Section (\ref{sect:EoSparam}) details some important features of DE parametrizations used in our study. In Section (\ref{sec:data}), we introduce the observational data and statistical tools used in this work. In Section (\ref{sec:data-analysis}), we use Bayesian inference to obtain constraints on our choice DE parametrizations using DESI BAO 2024,
Planck CMB anisotropic observations 2018, and three SNIa compilations: PantheonPlus, Union3, and DES-SN5YR. Finally, in Section (\ref{conlusion}), we summarize our findings and conclude the study.

\section{DE parametrizations} \label{sect:EoSparam}

As is widely recognized, CPL parametrization serves as a baseline model for dynamical DE parametrizations and is often regarded as a minimal framework in many analyses. 
In this work, in the context of two generalizations of the CPL parametrization, i.e., the rational Padé and the BA parametrizations, we explore the possibility of deviations from the $\Lambda$CDM cosmology, as is previously seen in CPL parametrization \cite{DESI:2024mwx}. Indeed, BA and Padé parametrizations allow us to test the sensitivity of the DESI results \cite{DESI:2024mwx} to the variation of the form of DE parametrization.
\subsection {Padé approximation}
For an arbitrary function $f(z)$, the Padé approximation of order $(m,n)$ is represented by the following rational function \citep{baker96}:
\begin{equation}\label{PadéO}
	f(z) = \frac{a_0 + a_1z + a_2z^2 + \ldots + a_nz^n}{b_0 + b_1z + b_2z^2 + \ldots + b_mz^m},
\end{equation}
where the exponents $(m,n)$ are positive integers, and the coefficients $a_i$, $b_i$ are constants. Notably, when $b_i = 0$ for $i \geq 1$, the Padé approximation simplifies to the standard Taylor expansion.

The Padé parametrization for the EoS parameter of DE is introduced as follows \citep{Wei:2013jya}:

\subsubsection{Padé parametrization}\label{sect:subPadé}
Based on Eq.(\ref{PadéO}), we first expand the EoS parameter $w_{\rm de}$ with respect to the scale factor up to order $(1,1)$ as follows \citep[see also][]{Wei:2013jya}:
\begin{equation}\label{Padé1}
	w_{\rm de}(a) = \frac{w_0 + w_1(1 - a)}{1 + w_2(1 - a)}.
\end{equation}
Henceforth, we refer to the above formula as Padé parametrization. In terms of redshift $z$, Eq.~(\ref{Padé1}) is expressed as
\begin{equation}\label{Padé1b}
	w_{\rm de}(z) = \frac{w_0 + \frac{w_1 z}{1 + z}}{1 + \frac{w_2 z}{1 + z}}.
\end{equation}
As expected, for $w_2 = 0$, Eq.~(\ref{Padé1}) reduces to the CPL parametrization.
Utilizing Eq.~(\ref{Padé1}) or Eq.~(\ref{Padé1b}), we identify the following special cases for the EoS parameter \citep[see also][]{Wei:2013jya}:
\begin{equation}\label{limit}
	w_{\rm de} =  \left\{ \begin{array}{rcl}
\frac{w_0 + w_1}{1 + w_2}, & \mbox{for}
&  a \rightarrow  0 (z \rightarrow  \infty, \mbox{early time}),\\
		w_0, & \mbox{for}
& a = 1 \quad (z = 0, \mbox{present}),\\
		\frac{w_1}{w_2}, & \mbox{for}
& a \rightarrow  \infty  (z \rightarrow  -1,\mbox{far future}),
\end{array}\right.
\end{equation}
where $w_2$ must not equal $0$ or $-1$. We observe that in all limiting cases of Eq.(\ref{limit}), the EoS parameter attains constant values. These values, particularly at the early and present epochs of the Universe, may coincide with the fixed $ w_{\Lambda} = -1$ or differ from it. This depends on how observational data put constraints on the parameters.
We can trace the evolution of the EoS from its early Universe value $\frac{w_0 + w_1}{1 + w_2}$ to its present value $w_0$ and assess whether the EoS deviates from $w_{\Lambda} = -1$. We can also observe that the CPL case evolves from the initial constant value in the early Universe is  $w_0 + w_a$, to the present constant value  $w_0$. It is noteworthy that, similar to the standard flat-$\Lambda$CDM cosmology in which the Universe will be $\Lambda$-dominated in the far future, in a Universe described by Padé parametrization, $w_{de}$ tends to a constant value $\frac{w_1}{w_2}$ in the far future. Consequently, the energy density of DE also tends to a constant value. This feature is not observed in the CPL parametrization, where $w_{de}$ diverges in the far future, leading to potentially more complex effects of DE on the dynamics of the future Universe.
\\
\subsubsection{Simplified Padé}
The Padé approximation has three free parameters  $w_0$, $w_1$, and $w_2$. By setting $w_1 = 0.0$, we obtain a simplified version of Padé, namely

\begin{equation}\label{Padésimp}
	w_{\rm de}(a) = \frac{w_0}{1 + w_2(1 - a)}.
\end{equation}

To avoid singularities in cosmic expansion, $w_2$ must not be equal to $-1.0$. This parametrization simplifies to $\frac{w_0}{1 + w_2}$ in the early Universe and converges to $w_0$ at the present time. Thus, the EoS can evolve from the initial constant value $\frac{w_0}{1 + w_2}$ to the present constant value $w_0$. 
\begin{figure}
	\centering
	\includegraphics[width=8cm]{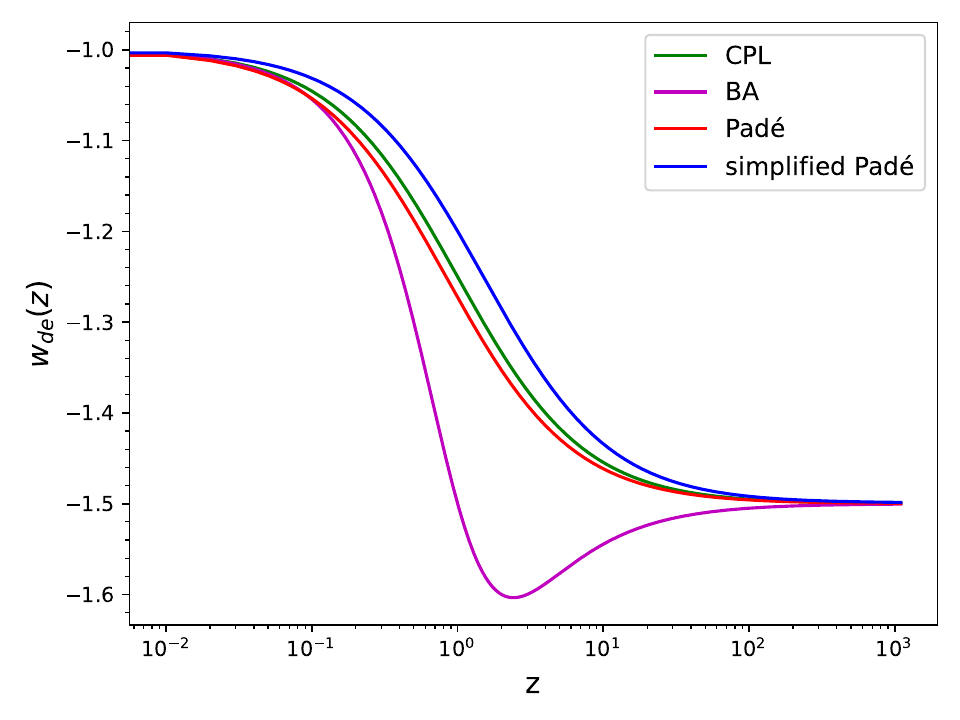}
	\includegraphics[width=8cm]{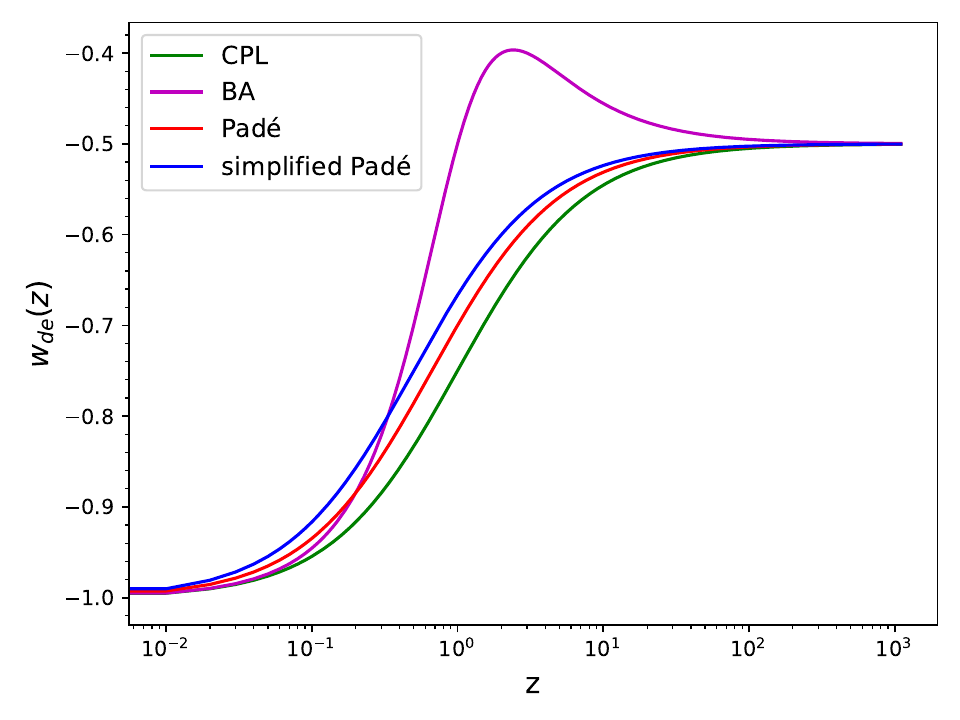}
	\caption{Evolution of the EoS parameters for CPL, BA, Simplified Padé, and Padé parametrizations across redshifts in both phantom (upper panel) and quintessence (lower panel) regimes. In the phantom regime, the numerical values are set as follows: $w_0 = -1.0$ and $w_a = -0.5$ for the CPL and BA parametrizations; $w_0 = -1.0$, $w_1 = -0.8$, and $w_2 = +0.2$ for the Padé parametrization; $w_0 = -1.0$ and $w_2 = -0.333$ for the simplified Padé parametrization. In the quintessence regime, the numerical values are set as $w_0 = -1.0$ and $w_a = +0.5$ for the CPL and BA parametrizations; $w_0 = -1.0$, $w_1 = +0.25$, and $w_2 = +0.5$ for the Padé parametrization; $w_0 = -1.0$ and $w_2 = +1.0$ for the simplified Padé parametrization.}
	\label{fig:EoS}
\end{figure}

\begin{figure}
	\centering
	\includegraphics[width=8cm]{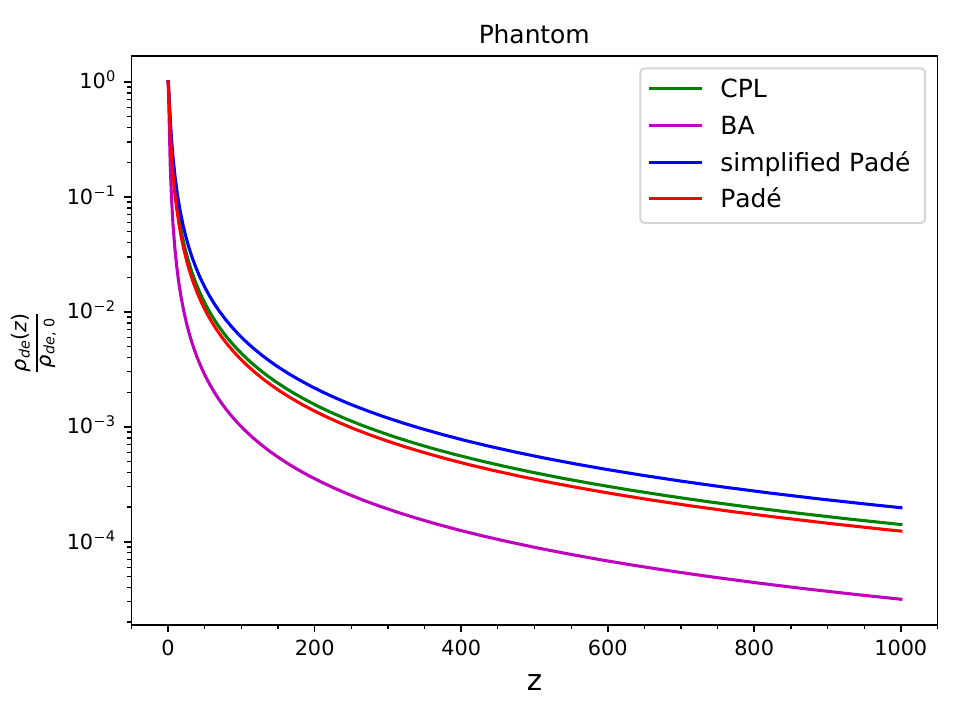}
	\includegraphics[width=8cm]{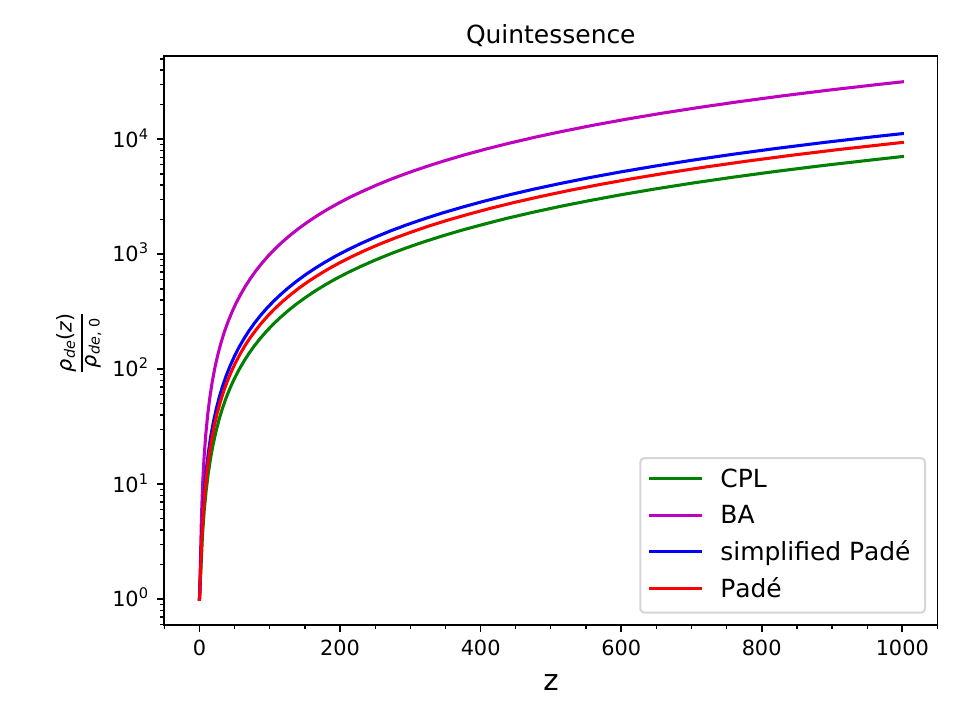}
	\caption{The redshift evolution of the relative energy density of DE for CPL, BA, Simplified Padé and Padé parametrizations in both phantom (upper panel) and quintessence (lower panel) regimes. The numerical values of the various DE parametrizations for both phantom and quintessence regimes are set as values used in Figure (\ref{fig:EoS}). }
	\label{fig:density}
\end{figure}
\subsection{BA parametrization} 
The CPL parametrization, given by $w_{de}(z) = w_0 + w_a \frac{z}{1+z}$, can be generalized as $w_{de}(z) = w_0 + w_a \frac{z(1+z)^{n-1}}{1+z^n}$. For the general EoS parameter, setting $n=1$ recovers the CPL form, while $n=2$ yields the BA parametrization, as introduced by \citep{Barboza_2008}. Figure (\ref{fig:EoS}) illustrates the evolution of the EoS parameter for all parametrizations as a function of redshift. In the upper (lower) panel, the numerical values, as described in the caption of the Figure, are set to show the phantom (quintessence) behavior of DE.
With these numerical values, the EoS parameter for all cases starts from the same initial values: $w_{de} = -1.5$ (upper panel) and $w_{de} = -0.5$ (lower panel) in the early universe and reaches the $\Lambda$CDM value $w_{de} = -1.0$ at the present time. This separates the phantom and quintessence regimes in the upper and lower panels, respectively. We explicitly observe the effect of the functional form $f(z)$, especially in the BA parametrization, at low redshift. 
In \cite{Colgain:2021pmf}, the authors demonstrated that the shape of the evolutionary function, $f(z)$, can affect the estimated value of $w_a$ when observational constraints are applied to these parameters.

\subsection{DE density and Hubble parameter}
The energy density of DE in each parametrization can be determined using the corresponding equation of state function, $w_{DE}(z)$, as follows:
\begin{equation}\label{eq:density}
	\rho_{DE}(z) = \rho_{DE,0} \exp \left( 3 \int_0^z \frac{1+w_{DE}(z')}{1+z'} dz' \right)
\end{equation}

Substituting the CPL EoS parameter into the equation, we get:
\begin{equation}
	\rho_{DE}^{CPL}(z) = \rho_{DE,0} (1+z)^{3(1+w_0+w_a)} \exp \left( \frac{-3w_a z}{1+z} \right)
\end{equation}

The energy density of DE, $\rho_{DE}(z)$, in the BA model can be expressed as:
\begin{equation}
	\rho_{DE}^{BA}(z) = \rho_{DE,0} (1+z)^{3(1+w_0)} (1+z^2)^{\frac{3w_a}{2}}
\end{equation}

Additionally, for the Padé parametrization and its simplified version, we obtain the following relationships:
\begin{equation}\label{eq:Rho_Padé}
	\begin{split}
		\rho_{DE}^{\text{Padé}}(z) &= \rho_{DE,0} (1+z)^{3\left(\frac{1+w_0+w_1+w_2}{1+w_2}\right)} \\
		&\quad \times \left[1+w_2\left(\frac{z}{1+z}\right)\right]^{-3\left(\frac{w_1-w_0w_2}{w_2(1+w_2)}\right)}
	\end{split}
\end{equation}

\begin{equation}
	\begin{split}
		\rho_{DE}^{simp \: \text{Padé}}(z) &= \rho_{DE,0} (1+z)^{3\left(\frac{1+w_0+w_2}{1+w_2}\right)} \\
		&\quad \times \left[1+w_2\left(\frac{z}{1+z}\right)\right]^{3\left(\frac{w_0}{(1+w_2)}\right)}
	\end{split}
\end{equation}
In general, dynamical DE models should be capable of diagnosing any deviation from the standard cosmological constant $\Lambda$ using observational data. In Figure (\ref{fig:density}), we plot the redshift evolution of the energy densities of the DE parametrizations studied in our analysis. The numerical values for all DE parametrizations are set to match those selected in Figure (\ref{fig:EoS}). We observe that the BA parametrization, in both the quintessence and phantom regimes, exhibits more dynamical behavior compared to the constant energy density $\rho_{\Lambda}$ and other parametrizations.\\
Ultimately, by applying the aforementioned equations along with the first Friedmann equation, we can compute the dimensionless Hubble parameter, $E(z) = \frac{H(z)}{H_0}$, for each parametrization as follows: 
\begin{equation}
	\begin{split}
		E^2_{CPL}(z) &= \Omega_{m,0} (1+z)^3  + \Omega_{r,0} (1+z)^4 + \Omega_{DE,0}\\
		&\quad \times (1+z)^{3(1+w_0+w_a)} \exp \left( \frac{-3w_a z}{1+z} \right)
	\end{split}
\end{equation}

\begin{equation}
	\begin{split}
		E^2_{BA}(z) &= \Omega_{m,0} (1+z)^3 + \Omega_{r,0} (1+z)^4 + \Omega_{DE,0}\\
		&\quad \times (1+z)^{3(1+w_0)} (1+z^2)^{\frac{3w_a}{2}}
	\end{split}
\end{equation}

\begin{equation}\label{eq:E_Padé}
	\begin{split}
		E^2_{\text{Padé}}(z) = \Omega_{m,0} (1+z)^3 + \Omega_{r,0} (1+z)^4 + \Omega_{DE,0}\\	
		\times(1+z)^{3\left(\frac{1+w_0+w_1+w_2}{1+w_2}\right)} 
		\left[1+w_2\left(\frac{z}{1+z}\right)\right]^{-3\left(\frac{w_1-w_0w_2}{w_2(1+w_2)}\right)}
	\end{split}
\end{equation}

\begin{equation}
	\begin{split}
		E^2_{simp \: \text{Padé}}(z) = \Omega_{m,0} (1+z)^3 + \Omega_{r,0} (1+z)^4 + \Omega_{DE,0}\\ 
		\times(1+z)^{3\left(\frac{1+w_0+w_2}{1+w_2}\right)} 
		\left[1+w_2\left(\frac{z}{1+z}\right)\right]^{3\left(\frac{w_0}{1+w_2}\right)}
	\end{split}
\end{equation}

where $\Omega_{m,0}$, $\Omega_{r,0}$ and $\Omega_{DE,0} = 1 - \Omega_{m,0} - \Omega_{r,0}$ are the present-day density parameters for matter, radiation, and DE respectively. In the next parts, we present the results of our analysis.

\section {Observational data and Statistical analysis}
\label{sec:data}
For our observational datasets, we incorporate the first year's baryon acoustic oscillation (BAO) measurements from DESI \citep{DESI:2024mwx}, Measurements of the Planck CMB temperature anisotropy and polarization power spectra \cite{Planck:2018vyg,Planck:2019nip}, their cross-spectra, and the combination of the ACT \cite{ACT:2023kun,ACT:2023dou} and Planck lensing power spectrum \cite{Carron:2022eyg}, and supernova luminosity distance data from the Dark Energy Survey (DES) Year 5 release \citep{DES:2024tys}, as well as from the PantheonPlus\citep{Scolnic:2021amr} and Union3 \citep{Rubin:2023ovl} compilations. To compare the theoretical predictions of various DE parametrizations against observations, we modify the publicly available cosmological code \texttt{CAMB} \cite{Howlett_2012}. To find the posterior distributions of cosmological parameters, we perform Markov Chain Monte Carlo (MCMC) sampling using the publicly available \texttt{Cobaya} sampler \cite{Lewis:2013hha}. The convergence of the MCMC chains is evaluated using the Gelman-Rubin statistic $R-1$ \cite{Gelman:1992zz}. We consider the chains to have converged if 
$R-1<0.01$ for all models and datasets.
\subsection{DESI BAO}
Baryonic acoustic oscillations (BAO) refer to the fluctuations in the density of baryonic matter, which are the result of acoustic density waves within the early Universe's primordial plasma. The related measurements are reported in reference \citep{DESI:2024mwx}. This source provides data and correlations for the comoving distance during the drag epoch, represented as $\frac{D_M}{r_d}$, and the distance variable $\frac{D_H}{r_d}$. The term $r_d$ signifies the drag epoch, indicating the maximum distance that sound waves could have traveled from the moment of the Big Bang until the point when baryons decoupled. In scenarios where the signal-to-noise ratio is low, the averaged quantity $\frac{D_V}{r_d}$ is utilized instead. It's important to note that when using DESI BAO data exclusively, only the combined value of $r_dH_0$ can be determined. However, by combining DESI BAO data with additional observational datasets, we gain the ability to distinguish between $r_d$ and $H_0$ independently.

In this study, we utilized the DESI DR1 BAO measurements within the redshift range of $0.3 \leq z \leq 2.33$. As indicated in Table (\ref{Tab:BAO_data}), there are two isotropic BAO datasets: Bright Galaxy Survey (BGS) at an effective redshift $z_{eff}=0.30$, and Quasar (QSO) at $z_{eff}=1.49$. Additionally, the anisotropic BAO datasets comprise ten data points, including Luminous Red Galaxy (LRG) at $z_{eff}$ of 0.51 and 0.71, LRG+ELG at $z_{eff}=0.93$, Emission Line Galaxies (ELG) at $z_{eff}=1.32$, and Lyman-$\alpha$ quasars (Lya QSOs) at $z_{eff}=2.33$.

\begin{table}
	\centering
	\caption{DESI BAO data used in this work. Note that for each sample, we have provided either both $\frac{D_M}{r_d}$ and $\frac{D_H}{r_d}$, which are correlated with a coefficient $r$, or $\frac{D_V}{r_d}$ \citep{DESI:2024mwx}.}
	\begin{tabular}{c c c c c}
		\hline
		tracer  & $z_{eff}$ & $D_M/r_d$ & $D_H/r_d$ & $r$ or $D_V/r_d$ \\
		\hline 
		BGS & $0.30$ & $-$ & $-$ & $7.93\pm 0.15$ \\
		LRG & $0.51$ & $13.62\pm 0.25$ & $20.98\pm 0.61$ & $-0.445$ \\
		LRG & $0.71$ & $16.85\pm 0.32$ & $20.08\pm 0.60$ & $-0.420$ \\
		LRG + ELG & $0.93$ & $21.71\pm 0.28$ & $17.88\pm 0.35$ & $-0.389$ \\
		ELG & $1.32$ & $27.79\pm 0.69$ & $13.82\pm 0.42$ & $-0.444$ \\
		QSO & $1.49$ & $-$ & $-$ & $26.07\pm 0.67$ \\
		Lya QSO & $2.33$ & $39.71\pm 0.94$ & $8.52\pm 0.17$ & $-0.477$ \\
		\hline
	\end{tabular}\label{Tab:BAO_data}
\end{table}

\subsection{Cosmic Microwave Background (CMB)}
The Cosmic Microwave Background (CMB) radiation provides a window into the Universe at the time of photon decoupling, acting as a crucial instrument for examining cosmological models. We perform comprehensive likelihood calculations using the final full-mission Planck measurements. The Planck CMB likelihoods employed in this work are as follows:
\begin{itemize}
\item[(i)] Measurements of the power spectra of temperature and polarization anisotropies, $C_{\ell}^{TT}$, $C_{\ell}^{TE}$, and $C_{\ell}^{EE}$, at small scales ($\ell > 30$), obtained using the Planck \texttt{plik} likelihood~\cite{Planck:2018vyg,Planck:2019nip};
\item[(ii)] Measurements of the temperature anisotropy spectrum, $C_{\ell}^{TT}$, at large scales ($2 \leq \ell \leq 30$), obtained using the Planck \texttt{Commander} likelihood~\cite{Planck:2018vyg,Planck:2019nip};
\item[(iii)] Measurements of the E-mode polarization spectrum, $C_{\ell}^{EE}$, at large scales ($2 \leq \ell \leq 30$), obtained using the Planck \texttt{SimAll} likelihood~\cite{Planck:2018vyg,Planck:2019nip};
\item[(iv)] Reconstruction of the lensing potential spectrum, obtained using the Planck PR4 \texttt{NPIPE} data release~\cite{Carron:2022eyg} in combination with the \texttt{ACT-DR6} lensing likelihood~\cite{ACT:2023kun,ACT:2023dou}.
\end{itemize}
\subsection{Type Ia Supernova (SNIa)}
SNIa serve as 'standard candles' due to their consistent intrinsic brightness. By measuring their apparent brightness and redshift, cosmologists can derive precise distance estimation and offer an alternative way to measure the expansion history of the Universe.
In this work, we analyze three samples of SNIa, each offering valuable data for our study. These include the PantheonPlus sample \citep{Scolnic:2021amr} and the Union3 sample \citep{Rubin:2023ovl}, both of which are compilations of SNIa events from various sources. While these two samples are almost identical, their analysis methodologies differ significantly. The third sample is the DES-SN5YR sample from the Dark Energy Survey \citep{DES:2024tys}. The DES-SN5YR sample is nearly independent of the PantheonPlus and Union3 samples, but it employs the same analysis method as PantheonPlus.
\subsubsection{PantheonPlus} 
The PantheonPlus sample represents the most recent collection of spectroscopically confirmed Type Ia supernovae (SNIa), analyzed across the entire redshift spectrum from $z = 0$ to $z = 2.3$. It comprises 1701 light curves from confirmed SNIa, with 77 originating from galaxies that are Cepheid hosts within the low redshift interval of $0.00122 \leq z \leq 0.01682$, utilized in deriving cosmological parameters within the Pantheon SNIa analysis and the SH0ES distance-ladder studies. The PantheonPlus compilation includes 18 distinct datasets, with all associated data, references, and their respective redshift ranges detailed in Table 1 of \citep{Scolnic:2021amr}. In comparison to the earlier Pantheon dataset \citep{Scolnic:2017caz}, the most substantial augmentation in SNIa data within the PantheonPlus catalogue is observed at lower redshifts, attributed to the inclusion of the LOSS1, LOSS2, SOUSA, and CNIa0.2 samples.

\subsubsection{DES-SN5YR} 
The Dark Energy Survey Supernova Program sample (DES-SN5YR) is the largest and deepest single sample survey consisting of 1635 SNe ranging in redshift from 0.10 to 1.13 from the full DES survey and is complemented by 194 spectroscopically confirmed low-redshift SNIa.
DES-SN5YR refers to the data collected over the first five years of the DES Supernova Program. This extensive dataset is critical for probing the properties of DE and its influence on the expansion history of the Universe. Over the five years, the survey detected over 3000 supernova candidates, with approximately 2000 confirmed as Type Ia supernovae. These supernovae span a redshift range of $0.01<z<1.2$, providing a robust dataset for cosmological analyses  which contains 1829 supernovae in total:
1635 genuine DES measurements augmented with 194 local SNIa measurements ($z < 0.1$) from the CfA/CSP foundation sample. The DES-SN5YR sample has better high redshift statistics compared to PantheonPlus or Union3 SN \citep{Rubin:2023ovl}.\\

\subsubsection{Union3 SN}
The Union3 Supernova compilation is a comprehensive dataset of 2087 Type Ia supernovae from 24 dataset samples ranging from redshift $0.01$ to $2.26$, standardized on a consistent distance scale using SALT3 light-curve fitting. The 22 binned distances of Union3 SNIa have been reported in \citep{Rubin:2023ovl}.
\begin{table*}
\caption{The best-fit values of cosmological parameters with their $1\sigma$ uncertainty obtained using the various combinations of the observational data in CPL parametrization. The values of $\Delta\chi^2$ and the corresponding deviation values indicating the preference for evolving DE scenarios are provided in the last two columns, respectively.}
\begin{tabular}{|l| c c c c c c|}
\hline \hline
Data  & $\Omega_{m0}$ & $H_0$[km/s/Mpc] & $w_0$ & $w_a$&$\Delta \chi^2$&Deviation \\
\hline
DESI BAO & $0.346^{+0.043}_{-0.027}$ & $-$ & $-0.53^{+0.37}_{-0.21}$ & $-1.72^{+0.23}_{-0.97}$ &$-3.6$&$1.2\sigma$\\ %
\hline 
DESI BAO + CMB & $0.332^{+0.027}_{-0.032}$ & $65.7^{+2.6}_{-3.2}$ & $-0.58^{+0.32}_{-0.29}$ & $-1.44^{+0.87}_{-0.76}$& $-9.0$&$2.5\sigma$\\
\hline
DESI BAO + Union3 & $0.335^{+0.022}_{-0.016}$ & $-$ & $-0.64\pm 0.13$ & $-1.47^{+0.67}_{-0.82}$ &$-10.0$&$2.7\sigma$\\
\hline
DESI BAO + DES-SN5YR & $0.328^{+0.020}_{-0.016}$ & $-$ & $-0.73^{+0.09}_{-0.10}$ & $-1.20\pm 0.65$&$-11.5$&$2.9\sigma$ \\
\hline
DESI BAO + PantheonPlus & $0.302^{+0.030}_{-0.014}$ & $-$ & $-0.87^{+0.06}_{-0.07}$ & $-0.38\pm 0.60$& $-3.3$&$1.1\sigma$\\
\hline
DESI BAO + CMB + Union3 & $0.3224\pm 0.0097$ & $66.56\pm 0.97$  & $-0.65\pm 0.10$ & $-1.28\pm 0.39$ &$-13.4$&$3.2\sigma$\\
\hline
DESI BAO + CMB + DES-SN5YR & $0.3160^{+0.0063}_{-0.0076}$ & $67.22\pm 0.69$ & $-0.74^{+0.07}_{-0.08}$ & $-0.98^{+0.35}_{-0.28}$ &$-18.5$&$3.9\sigma$\\
\hline
DESI BAO + CMB + PantheonPlus & $0.3071^{+0.0065}_{-0.0078}$ & $68.15\pm 0.73$ & $-0.83^{+0.05}_{-0.07}$ & $-0.77^{+0.32}_{-0.21}$&$-9.0$&$2.5\sigma$ \\
\hline \hline        
\end{tabular}\label{Tab:CPL}
\end{table*}
\begin{figure*} 
	\centering
	\includegraphics[width=7cm]{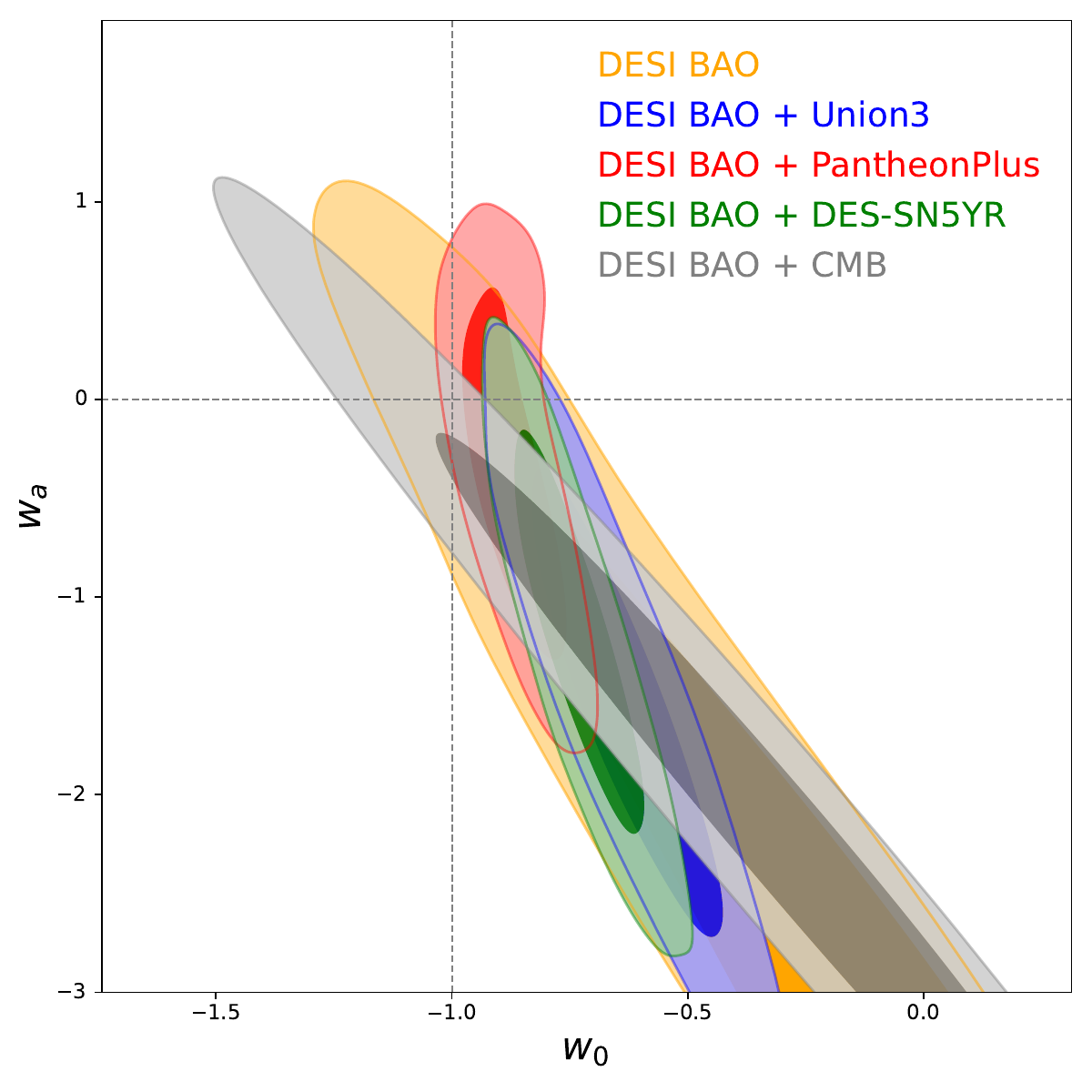}
        \includegraphics[width=7cm]{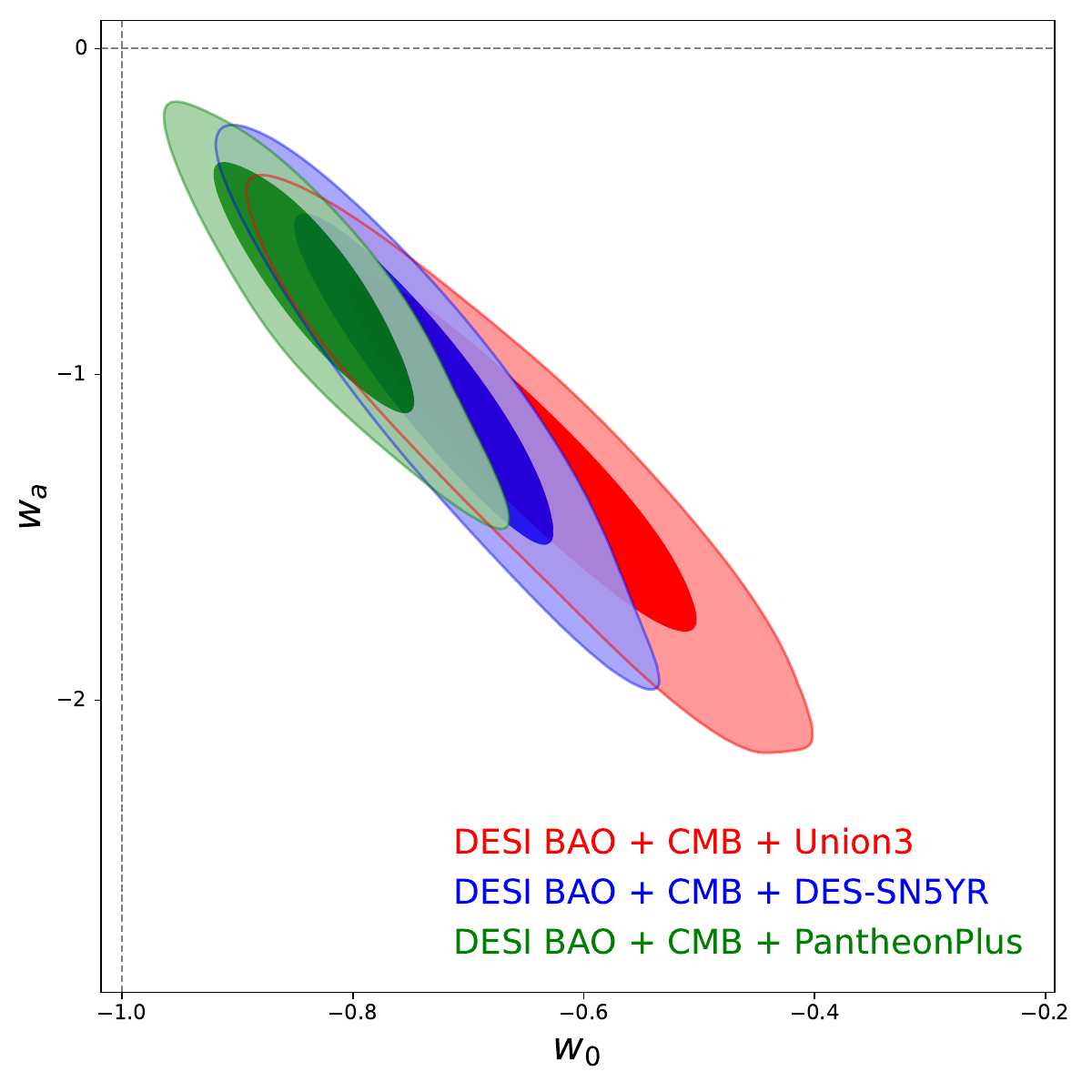}
	\caption{The $1\sigma$ and $2\sigma$ confidence regions of $w_0$ and $w_a$ in the CPL Parametrization using the various combinations of the observational data.}
	\label{fig:con_CPL}
\end{figure*}
\begin{figure*} 
	\centering
	\includegraphics[width=6cm]{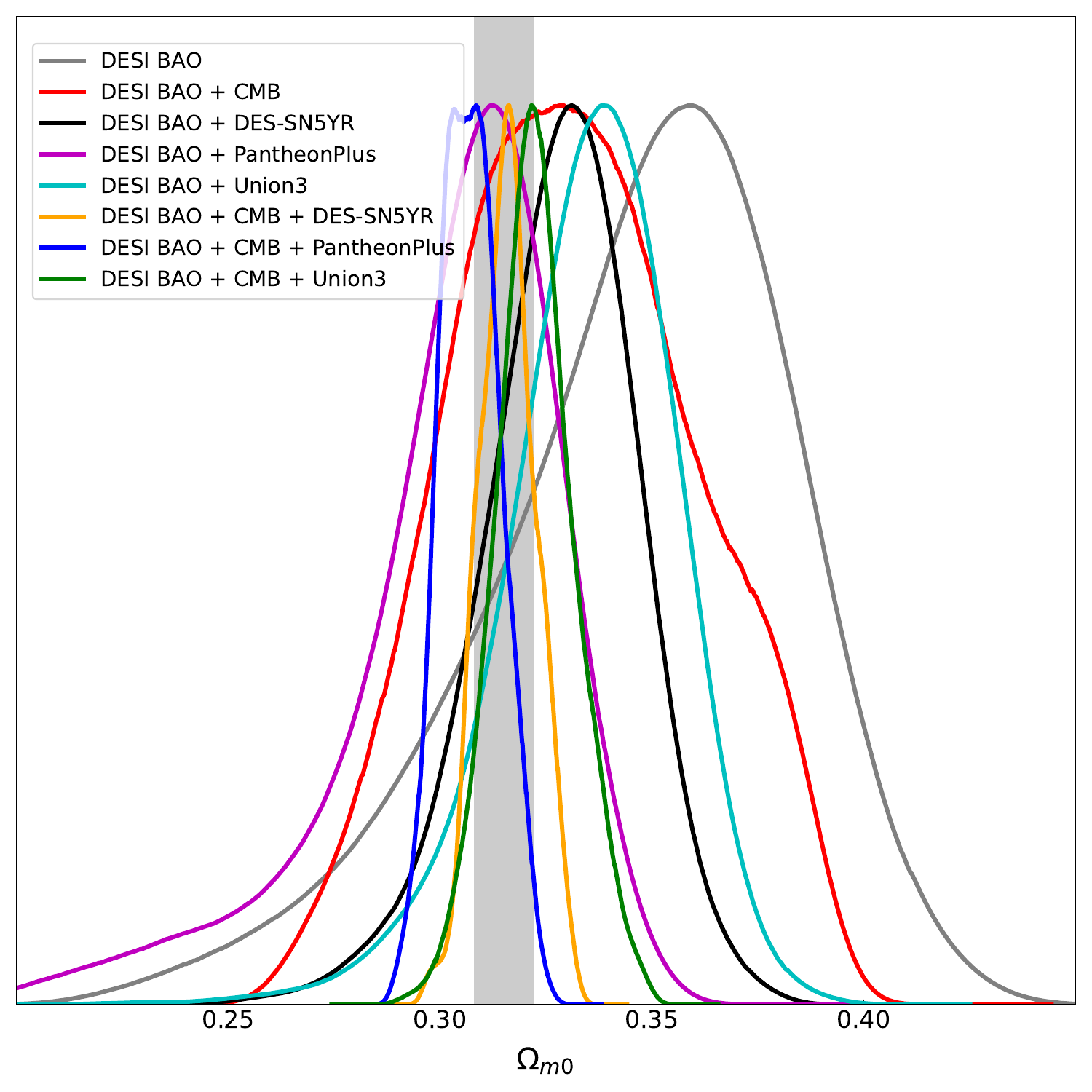}\includegraphics[width=6cm]{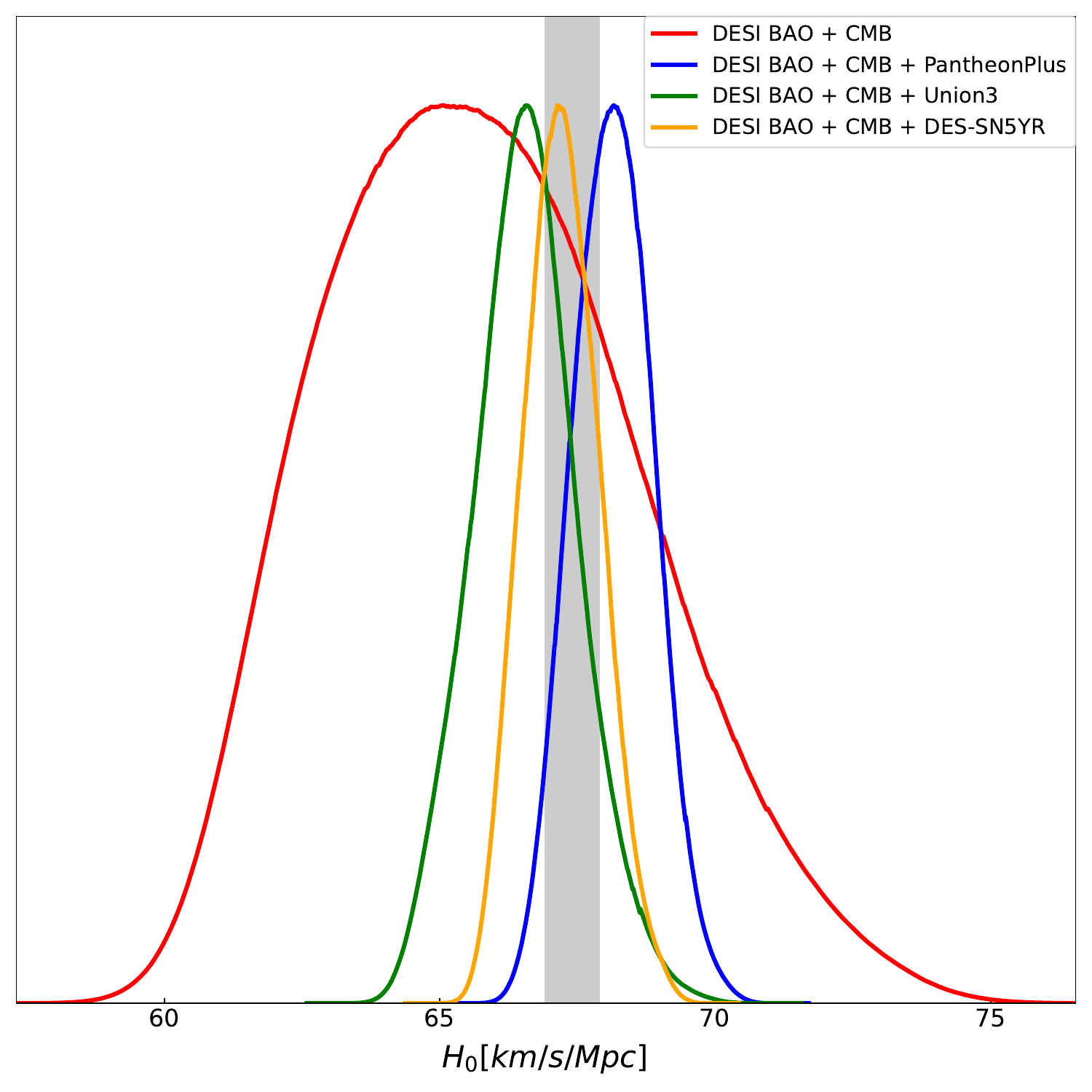}
	\caption{The distribution of $\Omega_{m0}$ in the CPL parametrization using different combinations of data (left). The distribution of $H_0$ using a combination of DESI BAO data with CMB and various SNIa samples (right). Additionally, we show the $1\sigma$ values from Planck measurements \cite{Planck:2018vyg} on $\Omega_{m0}$ (left panel) and $H_0$ (right panel) as vertical gray regions.}
	\label{fig:con_CPL1}
\end{figure*}
\section{Numerical Results}{\label{sec:data-analysis}}
In this section, we will put constraints on the cosmological parameters of the different parametrizations introduced in the section \ref{sect:EoSparam}. In addition, we use $\Delta \chi^2 = -2 \Delta \log \mathcal{L}$ as a statistical tool, defined as twice the difference between the maximum likelihood $\mathcal{L}$ of a given DE parametrization and that of the standard flat-$\Lambda$CDM cosmology. This metric indicates the statistical preference for evolving DE scenarios in the context of DE parametrizations.
To achieve this, we have used eight combinations of the observational data as follows:

	\begin{eqnarray}\label{r1}
		&& \text{DESI BAO}\nonumber\\
		&& \text{DESI BAO + Planck CMB}\nonumber \\
		&&\text{DESI BAO + Union3} \nonumber\\
		&&\text{DESI BAO + DES-SN5YR}\nonumber \\
		&&\text{DESI BAO + PantheonPlus} \\
		&& \text{DESI BAO + Planck CMB + Union3} \nonumber\\
		&&  \text{DESI BAO + Planck CMB + DES-SN5YR} \nonumber\\
		&& \text{DESI BAO + Planck CMB + PantheonPlus} \nonumber
	\end{eqnarray}
We consider flat priors for all cosmological parameters assumed to be free in our analysis, as mentioned in Table 2 of \cite{DESI:2024mwx}. The extended parameters in our study are $w_1$ and $w_2$ in Padé and simplified Padé parametrizations. We chose a flat prior for $w_1$ as $[-3, 2]$, similar to what was considered for $w_a$ in \cite{DESI:2024mwx}, and for $w_2$ as $[-0.99, 2]$ for simplified Padé. Since $w_2$ appears in the denominator of Eq.~\ref{Padésimp}, we avoid considering priors crossing $-1$. Additionally, in the case of Padé parametrization, we limit the prior of $w_2$ to $-1.0 < w_2 < 0.0$ to avoid divergence in Equations \ref{eq:Rho_Padé} and \ref{eq:E_Padé}.
The results for each DE parametrization are presented below. Note that since studying DE parametrizations at the perturbation level is not our main goal, we report results for background expansion parameters, including the pressureless matter density $\Omega_{m0}$, Hubble constant $H_0$, and parameters related to the equation of state (EoS) of various DE parametrizations, i. e., $w_0$, $w_a$ (for CPL and BA), $w_1$ (for Padé), and $w_2$ (for Padé and simplified Padé). We report the $H_0$ parameter when CMB data is included in our combinations.
\subsection{CPL parametrization}
We first analyze the CPL parametrization to validate our methodology against the DESI results reported in \cite{DESI:2024mwx}. This consistency check demonstrates the robustness of our analysis framework and provides confidence in our subsequent investigations of different dark energy parameterizations.
As we know, the CPL parametrization converges to the $\Lambda$CDM model when $w_0 = -1$ and $w_a = 0$. The numerical values of the cosmological parameters in CPL parametrization obtained from different constraints of various dataset combinations are reported in Table (\ref{Tab:CPL}). 
In addition, Figure (\ref{fig:con_CPL}) illustrates the $1\sigma-2\sigma$ confidence region for $w_0$ and $w_a$ within the CPL parametrization, utilizing various combinations of observational data. As we observe, DESI alone does not have sufficient power to impose tight constraints on $w_0$ and $w_a$, such that the corresponding contour intersects the lower limit of our prior for $w_a$. However, when combining DESI with other observational data, we can achieve tighter constraints on $w_0$ and $w_a$. In particular, the combination of DESI BAO with various SNIa samples yields tighter constraints on $w_0$ and $w_a$.
Combining DESI BAO data with CMB and different SNIa samples results in the best-fit values of $w_0$ and $w_a$ deviating from the $\Lambda$CDM model values.
Our constraints on the parameter $w_a$ are crucial because $w_a \neq 0$ indicates the variation of the EoS parameter $w_{de}$ with redshift $z$. 
Indeed, the variation of $w_{de}(z)$ is determined by $w_a$ and modulated by the function $f(z)$. 
The statistical deviation of the CPL parametrization from the standard flat-$\Lambda$CDM cosmology is reported in the last two columns of Table (\ref{Tab:CPL}). We observe small negative values of $\Delta \chi^2$ for the DESI BAO and DESI BAO + PantheonPlus datasets, indicating a weak statistical preference (less than $2\sigma$) for an evolving DE scenario over the $\Lambda$CDM cosmology. In other words, there is no significant evidence for deviations from the standard flat-$\Lambda$CDM model. We also observe moderate evidence ($2\sigma$ to $3\sigma$ level) for potential deviations from the standard flat-$\Lambda$CDM cosmology in the cases of DESI BAO + CMB, and DESI BAO + SNIa (Union3 and DES-SN5YR samples).
This implies that we cannot observe an explicit evolution for $w_{de}$. However, as shown in the right panel of Figure (\ref{fig:con_CPL}), for data combinations including all three datasets, there are significant deviations from the $\Lambda$CDM point $\{w_0=-1.0, w_a=0.0\}$ in the $w_0-w_a$ plane. Statistically, the deviations from flat-$\Lambda$CDM cosmology are $2.5\sigma$ ($\Delta \chi^2=-9.0$), $3.2\sigma$ ($\Delta \chi^2=-13.4$), and $3.9\sigma$ ($\Delta \chi^2=-18.5$), respectively, for DESI BAO + CMB + PantheonPlus, DESI BAO + CMB + Union3, and DESI BAO + CMB + DES-SN5YR combinations. These results are in agreement with \cite{DESI:2024mwx}. We observe a strong statistical preference for evolving DE scenarios over the standard flat-$\Lambda$CDM cosmology for the DESI BAO + CMB + Union3 and DESI BAO + CMB + DES-SN5YR combinations (see also \citep{DESI:2024mwx}). Notice that throughout our analysis of all DE parametrizations, we have quantified deviations using the $\Delta\chi^2$ between each DE parametrization and the $\Lambda$CDM cosmology. This method is more robust than simply measuring the distance of $\{w_0, w_a\}$ from the $\Lambda$CDM point $(-1.0, 0.0)$ in the two-dimensional parameter space. The $\Delta\chi^2$ approach accounts for the full multi-dimensional parameter space, incorporating all degeneracies and correlations among parameters. In contrast, the 2D contours in the $\{w_0, w_a\}$ plane represent only a projected likelihood.  
When marginalized to two dimensions, correlations with other cosmological parameters, such as $\Omega_{m0}$ and $H_0$, can distort the contours, either flattening or shifting them. As a result, deviations from $\Lambda$CDM may appear less significant in the 2D projection than they truly are in the full parameter space. For instance, Table~\ref{Tab:CPL} reports a $2.5\sigma$ deviation for the CPL parametrization from $\Lambda$CDM using the DESI BAO + CMB dataset, whereas Figure~\ref{fig:con_CPL} shows a deviation of less than $2\sigma$ due to this projection effect.
We emphasize that the parameter $w_0$ alone indicates a possible deviation from $\Lambda$CDM cosmology at the present time. Moreover, a significant deviation of $w_a$ from zero indicates the evolution of the EoS parameter with redshift. For completeness, in Appendix \ref{app:apx1} (see the CPL case), we show the redshift evolution of the EoS parameter, $w_{de}$, for various DE parametrizations within the $2\sigma$ confidence level for all dataset combinations considered in our analysis. 
We note that while SNIa data and DESI BAO observations are sensitive to $\Omega_{m0}$, they do not constrain the Hubble constant ($H_0$) as precisely as CMB observations. This is evident from the constraints on $\Omega_{m0}h^2$ in Figure~7 of \cite{DESI:2024mwx}, where DESI BAO alone, DES-SN5YR alone, and their combination yield broad distributions due to the multiplication of $\Omega_{m0}$ by a loose $H_0$ prior (i.e., $[20-100]$ km/s/Mpc). However, when combined with CMB data, these constraints tighten significantly, allowing for precise measurements of both $\Omega_{m0}h^2$ and $H_0$ (see Figure~7 of \cite{DESI:2024mwx}). Accordingly, in our analysis, we report observational constraints on $\Omega_{m0}$ for all datasets, while $H_0$ constraints are provided only for dataset combinations that include CMB anisotropy measurements (see Table~\ref{Tab:CPL} and Figure~\ref{fig:con_CPL1}). The left panel of Figure~\ref{fig:con_CPL1} displays the $\Omega_{m0}$ distributions for all dataset combinations, whereas the right panel shows the $H_0$ distributions for cases involving CMB data.
As evident from the results, the uncertainties on the $\Omega_{m0}$ constraints are larger for DESI BAO alone and for DESI BAO combined with various SNIa compilations. However, when SNIa, BAO, and CMB data are jointly analyzed, the uncertainties on $\Omega_{m0}$ decrease significantly, aligning closely with the Planck-inferred range   
Moreover, all $H_0$ constraints remain consistent with the Planck-inferred values \cite{Planck:2018vyg}. Notably, the combination of SNIa, BAO, and CMB data leads to a substantial reduction in the uncertainties on $H_0$ within the CPL model.
\begin{table*}
\centering
\caption{Same as Table (\ref{Tab:CPL}), but for BA parametrization. }
\begin{tabular}{|l| c c c c c c|}
\hline \hline
Data  & $\Omega_{m0}$ & $H_0$[Km/s/Mpc] & $w_0$ & $w_a$ & $\Delta \chi^2$ & Deviation\\
\hline
DESI BAO & $0.415^{+0.053}_{-0.035}$ & $-$ & $-0.02^{+0.46}_{-0.23}$ & $-2.03^{+0.23}_{-0.97}$&$-5.8$&$1.8\sigma$\\ 
\hline
DESI BAO + CMB & $0.324^{+0.030}_{-0.048}$ & $66.7\pm 3.8$ & $-0.70^{+0.28}_{-0.42}$ & $-0.62^{+0.60}_{-0.41}$&$-7.7$&$2.2\sigma$\\
\hline
DESI BAO + Union3 & $0.337^{+0.021}_{-0.017}$ & $-$ & $-0.70^{+0.10}_{-0.12}$ & $-0.79^{+0.43}_{-0.37}$&$-9.9$&$2.7\sigma$\\
\hline
DESI BAO + DES-SN5YR & $0.326^{+0.022}_{-0.015}$ & $-$ & $-0.78^{+0.06}_{-0.08}$ & $-0.57\pm 0.35$&$-11.2$&$2.9\sigma$\\
\hline
DESI BAO + PantheonPlus & $0.307^{+0.026}_{-0.014}$ & $-$ & $-0.88\pm 0.06$ & $-0.26\pm 0.32$&$-3.3$&$1.1\sigma$\\
\hline
DESI BAO + CMB + Union3 & $0.321\pm 0.010$ & $66.79^{+0.9}_{-1.1}$ & $-0.72^{+0.09}_{-0.08}$ & $-0.59\pm 0.18$&$-11.4$&$3.0\sigma$\\
\hline
DESI BAO + CMB + DES-SN5YR & $0.315\pm 0.007$ & $67.31\pm 0.67$ & $-0.78\pm 0.06$ & $-0.50^{+0.16}_{-0.14}$&$-18.1$&$3.9\sigma$\\
\hline	
DESI BAO + CMB + PantheonPlus & $0.307\pm 0.007$ & $68.18\pm 0.71$ & $-0.86^{+0.05}_{-0.06}$ & $-0.36^{+0.15}_{-0.12}$&$-8.1$&$2.3\sigma$\\
\hline \hline
\end{tabular}\label{Tab:BA}
\end{table*}
\begin{figure*} 
	\centering
	\includegraphics[width=7cm]{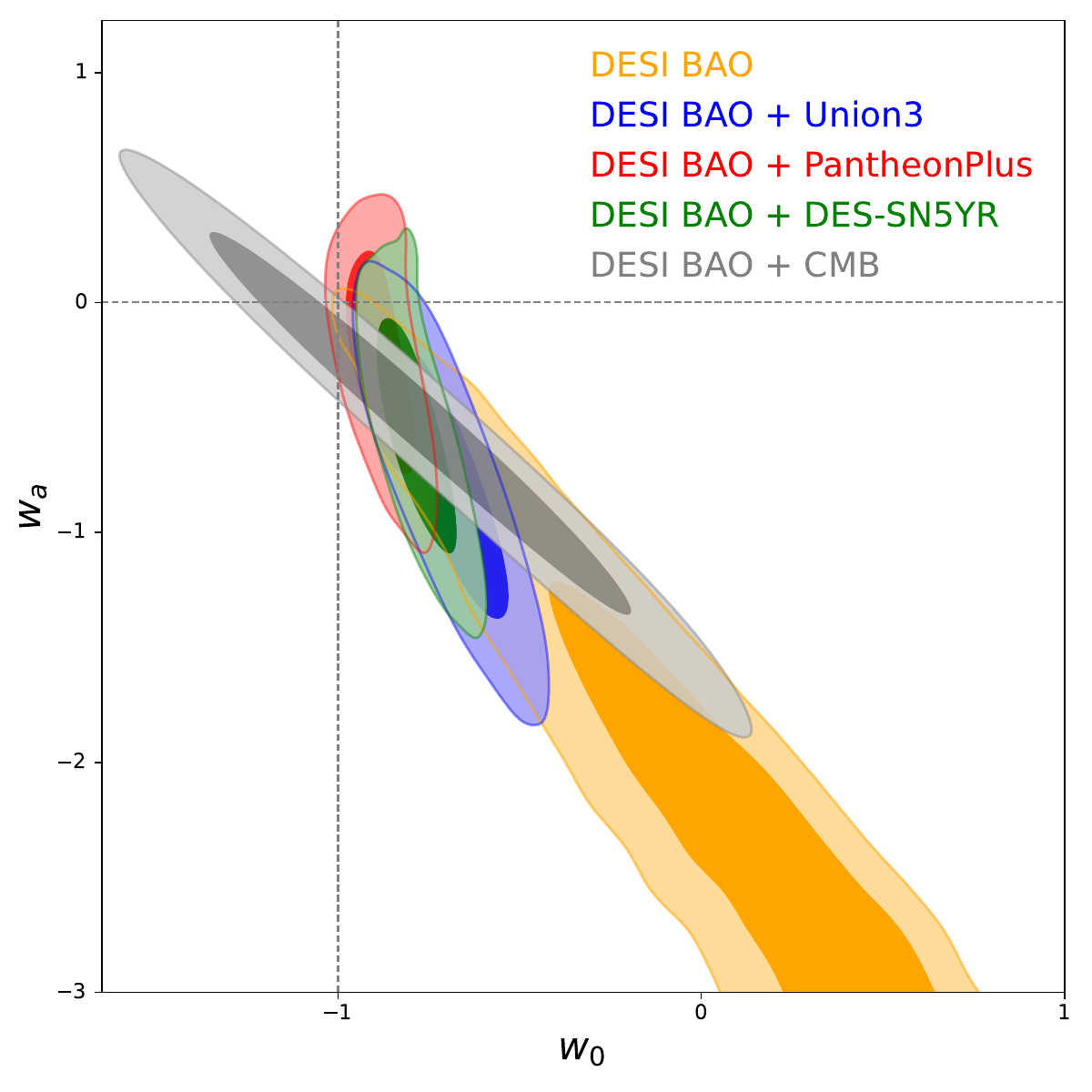}
         \includegraphics[width=7cm]{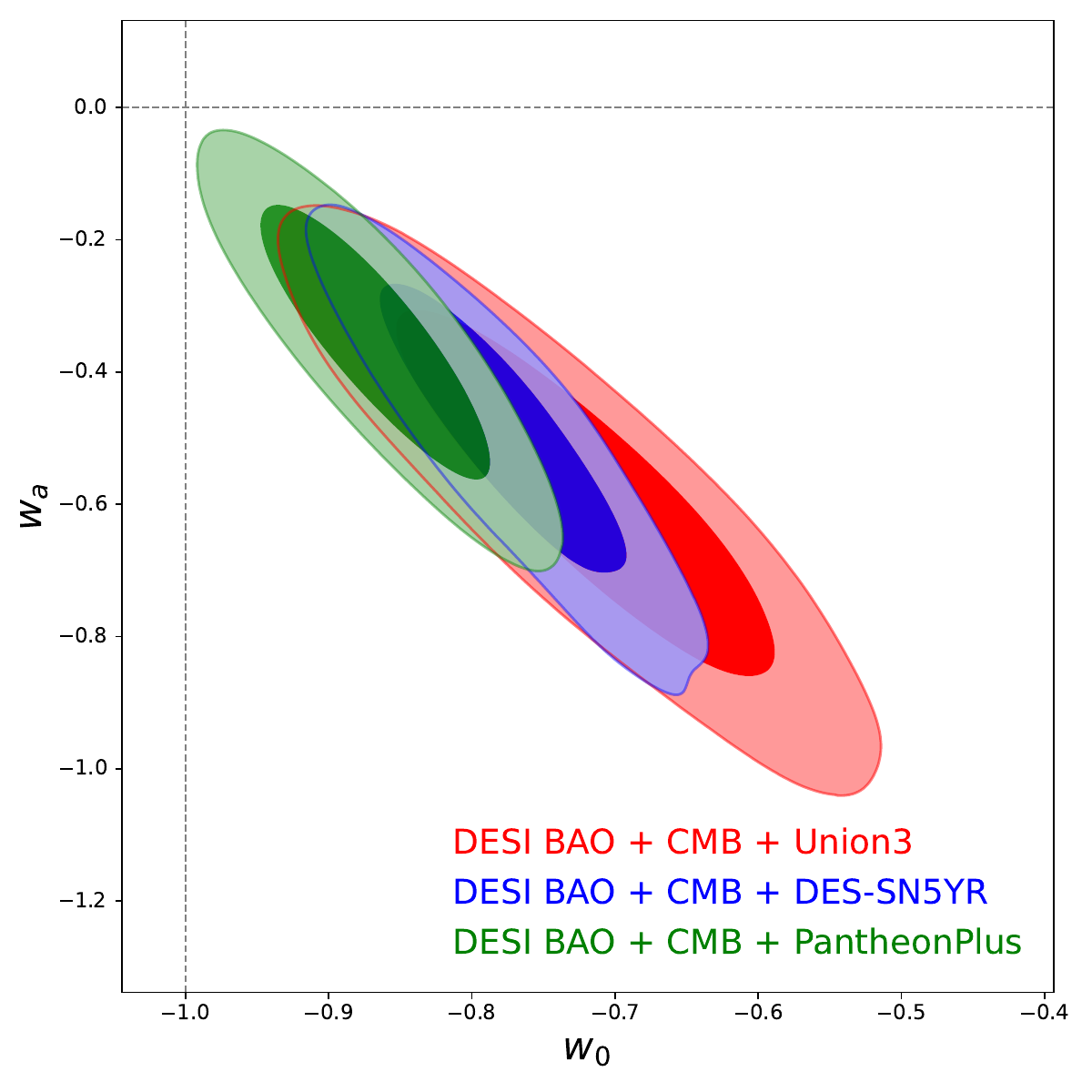}
	\caption{The $1\sigma$ and $2\sigma$ confidence regions of $w_0$ and $w_a$ in the BA parametrization using the various combinations of the observational data.}
	\label{fig:con_BA}
\end{figure*}

\begin{figure*} 
	\centering
	\includegraphics[width=6cm]{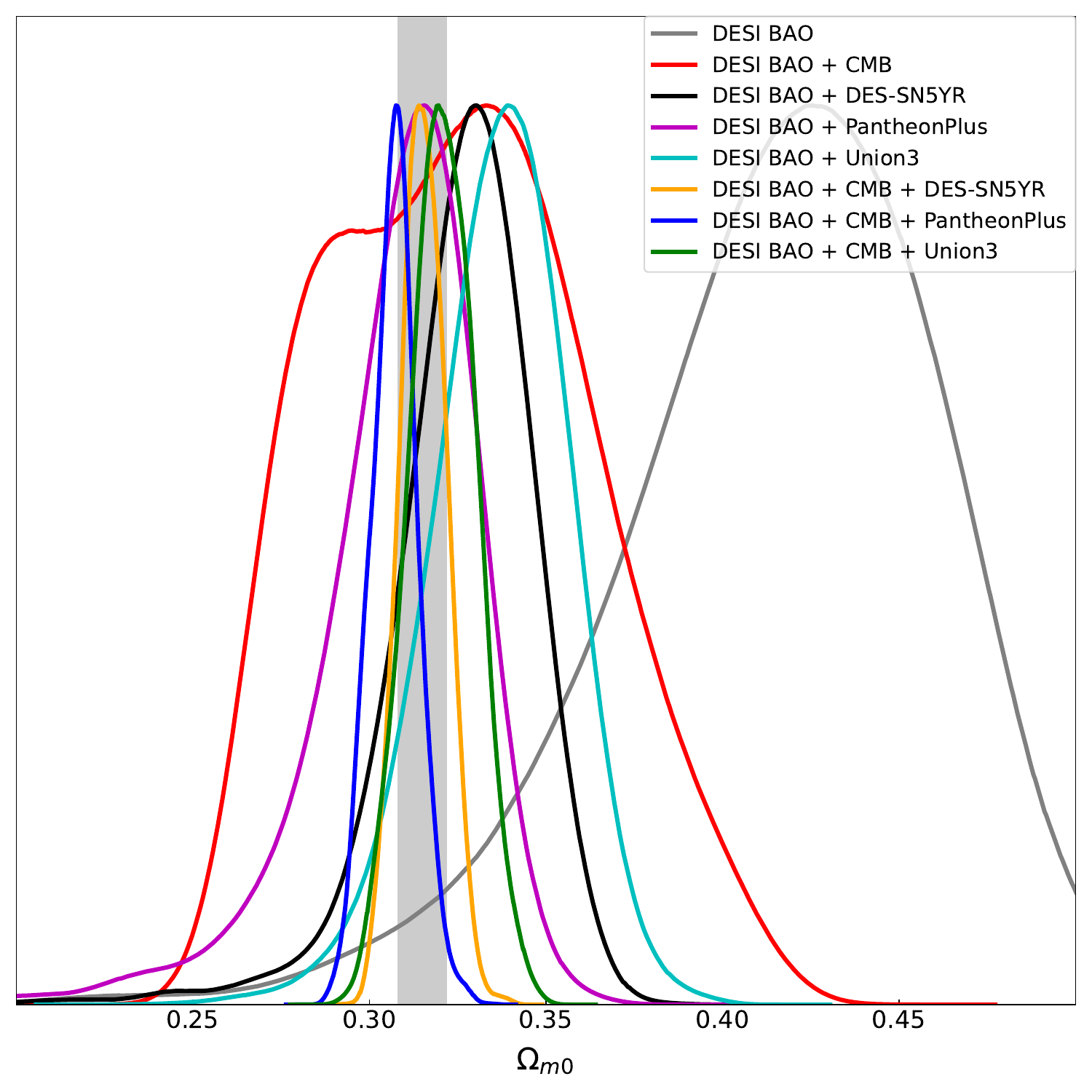}\includegraphics[width=6cm]{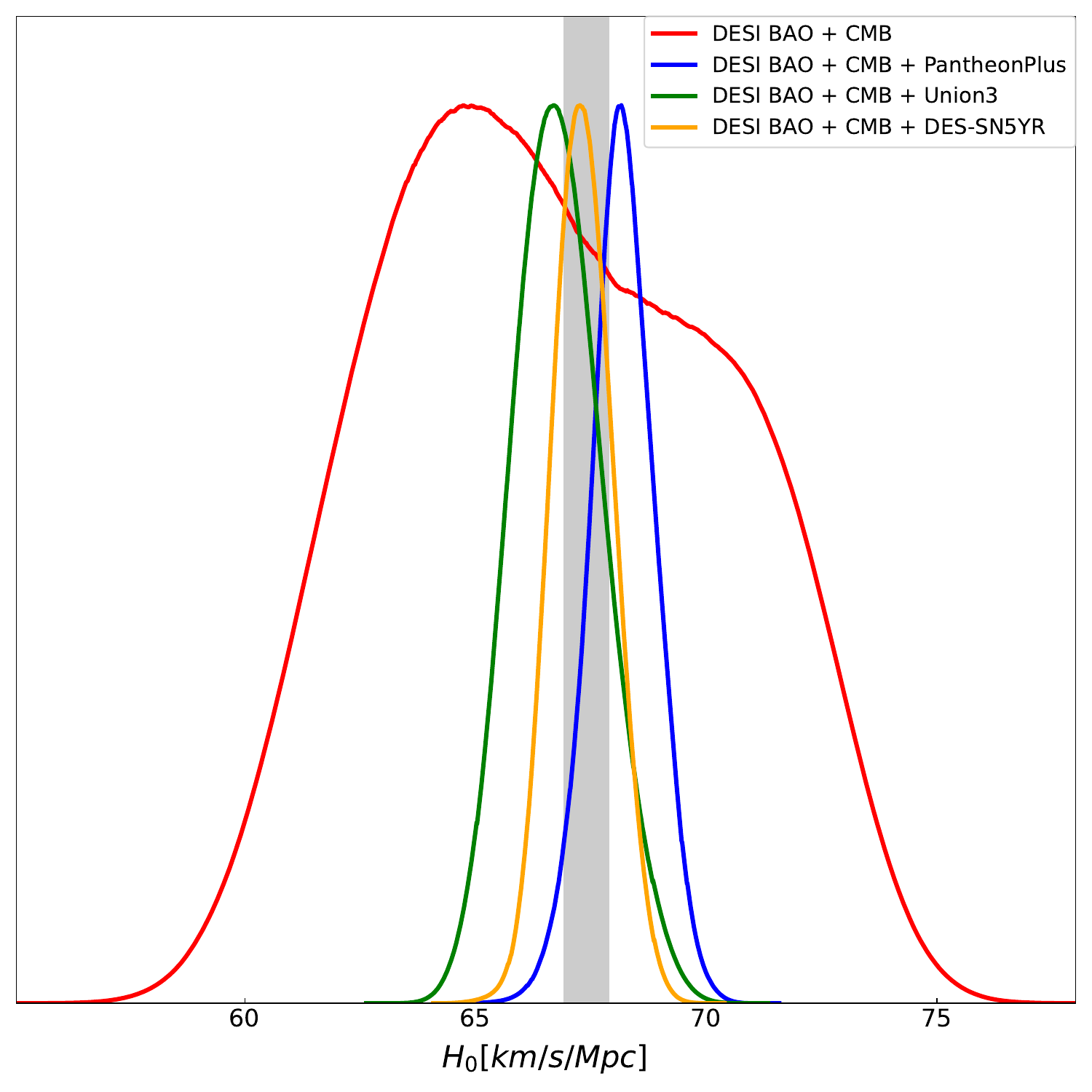}
	\caption{Same as Figure (\ref{fig:con_CPL1}), but for BA parametrization.}
	\label{fig:con_BA1}
\end{figure*}
\subsection{BA Parametrization}
Our numerical results for the cosmological parameters in the BA parametrization, obtained from constraints utilizing different combinations of cosmological datasets, are reported in Table (\ref{Tab:BA}). Similar to the CPL case, the BA parametrization supports the $\Lambda$CDM model with constant energy density ($w_{0}=-1.0, w_a=0.0$) when using the DESI BAO alone. In this case, DESI BAO alone is insufficient to determine $w_a$ and yields enough tight constraints on $w_a$ to exclude zero without the inclusion of additional observational constraints. In this scenario, we observe less than $1.8\sigma$ deviation from the standard flat-$\Lambda$CDM model (see also the left panel of Figure (\ref{fig:con_BA})). 
In the cases of DESI BAO + CMB and DESI BAO + SNIa (Union3 and DES-SN5YR compilations), similar to the CPL parametrization, we measure small values for $\Delta \chi^2$, indicating a moderate deviation ($2\sigma - 3\sigma$) from the standard flat-$\Lambda$CDM cosmology. In the case of DESI BAO + PantheonPlus, similar to the CPL case, we observe weak evidence (less than $2\sigma$) for deviation from the standard model (see the last two columns of Table (\ref{Tab:BA}) and the left panel of Figure (\ref{fig:con_BA})).
Note that there are slight discrepancies between the deviation values reported in Table \ref{Tab:BA} and the 2D contours in Figure \ref{fig:con_BA}, which arise from projection effects.
This result for the above dataset combinations implies that there is no strong evidence for preferring the evolving EoS of DE, $w_{de}$. However, when combining all three datasets, we observe deviations from the standard flat-$\Lambda$CDM cosmology at $2.3\sigma$ ($\Delta \chi^2=-8.1$), $3.0\sigma$ ($\Delta \chi^2=-11.4$), and $3.9\sigma$ ($\Delta \chi^2=-18.1$) for DESI BAO + CMB + PantheonPlus, DESI BAO + CMB + Union3, and DESI BAO + CMB + DES-SN5YR combinations, respectively. In the last two cases, we observe a significant potential for the statistical preference to support the evolving DE scenarios in the context of BA parametrization and consequently strong evidence for the deviation from the standard flat-$\Lambda$CDM scenario (see the right panel of Figure (\ref{fig:con_BA})).
The deviations of $w_0$ and $w_a$ from the $\Lambda$CDM point $\{w_0 = -1.0, w_a = 0.0\}$ in the BA parametrization are similar to those obtained from the CPL parametrization. However, we observe that the best-fit values of $w_a$ in the BA parametrization are larger (closer to zero) with smaller uncertainties compared to the CPL parametrization. This result is due to the correlation between the uncertainty of $w_a$ and the evolutionary behavior of $f(z)$, as discussed in \cite{Colgain:2021pmf}. In addition, we observe that larger values of $w_a$ in the BA parametrization are multiplied by a rapidly decreasing function $f(z)$ so that, similar to what we have in the CPL parametrization, the redshift evolution of $w_{de}(z)$ transitions from the phantom regime at high redshifts to the quintessence regime at the present time (see Appendix \ref{app:apx1} for the BA case).
Similar to the CPL parametrization, in the BA parametrization, the combinations of all datasets in our analysis results in tight constraints on the distribution of $\Omega_{m0}$ parameter, as shown in the left panel of Figure (\ref{fig:con_BA1}) and Table (\ref{Tab:BA}). In the right panel of Figure (\ref{fig:con_BA1}), we show that the distribution of $H_0$ for all dataset combinations, including CMB observations, supports the Planck-inferred value \cite{Planck:2018vyg}. As with the CPL case, our measurement of $H_0$ in the BA parametrization for combinations of all three datasets is precise with smaller uncertainties.
\begin{table*}
\centering
\renewcommand{\arraystretch}{0.75} 
\setlength{\tabcolsep}{0.45pt} 
\caption{Same as Table (\ref{Tab:CPL}), but for Padé parametrization.}
\begin{tabular}{|l| c c c c c c c|}
\hline \hline
Data  & $\Omega_{m0}$ & $H_0$ $[Km/s/Mpc]$ & $w_0$ & $w_1$ & $w_2$&$\Delta\chi^2 $&Deviation \\
\hline 
DESI BAO & $0.347^{+0.053}_{-0.033}$ & $-$ & $-0.50^{+0.41}_{-0.29}$ & $-1.57^{+0.43}_{-1.43}$ & $-0.29^{+0.27}_{-0.06}$&$-4.3$&$0.9\sigma$ \\ 
\hline
DESI BAO + CMB & $0.328^{+0.033}_{-0.039}$ & $66.2^{+2.9}_{-4.2}$ & $-0.56^{+0.31}_{-0.30}$ & $-1.14^{+0.91}_{-1.14}$ & $-0.28^{+0.27}_{-0.10}$&$-9.4$&$2.2\sigma$ \\ 
\hline
DESI BAO + Union3 & $0.338^{+0.021}_{-0.017}$ & $-$ & $-0.66^{+0.11}_{-0.14}$ & $-1.03^{+0.84}_{-0.75}$ & $-0.03^{+0.01}_{-0.02}$&$-10.1$&$2.3\sigma$ \\ 
\hline
DESI BAO + DES-SN5YR & $0.330^{+0.018}_{-0.014}$ & $-$ & $-0.75^{+0.08}_{-0.06}$ & $-0.72^{+0.73}_{-0.65}$ & $-0.38^{+0.37}_{-0.14}$ &$-11.6$&$2.6\sigma$\\ 
\hline
DESI BAO + PantheonPlus & $0.310^{+0.022}_{-0.017}$ & $-$ & $-0.88^{+0.06}_{-0.07}$ &  $-0.45^{+0.35}_{-0.15}$&$-0.03^{+0.68}_{-0.39}$ &$-3.5$& $0.7\sigma$\\ 
\hline
DESI BAO + CMB + Union3 & $0.330^{+0.007}_{-0.011}$ & $66.0^{+1.10}_{-0.71}$ & $-0.68\pm 0.10$ & $-1.01^{+0.35}_{-0.43}$ & $-0.03\pm0.02$ &$-11.7$&$2.6\sigma$\\	
\hline
DESI BAO + CMB + DES-SN5YR & $0.318^{+0.006}_{-0.004}$ & $67.03^{+0.48}_{-0.60}$ & $-0.74^{+0.08}_{-0.06}$ & $-0.99\pm0.35$ & $-0.04\pm0.02$&$-20.0$&$3.8\sigma$ \\
\hline
DESI BAO + CMB + PantheonPlus & $0.306^{+0.006}_{-0.005}$ & $68.19^{+0.36}_{-0.62}$ & $-0.84\pm 0.06$ & $-0.66\pm 0.24$ & $-0.08\pm0.02$&$-9.1$ &$2.1\sigma$\\
\hline \hline
\end{tabular}\label{Tab:PadéI}
\end{table*}
\begin{figure*} 
	\centering
		\includegraphics[width=7cm]{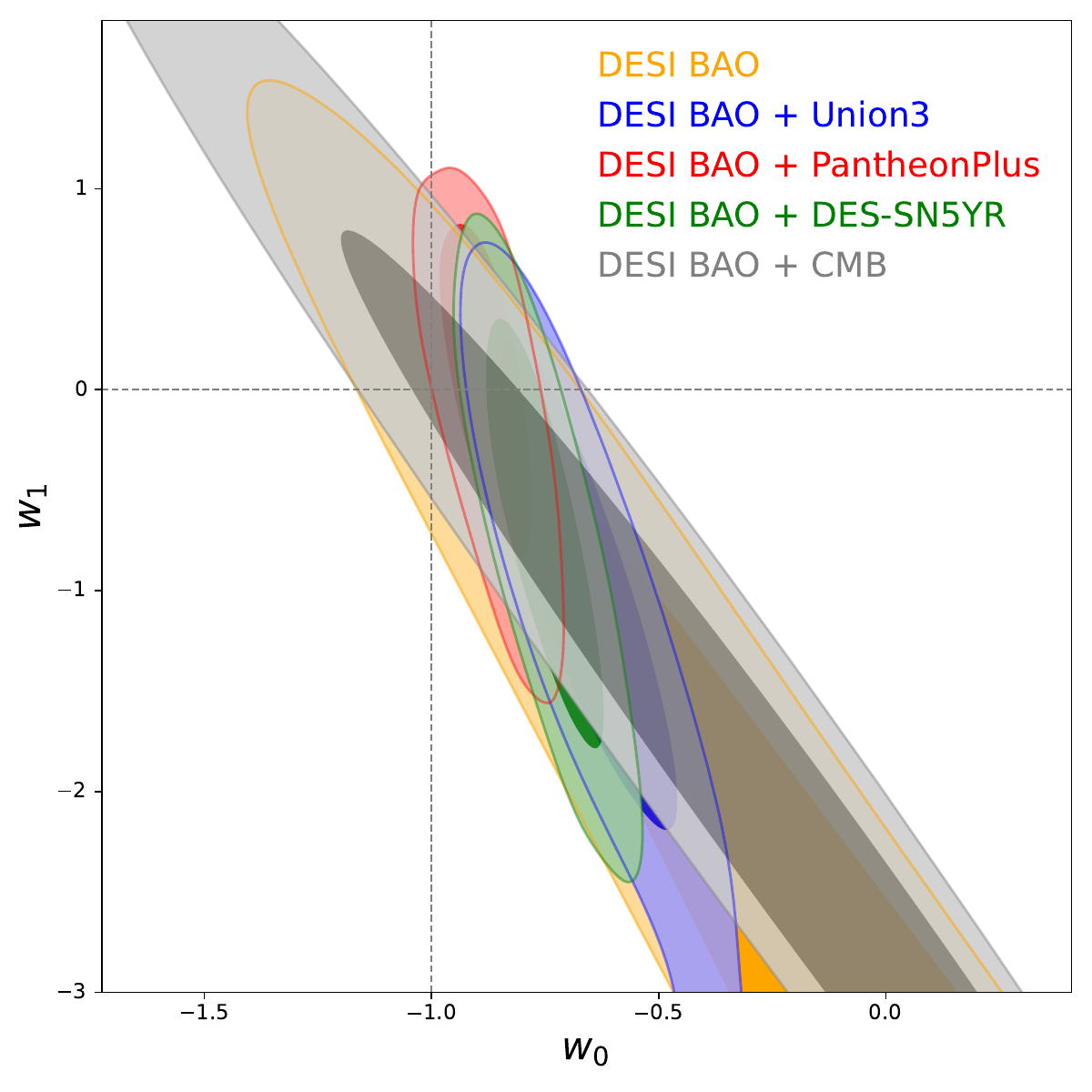}\includegraphics[width=7cm]{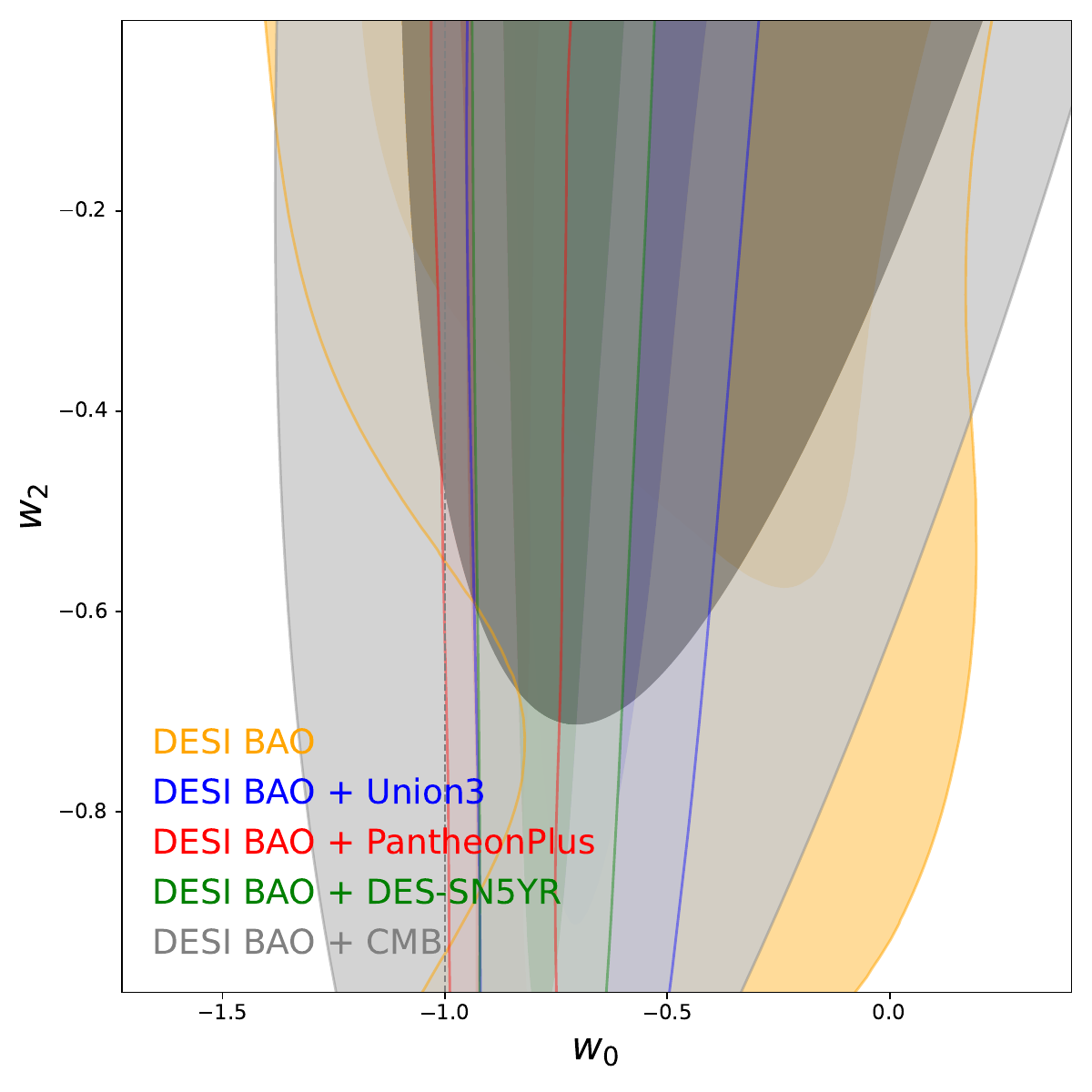}
    \includegraphics[width=7cm]{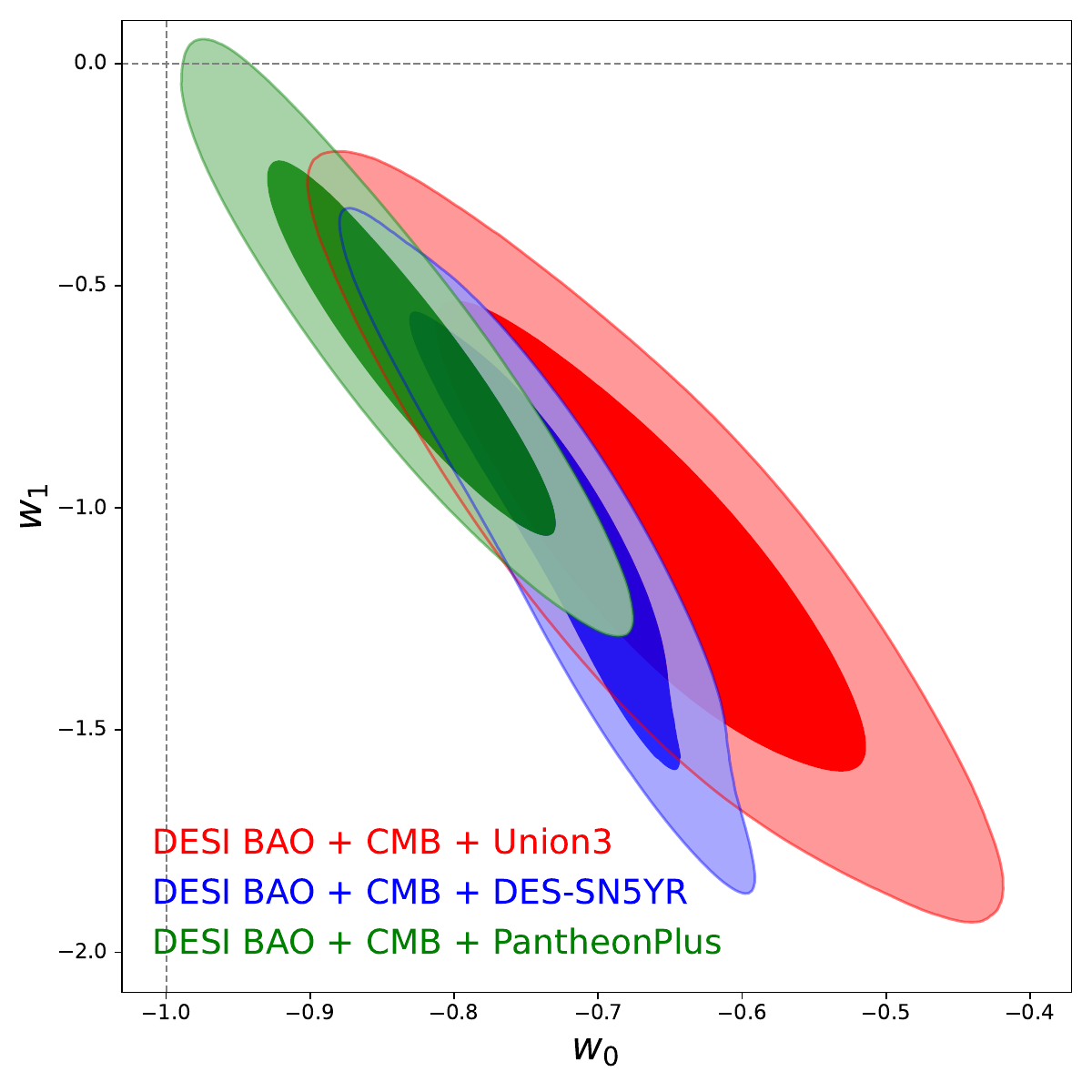}\includegraphics[width=7cm]{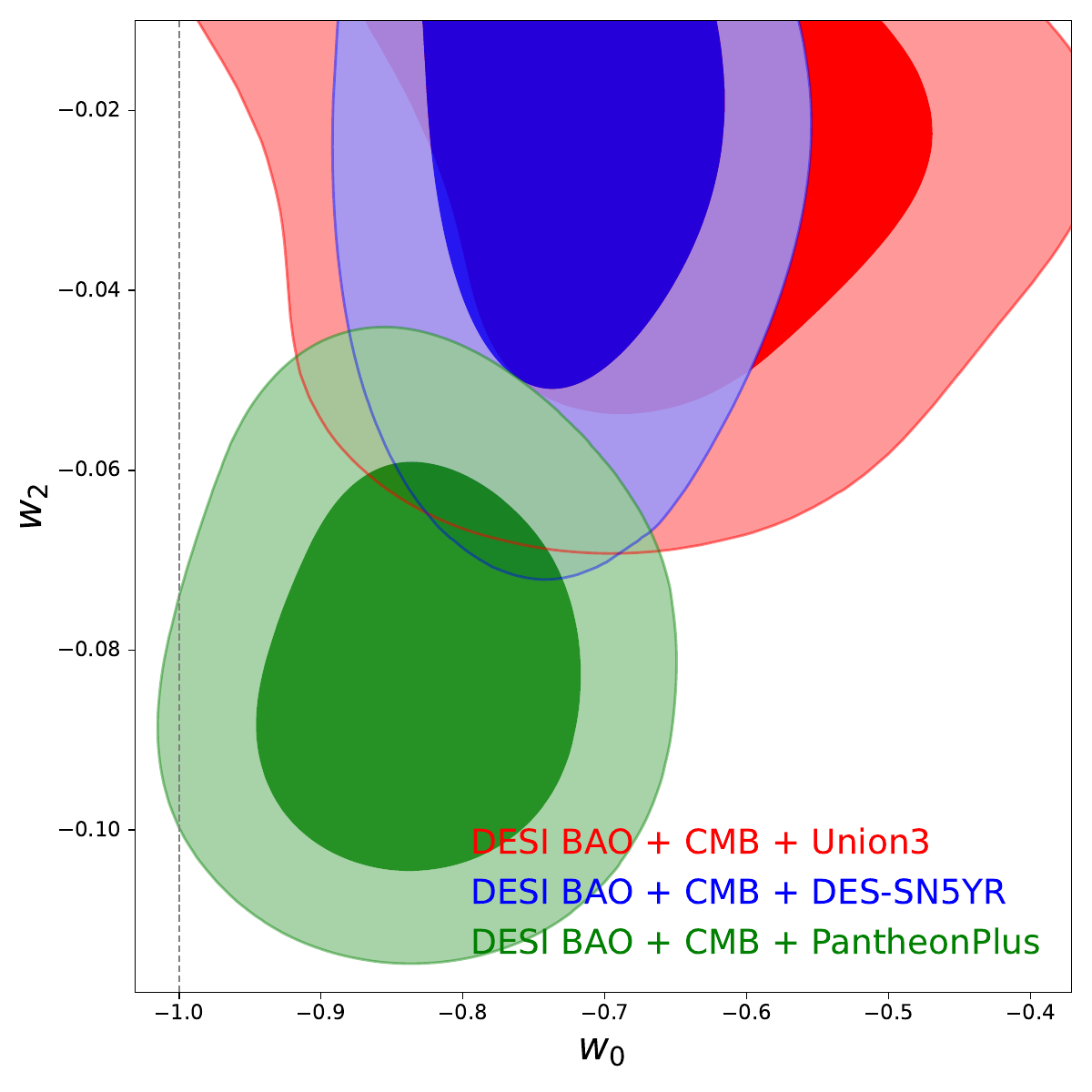}
	\caption{The $1\sigma$ and $2\sigma$ confidence regions of $w_0$, $w_1$ and $w_2$ in the Padé parametrization using the various combinations of the observational data.}
	\label{fig:con_PadéI}
\end{figure*}
\begin{figure*} 
	\centering
	\includegraphics[width=6cm]{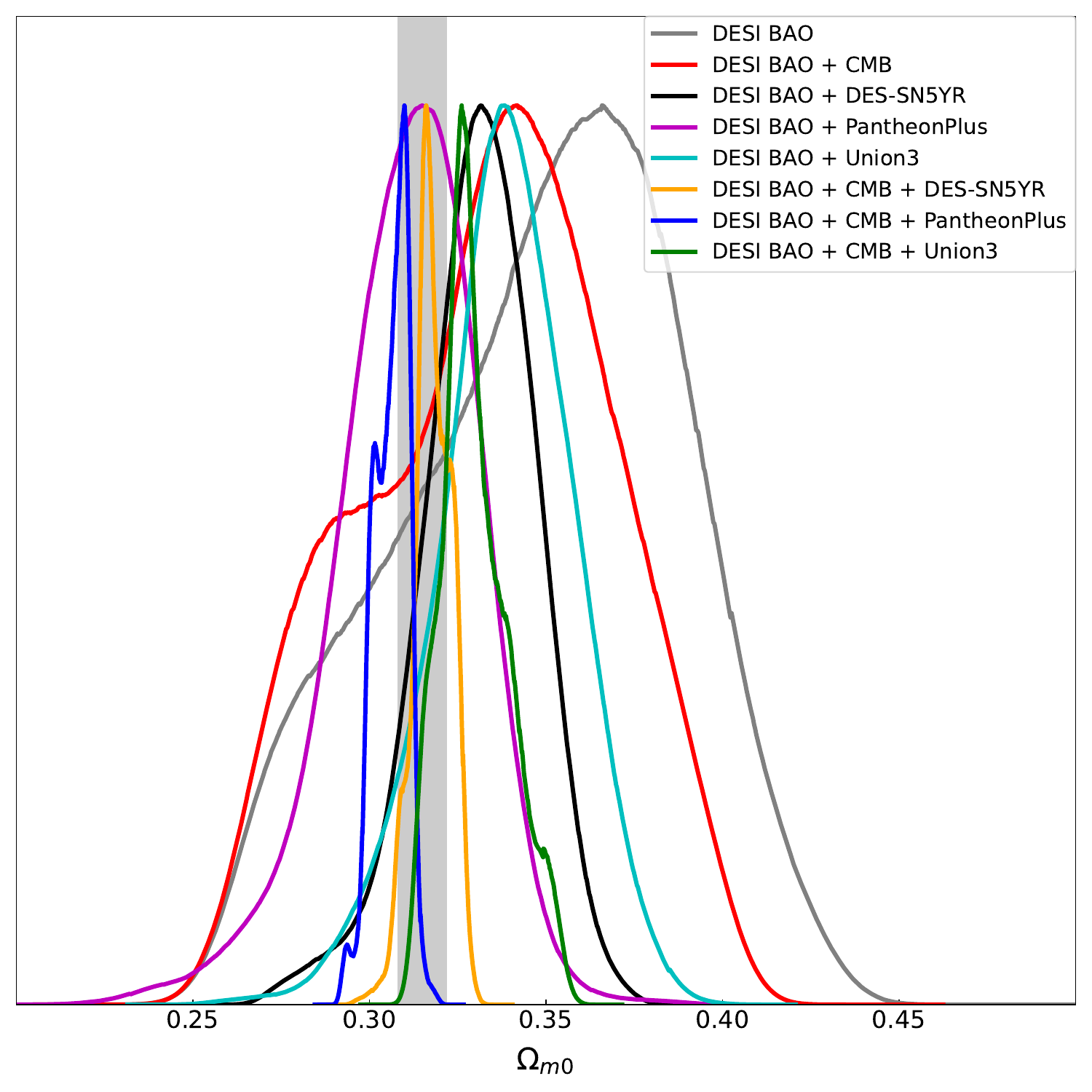}\includegraphics[width=6cm]{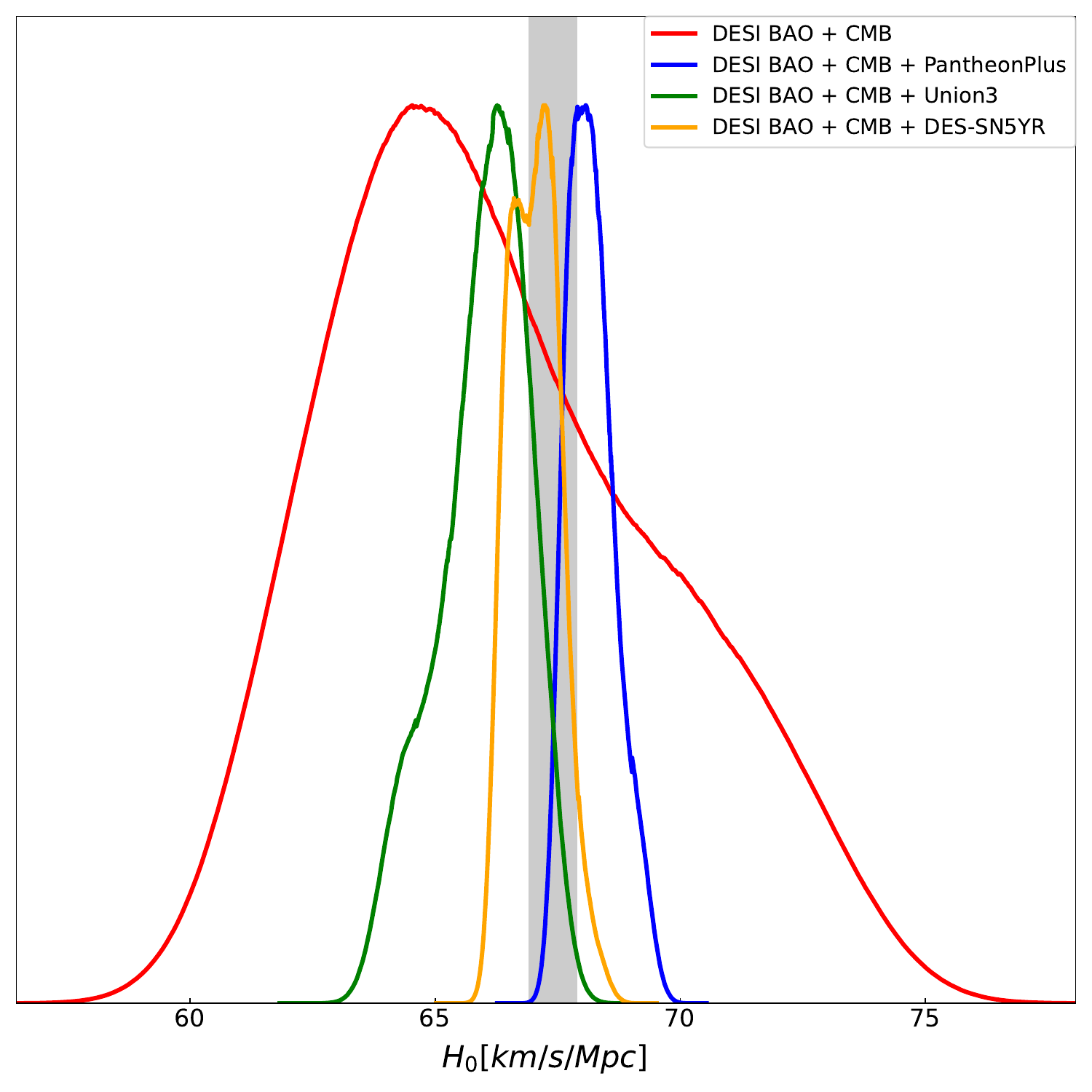}
	\caption{Same as Figure (\ref{fig:con_CPL1}), but for Padé parametrization.}
	\label{fig:con_Padé1}
\end{figure*}
\begin{table*}
\centering
\caption{Same as Table (\ref{Tab:CPL}), but for simplified Padé parametrization.}
\begin{tabular}{|l| c c c c c c|}
\hline \hline
Data  & $\Omega_{m0}$ & $H_0$ $[Km/s/Mpc]$ & $w_0$ & $w_2$ &$\Delta\chi^2 $&Deviation\\
\hline 
DESI BAO & $0.295\pm 0.026$ & $-$ & $-0.99^{+0.25}_{-0.14}$ & $-0.03^{+0.32}_{-0.96}$&$-1.5$ &$0.3\sigma$\\ 
\hline
DESI BAO + CMB & $0.317^{+0.010}_{-0.006}$ & $67.1^{+1.10}_{-0.79}$ & $-0.83^{+0.13}_{-0.09}$ & $-0.63^{+0.12}_{-0.36}$&$-10.0$&$2.7\sigma$ \\
\hline
DESI BAO + Union3 & $0.314^{+0.026}_{-0.017}$ & $-$ & $-0.82^{+0.08}_{-0.06}$ & $-0.51^{+0.13}_{-0.28}$&$-8.3$&$2.4\sigma$ \\
\hline
DESI BAO + DES-SN5YR & $0.315^{+0.025}_{-0.016}$ & $-$ & $-0.83^{+0.06}_{-0.05}$ & $-0.44^{+0.20}_{-0.47}$&$-10.8$ &$2.8\sigma$\\
\hline
DESI BAO + PantheonPlus & $0.301^{+0.025}_{-0.019}$ & $-$ & $-0.90^{+0.06}_{-0.05}$ & $-0.14^{+0.33}_{-0.62}$ &$-3.6$&$1.2\sigma$\\
\hline
DESI BAO + CMB + Union3 & $0.310^{+0.010}_{-0.013}$ & $67.8^{+1.2}_{-1.0}$ & $-0.80\pm0.10$ & $-0.48^{+0.27}_{-0.31}$&$-14.0$ &$3.3\sigma$\\
\hline
DESI BAO + CMB + DES-SN5YR & $0.311^{+0.009}_{-0.011}$ & $67.58\pm 0.93$ & $-0.81^{+0.07}_{-0.06}$ & $-0.55^{+0.15}_{-0.44}$ &$-18.0$&$3.9\sigma$\\
\hline
DESI BAO + CMB + PantheonPlus & $0.301^{+0.009}_{-0.012}$ & $68.7^{+1.20}_{-0.92}$ & $-0.86^{+0.06}_{-0.04}$ & $-0.59^{+0.14}_{-0.28}$&$-11.0$ &$2.8\sigma$\\
\hline \hline
\end{tabular}\label{Tab:sim_Padé}
\end{table*}

\begin{figure*} 
	\centering
	\includegraphics[width=7cm]{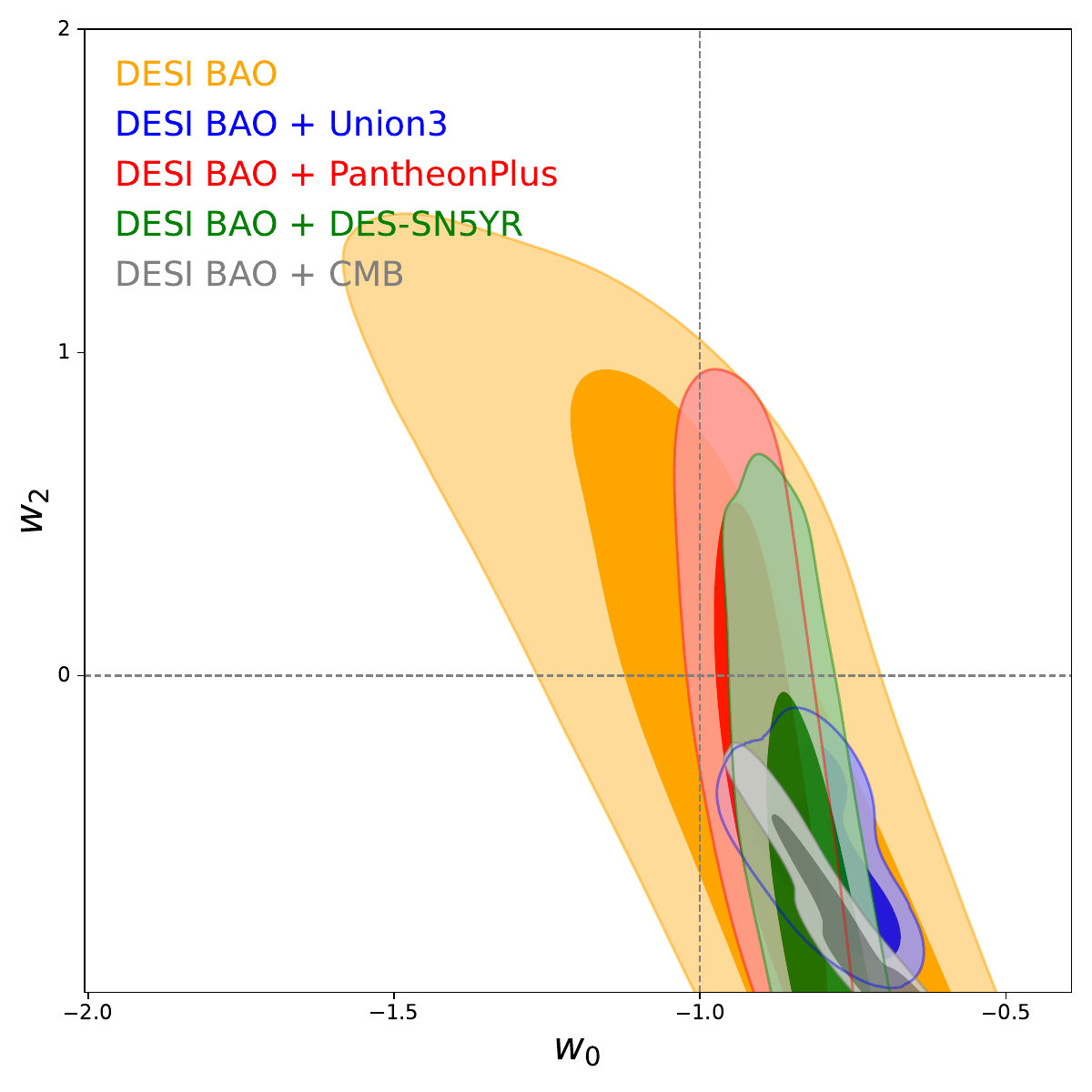}\includegraphics[width=7cm]{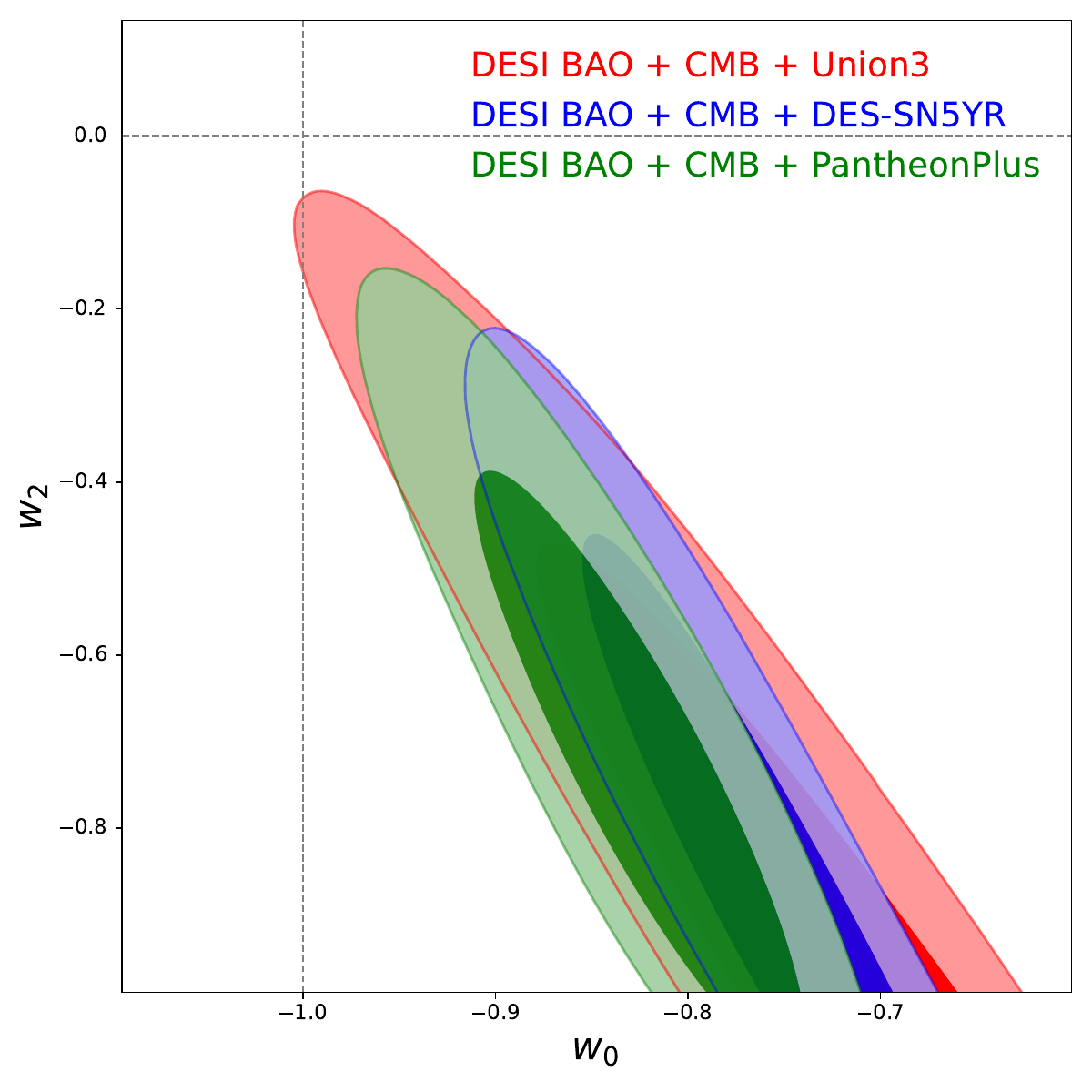}
	\caption{The $1\sigma$ and $2\sigma$ confidence regions of $w_0$ and $w_2$ in the simplified Padé parametrization using the various combinations of the observational data.}
	\label{fig:con_SimPadé}
\end{figure*}

\begin{figure*} 
	\centering
	\includegraphics[width=6cm]{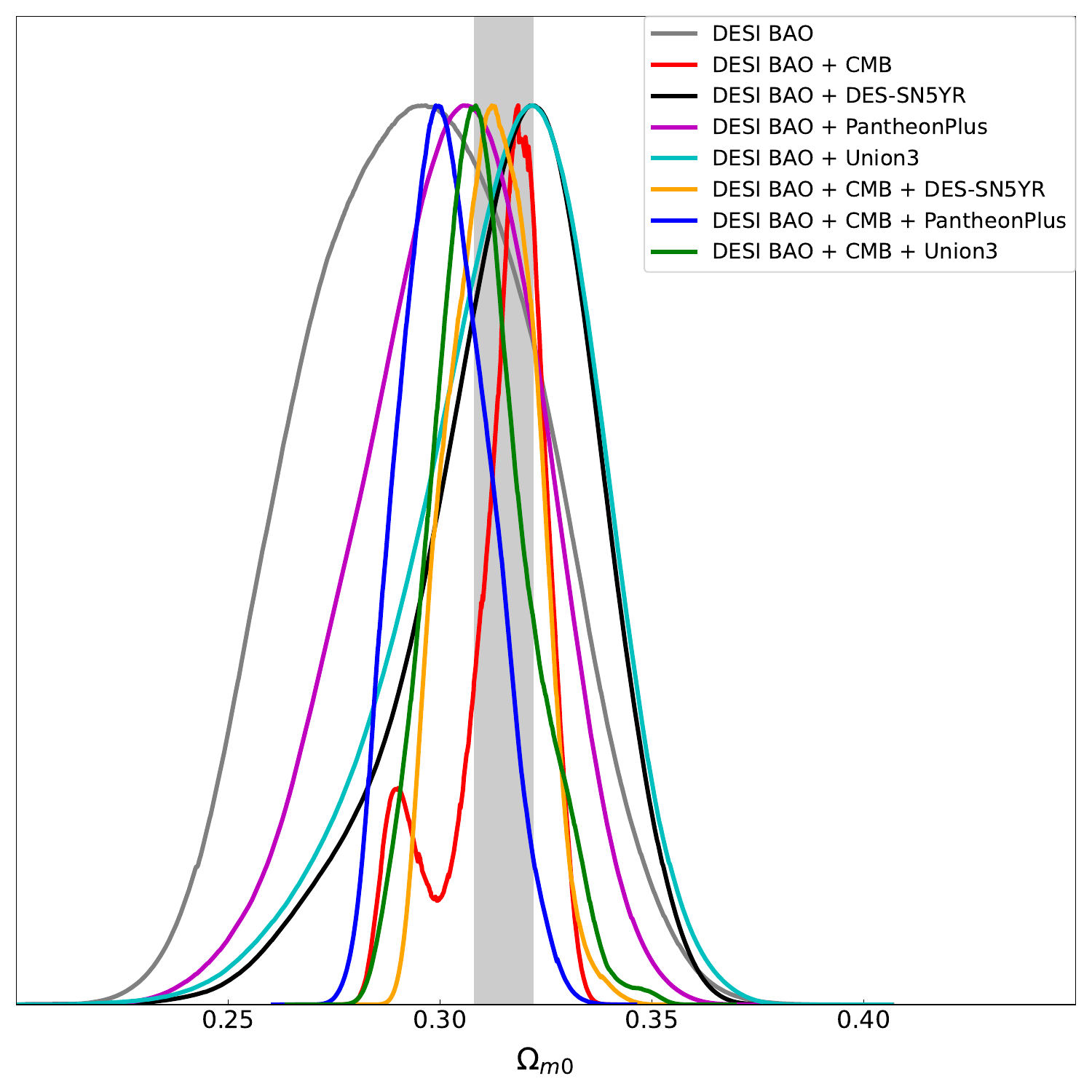}\includegraphics[width=6cm]{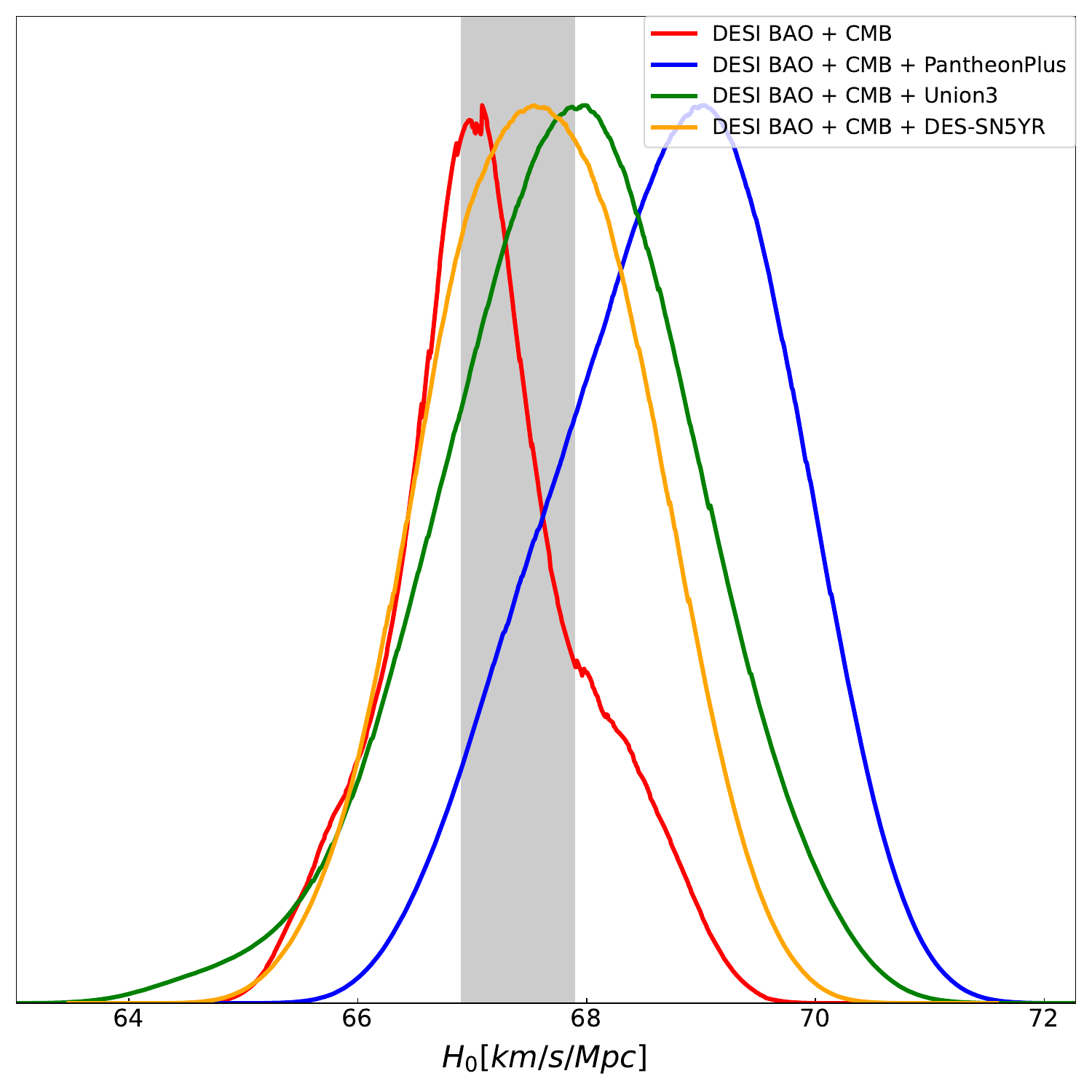}
	\caption{Same as Figure (\ref{fig:con_CPL1}), but for simplified Padé parametrization.}
	\label{fig:con_SimPadé1}
\end{figure*}

\subsection{Padé parametrization}
In the Padé parametrization, based on Eq.\ref{Padé1}, the two parameters $w_1$ and $w_2$ govern the evolution of the EoS parameter. Both $w_1$ and $w_2$ significantly influence the evolution of $w_{de}$, providing a diagnostic for observing a possible deviation from the constant $w_{\Lambda}=-1$. In particular, the same positive (negative) signs of these parameters cause a more rapid evolution into the quintessence (phantom) regime.
Our numerical results obtained from observational constraints for the Padé parametrization are presented in Table (\ref{Tab:PadéI}) and Figures (\ref{fig:con_PadéI}). 
For DESI BAO alone and combinations of two datasets, including: DESI BAO + CMB and DESI BAO + SNIa for various SNIa samples, our results for $w_0$ and $w_1$ are similar to those observed in the same cases for BA and CPL parametrizations for $w_0$ and $w_a$. Additionally, we cannot place tight constraints on the parameter $w_2$, as shown in the upper-right panel of Figure (\ref{fig:con_PadéI}). Therefore, similar to the CPL and BA parametrizations for the aforementioned data combinations, the Padé parametrization does not significantly deviate from standard flat-$\Lambda$CDM cosmology. In this line, for the datasets mentioned above, we measure small negative values for $\Delta \chi^2$ and, consequently, observe no significant support (less than $3\sigma$) for the statistical preference of the DE scenario over the standard flat-$\Lambda$CDM cosmology in the context of the Padé parametrization. Hence, the standard flat-$\Lambda$CDM cosmology is recovered. 
Finally, we combine all three datasets as DESI BAO + CMB + SNIa (for various SNIa samples). In continuation, we observe that combining all datasets enables us to place tight constraints on all model parameters. The numerical values are presented in the last three rows of Table (\ref{Tab:PadéI}), and constraints on the $\{w_0, w_1, w_2\}$ parameters are plotted in the lower panels of Figure (\ref{fig:con_PadéI}). We measure $\Delta \chi^2 = -9.1$ ($2.1\sigma$), $\Delta \chi^2 = -11.7$ ($2.6\sigma$), and $\Delta \chi^2 = -20.0$ ($3.8\sigma$) for the DESI BAO + CMB + PantheonPlus, DESI BAO + CMB + Union3, and DESI BAO + CMB + DES-SN5YR combinations, respectively. Generally, the deviation from the flat $\Lambda$CDM cosmology is reduced in the Padé parametrization due to the inclusion of an additional free parameter compared to the CPL and BA parametrizations.  
Notably, we only observe a significant deviation from the flat-$\Lambda$CDM cosmology and strong evidence for preferring the evolving DE scenario when using the DESI BAO + CMB + DES-SN5YR combination.
In contrast to the combinations of two datasets, we can place tight constraints on the $w_2$ parameter. Interestingly, our constraints on $w_2$ show values close to zero, supporting the idea that our constraints on the Padé parametrization are similar to those on the CPL parametrization. Particularly, from the lower panels of Figure (\ref{fig:con_PadéI}), we observe correlations between our constraints on $w_1$ and $w_2$. Specifically, for the DESI BAO + CMB + Union3 and DESI BAO + CMB + DES-SN5YR combinations, our constraints on $w_1$ deviate significantly from zero while $w_2$ approaches zero. Conversely, for the DESI BAO + CMB + PantheonPlus combination, $w_1$ approaches zero, and $w_2$ is completely separated from zero.
Finally, as illustrated in Figure~\ref{fig:con_Padé1}, we find that the Padé parametrization - consistent with previous parametrizations in our analysis - does not yield tight constraints on $\Omega_{m0}$ when using DESI BAO data alone. Moreover, low-redshift observations are unable to provide stringent bounds on the $H_0$ parameter, resulting in significantly larger uncertainties compared to dataset combinations that incorporate CMB observations.

\subsection{Simpliﬁed Padé}
In this case, similar to other parametrizations, $w_0$ controls the EoS parameter at the present time. Moreover, based on Eq. \ref{Padésimp}, $w_2$ governs the variation of $w_{de}$ in terms of redshift. Our numerical results for different dataset combinations are presented in Table (\ref{Tab:sim_Padé}) and in Figure (\ref{fig:con_SimPadé}). As we observe, DESI BAO alone does not have sufficient influence to place tight constraints on the $w_0$ and $w_2$ parameters. Therefore, we expect that 
our constraints support the $\Lambda$CDM point $\{w_0 = -1.0, w_2 = 0.0\}$ very well within the $1\sigma$ level using DESI BAO alone.
Adding CMB data to DESI BAO reduces the uncertainty of the measurements significantly. In this case, the simplified Padé parametrization deviates from the standard flat-$\Lambda$CDM cosmology with $2.7\sigma$ ($\Delta \chi^2 = -10.0$). For the DESI BAO + Union3 and DESI BAO + DES-SN5YR combinations, we measure $2.4\sigma$ ($\Delta \chi^2 = -8.3$) and $2.8\sigma$ ($\Delta \chi^2 = -10.8$), respectively. These results indicate mild deviation from the standard flat-$\Lambda$CDM model and, consequently, moderate support for the dynamical DE scenario. It should be noted that slight variations occur between the deviation values reported in Table \ref{Tab:BA} and the corresponding 2D contours in Figure \ref{fig:con_BA}, attributable to projection effects.
For the DESI BAO + PantheonPlus combination, we observe weak evidence ($1.2\sigma$) against the flat-$\Lambda$CDM cosmology. All the above results indicate that using DESI BAO alone, DESI BAO + CMB, and DESI BAO + SNIa (Union3, DES-SN5YR, and PantheonPlus compilations) does not provide significant evidence for the possibility of dynamical behavior of DE in the simplified Padé parametrization, similar to other DE parametrizations studied in this work.
Finally, we extend our analysis to all dataset combinations. We measure the possibility of deviation from flat-$\Lambda$CDM cosmology as $3.3\sigma$ ($\Delta \chi^2 = -14.0$), $3.9\sigma$ ($\Delta \chi^2 = -18.0$), and $2.8\sigma$ ($\Delta \chi^2 = -11.0$) for the DESI BAO + CMB + Union3, DESI BAO + CMB + DES-SN5YR, and DESI BAO + CMB + PantheonPlus combinations, respectively. We observe significant evidence (deviation from the standard model more than $3\sigma$) for the evolution of DE in the simplified Padé parametrization when we include the Union3 and DES-SN5YR compilations. This result is close to the result we obtained using the CPL parameterization.
Consistent with other parametrizations, the simplified Padé parametrization provides tight constraints on $\Omega_{m0}$ that align well with the Planck-inferred values \cite{Planck:2018vyg} when using combined datasets. Furthermore, it yields precise determinations of $H_0$ when CMB observations are incorporated in the analysis (see Figure~\ref{fig:con_SimPadé1}).

\section{Conclusion}\label{conlusion}
DE parametrizations can be expressed as $ w_{de}(z) = w_0 + w_1 f(z)$, where $f(z)$ is a function that defines the evolution of the equation of state (EoS) of DE. Recent studies \citep{Colgain:2021pmf} have highlighted the constraints of dynamical DE parametrizations, noting that the precision of the $w_1$ parameter measurement from observational data, which governs the EoS variation, is contingent on the behavior of the function $f(z)$ that modulates the EoS evolution. The negative correlation between $w_1$ and $f(z)$ could potentially influence our conclusions within specific DE parametrizations. Consequently, it is crucial to validate the findings from one DE parametrization against others to confirm their robustness. This method aids in reducing the risk of misinterpretation due to the dependency of measurement precision on the behavior of $f(z)$.

In this study, we have employed well-known parametrizations of the EoS for DE, extending beyond the CPL parametrization, to investigate any potential deviations from the standard $\Lambda$CDM constant EoS parameter $w_{\Lambda} = -1$, following the one-year release of Dark Energy Spectroscopic Instrument (DESI) Baryon Acoustic Oscillations (BAO) observations in 2024. We utilized the Barboza \& Alcaniz (BA) and Padé parametrizations to cross-verify the cosmological constraint results from recent work \citep{DESI:2024mwx} by the DESI collaboration, which suggests a possible deviation from the constant $w_{\Lambda} = -1$ when combining DESI BAO, CMB, and various SNIa compilations. 

Our findings are summarized as follows:\\
\begin{itemize}
	\item {\bf DESI BAO alone}:\\
	  In all DE parametrizations, we observed that DESI BAO alone does not have the power to place tight constraints on the EoS parameter of DE. Therefore, as expected, we cannot measure the deviation of the EoS parameter of DE from the constant EoS $w_{\Lambda} = -1.0$ in our analysis, thus supporting the standard flat-$\Lambda$CDM cosmology in all DE parametrizations.
      
      \item {\bf DESI BAO + CMB:}\\
      Combining the DESI BAO with CMB observations results in tighter constraints on the EoS parameter of DE. This behavior is observed for all DE parametrizations. The tightening of our constraints leads to mild tension with the standard flat-$\Lambda$CDM cosmology. In particular, we measure a $ 2\sigma-3\sigma$ deviation from the standard model, indicating a mild statistical preference for the existence of dynamical behavior in DE scenarios.

      \item{\bf DESI BAO + SNIa (various PantheonPlus, Union3 and DES-SN5YR compilations):}\\
      As expected, adding the SNIa data to DESI BAO results in tighter constraints compared to DESI BAO alone. In this context, we observe that for all DE parametrizations, the DESI BAO + Union3 and DESI BAO + DES-SN5YR combinations show a mild deviation ($2\sigma - 3\sigma$) from the standard flat-$\Lambda$CDM cosmology, indicating a preference for evolving DE scenarios. However, for the DESI BAO + PantheonPlus combination, we observe weak evidence for preferring dynamical DE scenarios for all DE parametrizations against the standard model.

    \item{\bf DESI BAO + CMB + SNIa:}\\
 Our analysis of all DE parametrizations reveals strong evidence for dynamical DE behavior, favoring it over the constant energy density predicted by the standard flat-$\Lambda$CDM cosmology. In particular, the DESI BAO + CMB + Union3 and DESI BAO + CMB + DES-SN5YR combinations show deviations exceeding $3\sigma$ from $\Lambda$CDM, strongly supporting dynamical DE scenarios. The only exception is the Padé parametrization with DESI BAO + CMB + Union3, which shows weaker deviation due to its additional free parameter that reduces its distinction from $\Lambda$CDM model.
These findings extend the initial evidence for deviations from $\Lambda$CDM reported by \citet{DESI:2024mwx} using the CPL parametrization to both BA and Padé parametrizations. Our combined analysis of DESI BAO, CMB, and SNIa data shows a consistent evolutionary pattern across all parametrizations: the DE equation of state transitions from a phantom regime ($w_{\rm de} < -1$) at high redshifts to a quintessence regime ($w_{\rm de} > -1$) at low redshifts (see Appendix~\ref{app:apx1}). This transition further strengthens the case for dynamical DE behavior originally suggested by DESI results.
\end{itemize} 
\section{ Data availability Statement}
The data that support the findings of this study are available from the corresponding author upon reasonable request.
\section{Acknowledgments}
We would like to express our gratitude to the anonymous referee for their careful review of our manuscript and for providing constructive and valuable comments that significantly improved our paper. The work of SP is based upon research funded by the Iran National Science Foundation (INSF) under project No. 4024802. ZD was supported by the Korea Institute for Advanced Study Grant No. 6G097301.
 \bibliographystyle{apsrev4-1}
  \bibliography{ref}
  \label{lastpage}
\appendix
\setcounter{figure}{0} 
\renewcommand{\thefigure}{A\arabic{figure}} 
\section{Redshift evolution of EoS parameter}
\label{app:apx1}

\begin{figure*}
\centering
    \includegraphics[width=5cm]{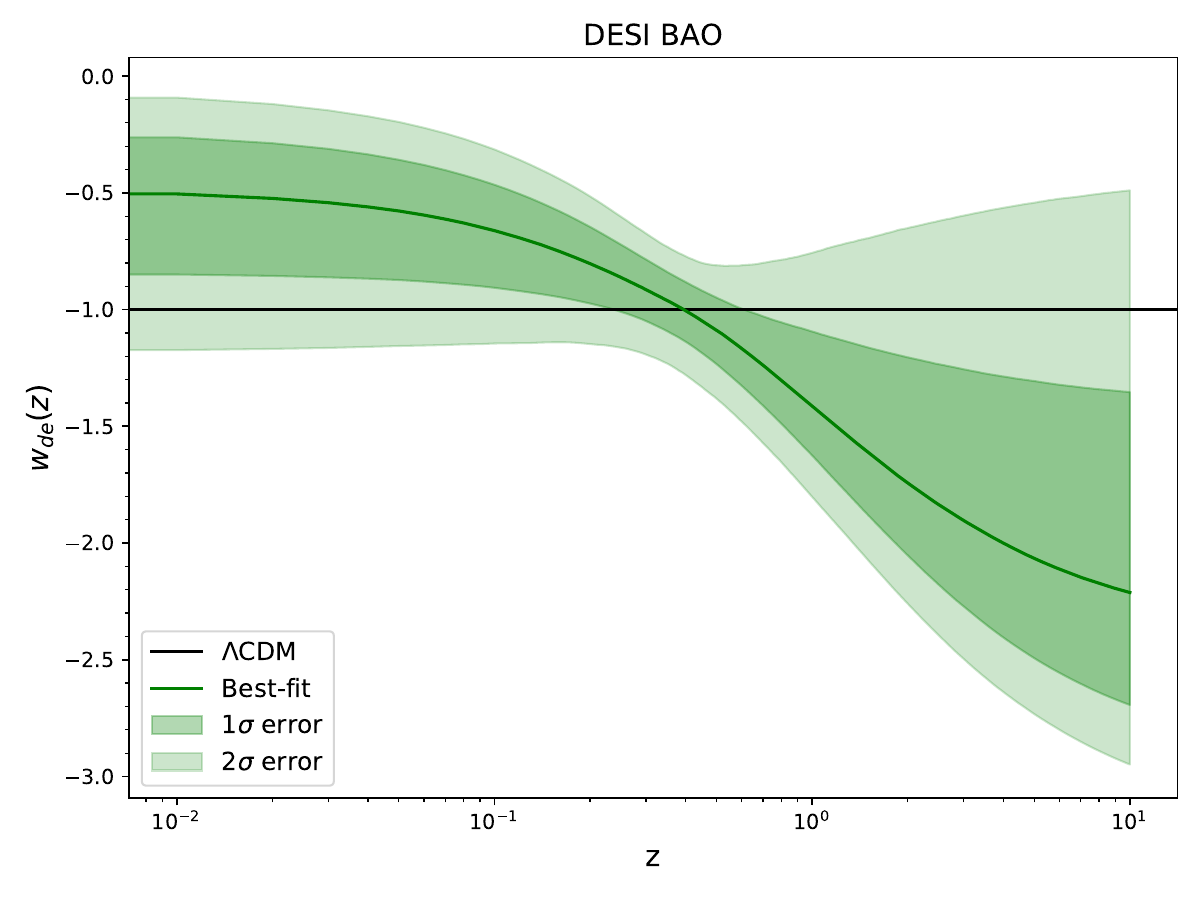}\includegraphics[width=5cm]{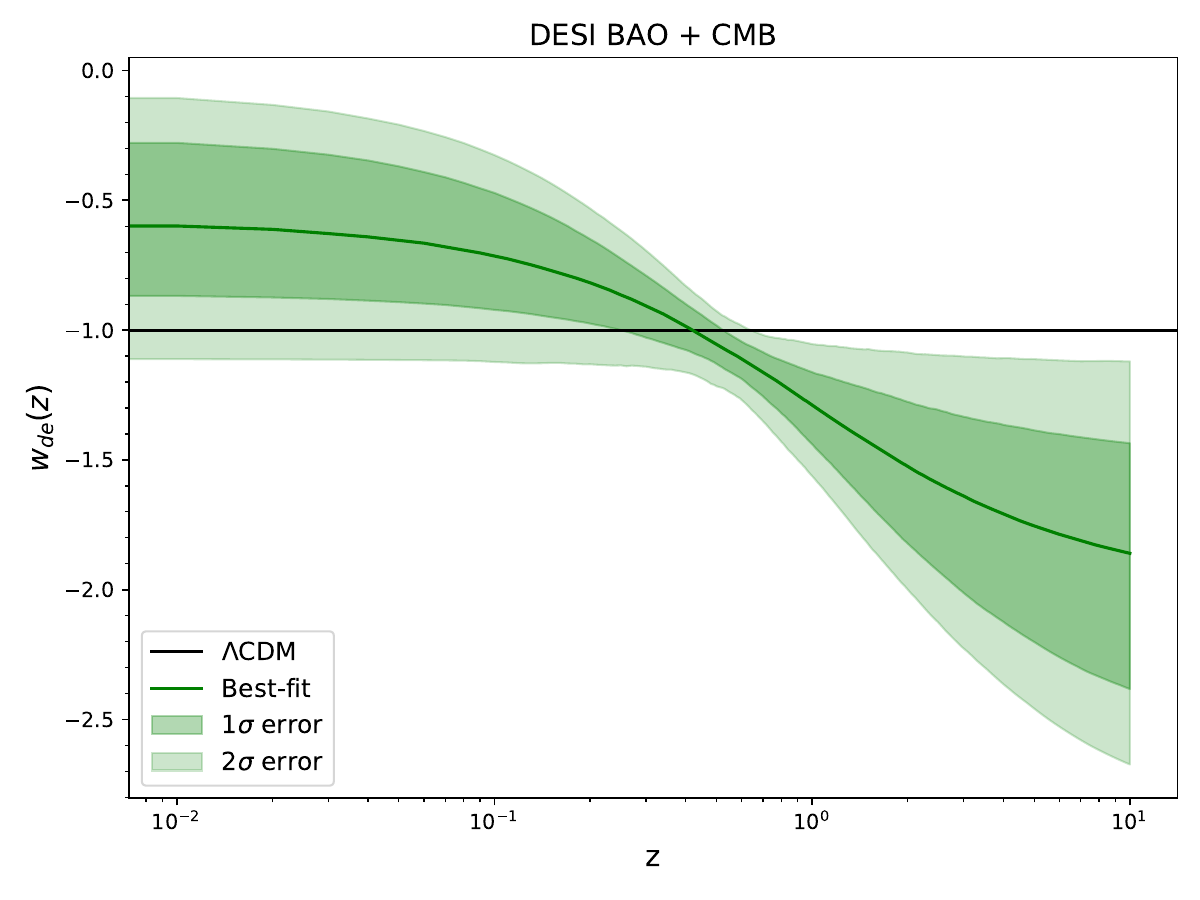}
\includegraphics[width=5cm]{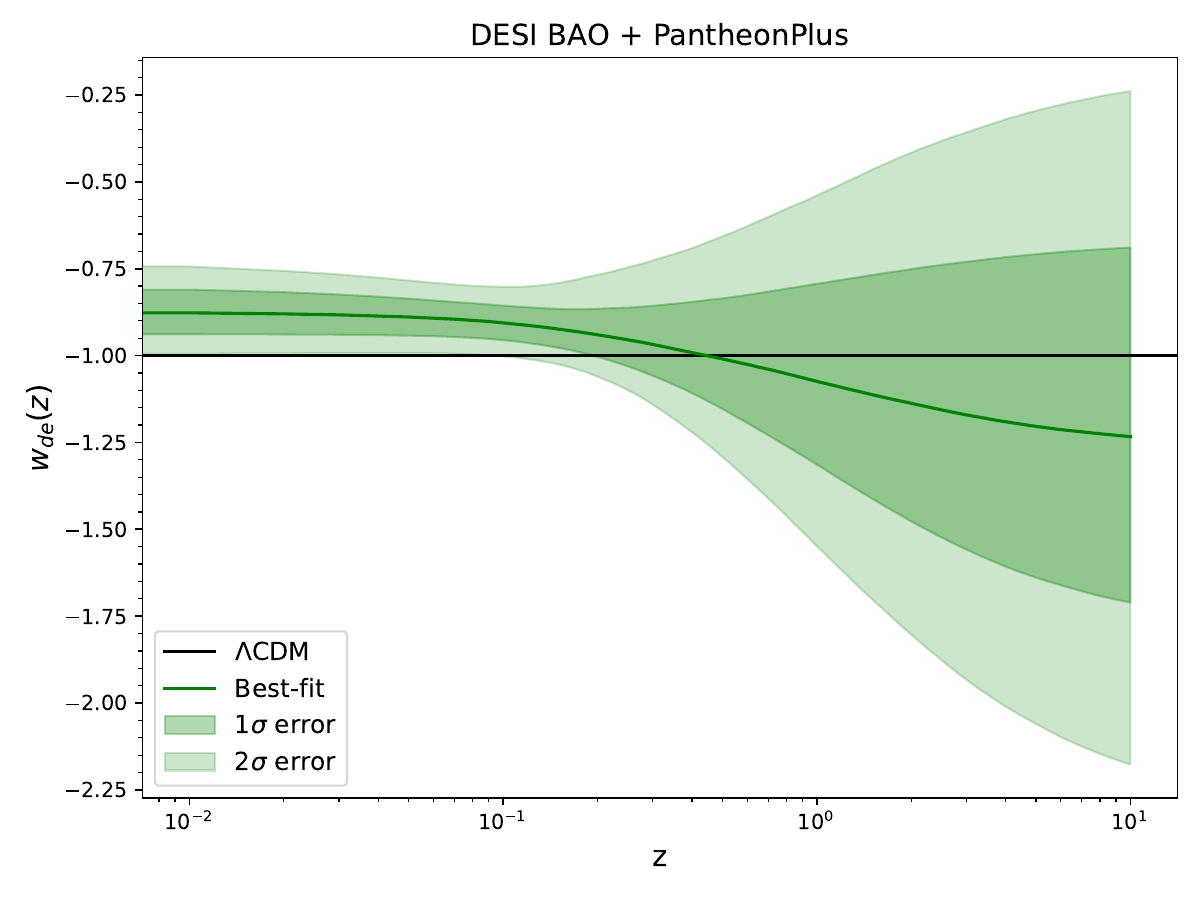}\includegraphics[width=5cm]{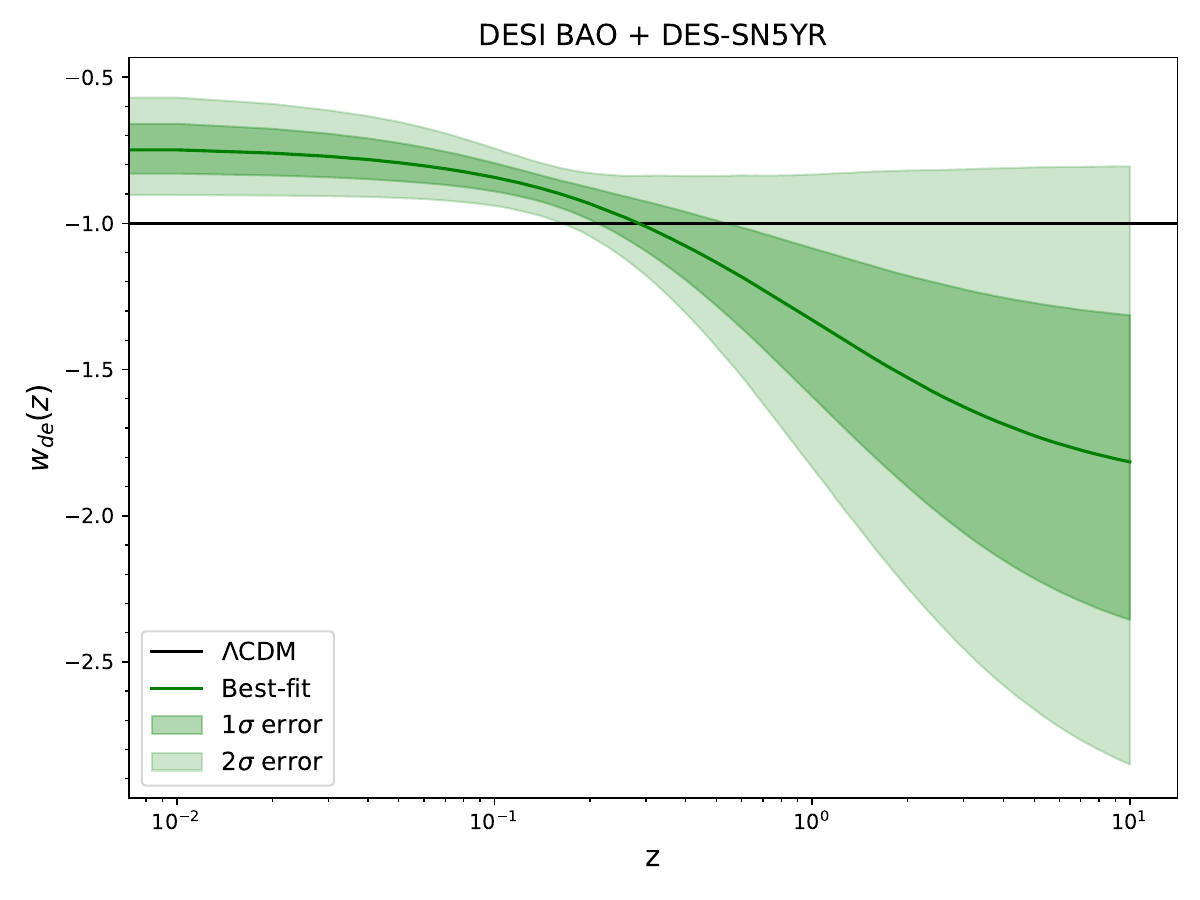}
\includegraphics[width=5cm]{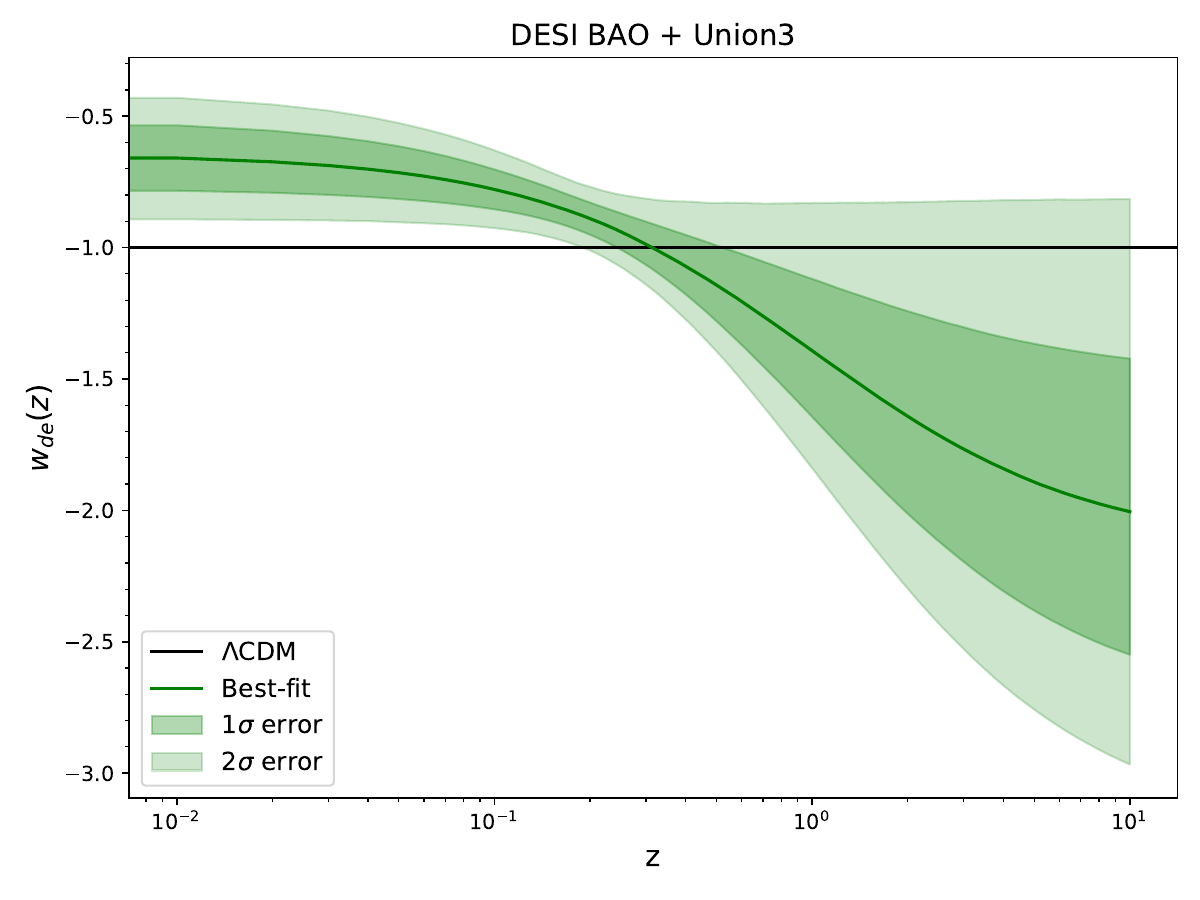}
   \includegraphics[width=5cm]{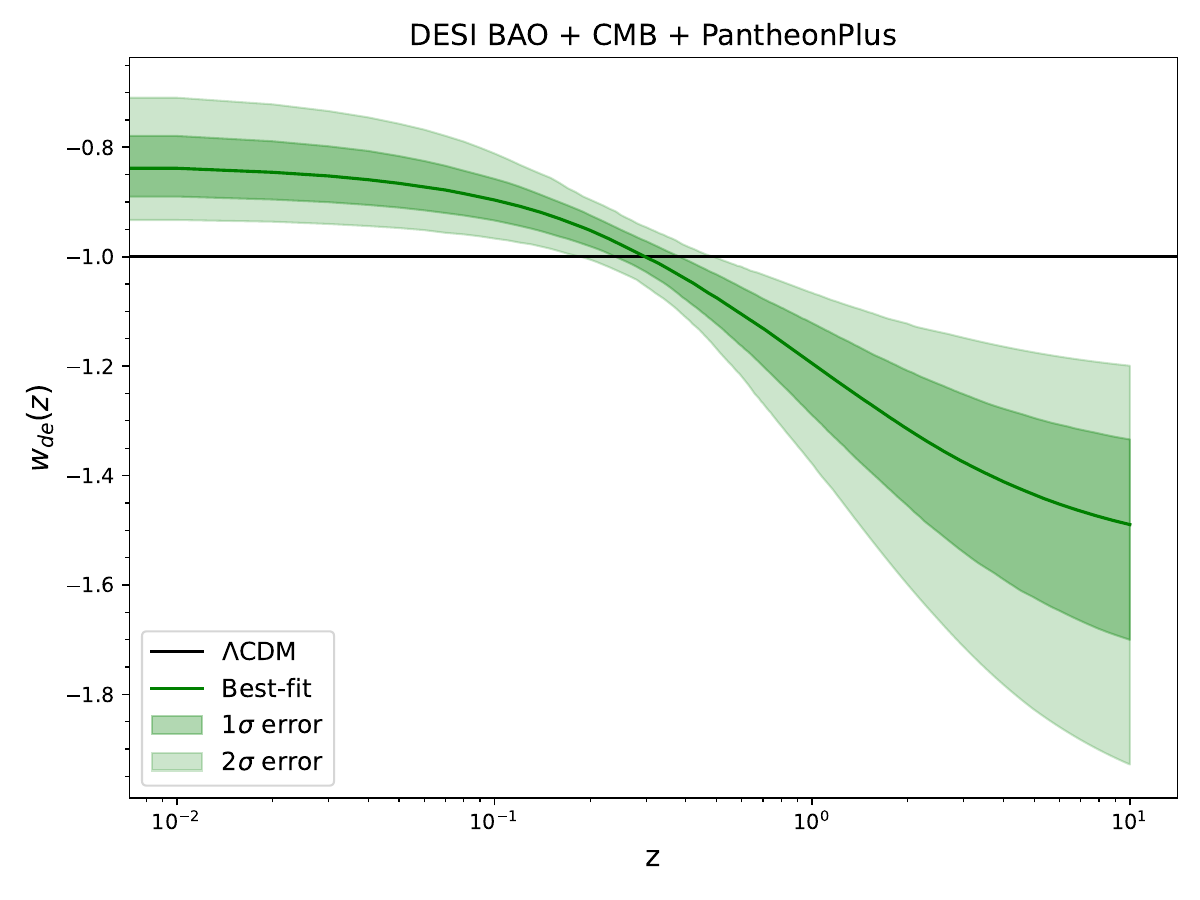}\includegraphics[width=5cm]{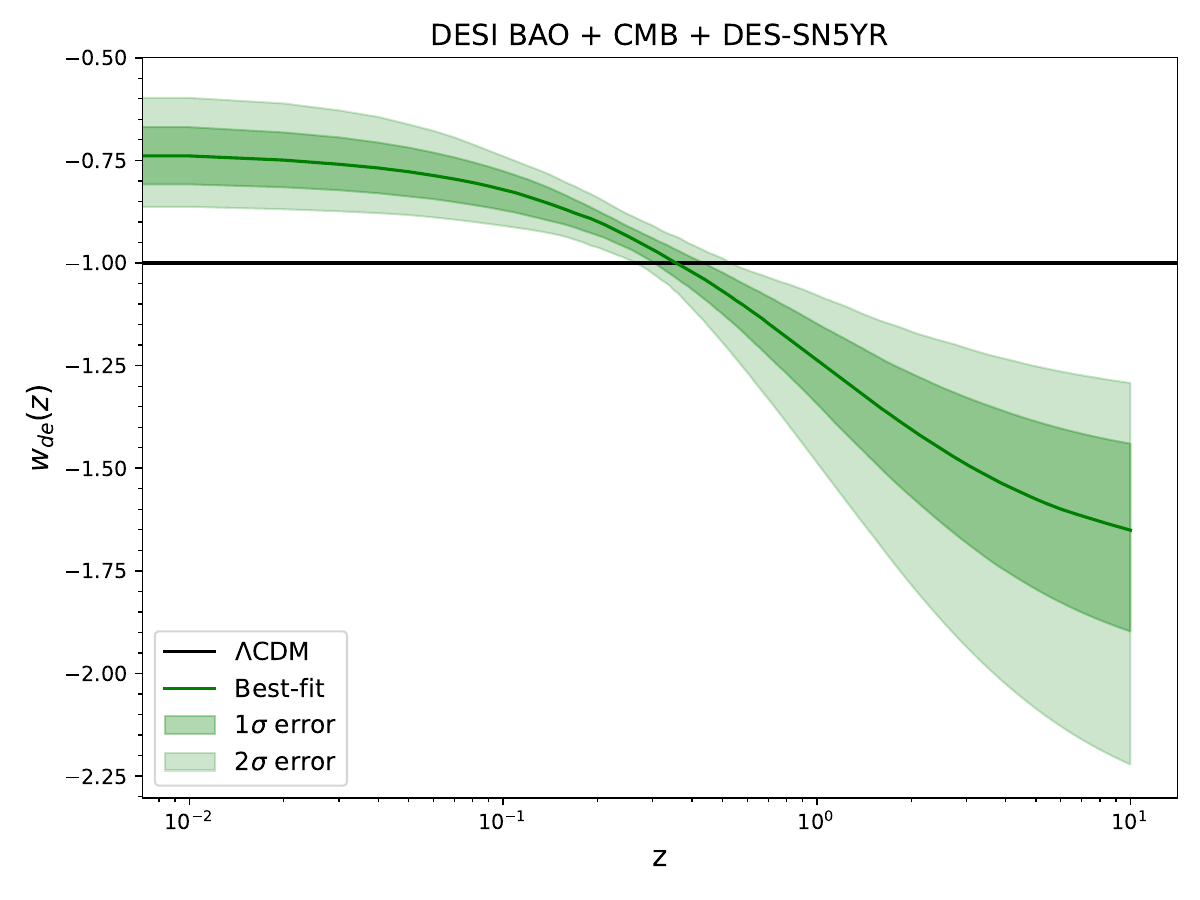}\includegraphics[width=5cm]{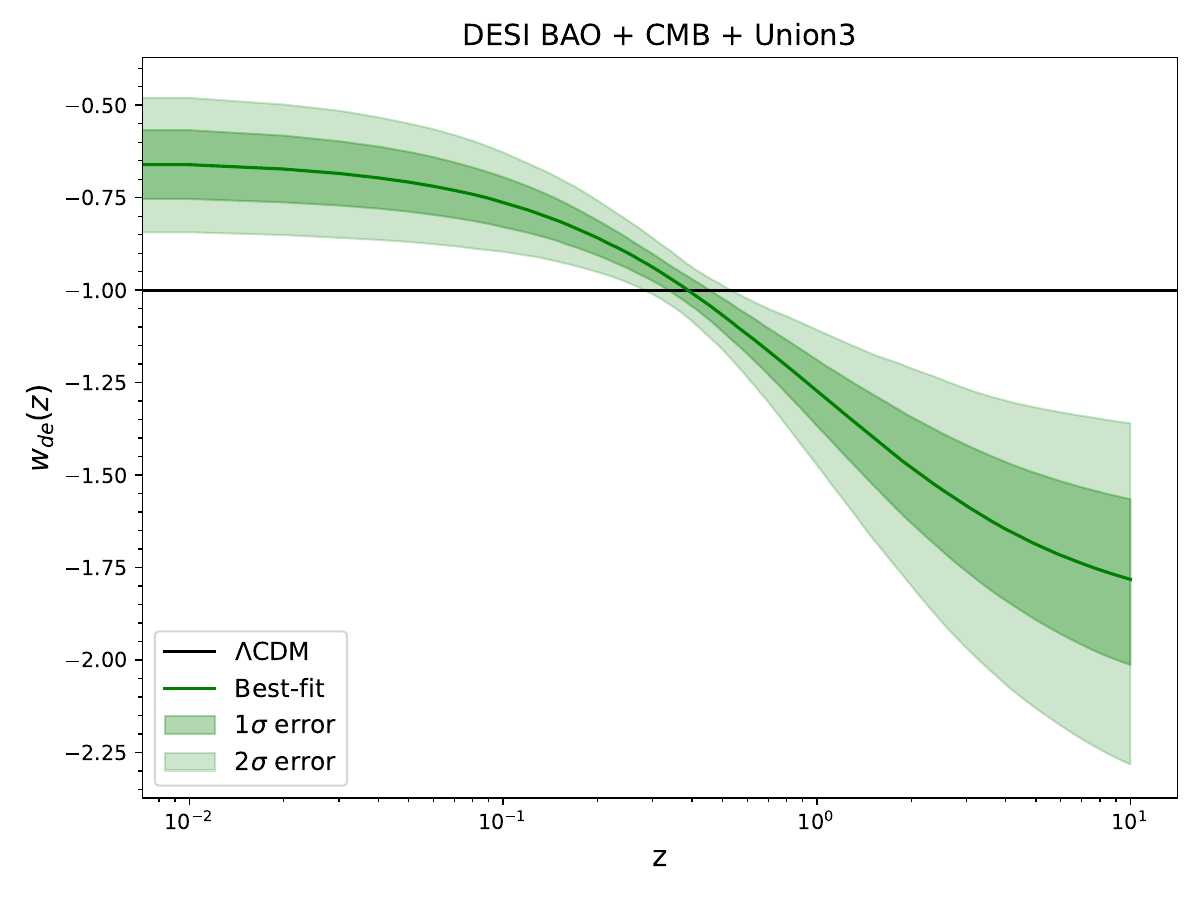}
\caption{Evolution of $w_{de}(z)$ in CPL parametrization utilizing the various combinations of the observational data.}
	\label{fig:w_CPL}
\end{figure*}

\begin{figure*} 
	\centering
    \includegraphics[width=5cm]{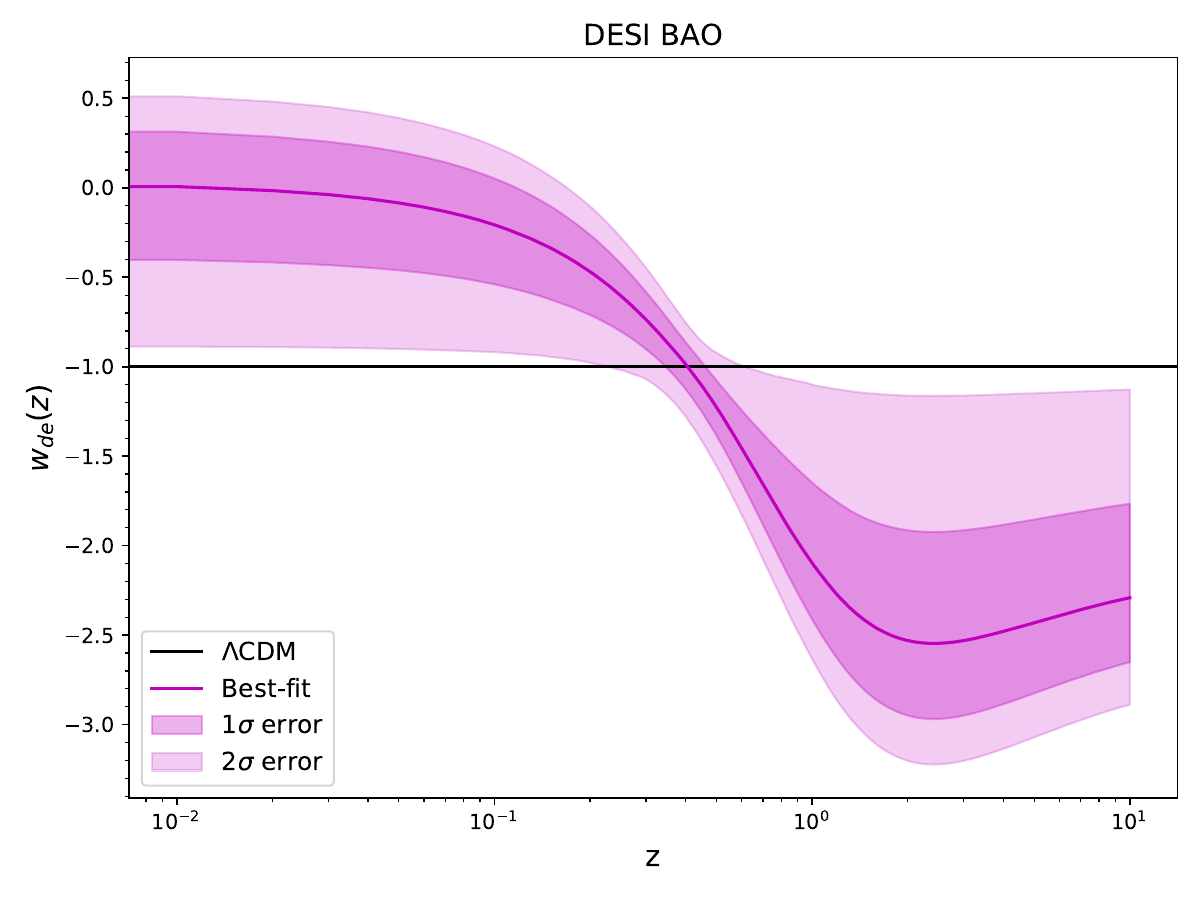}\includegraphics[width=5cm]{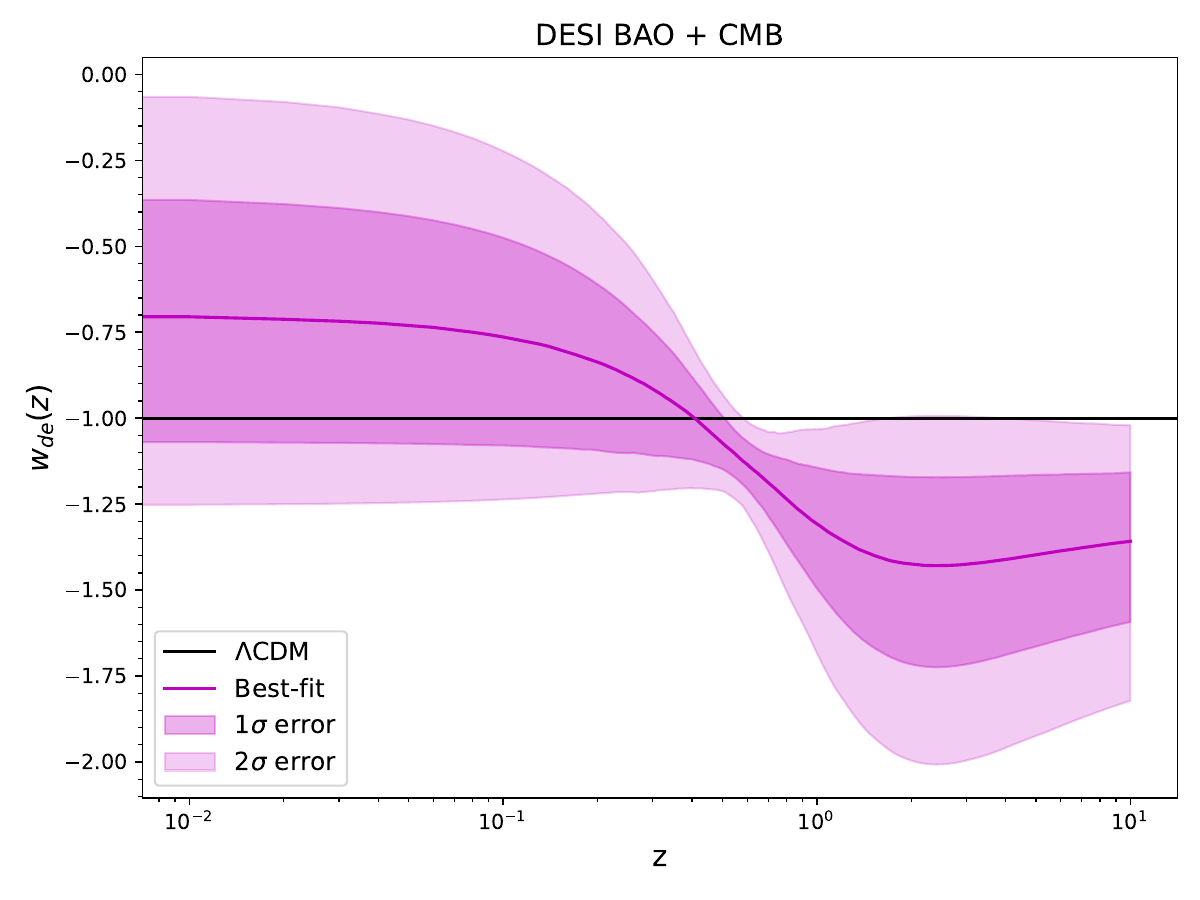}
\includegraphics[width=5cm]{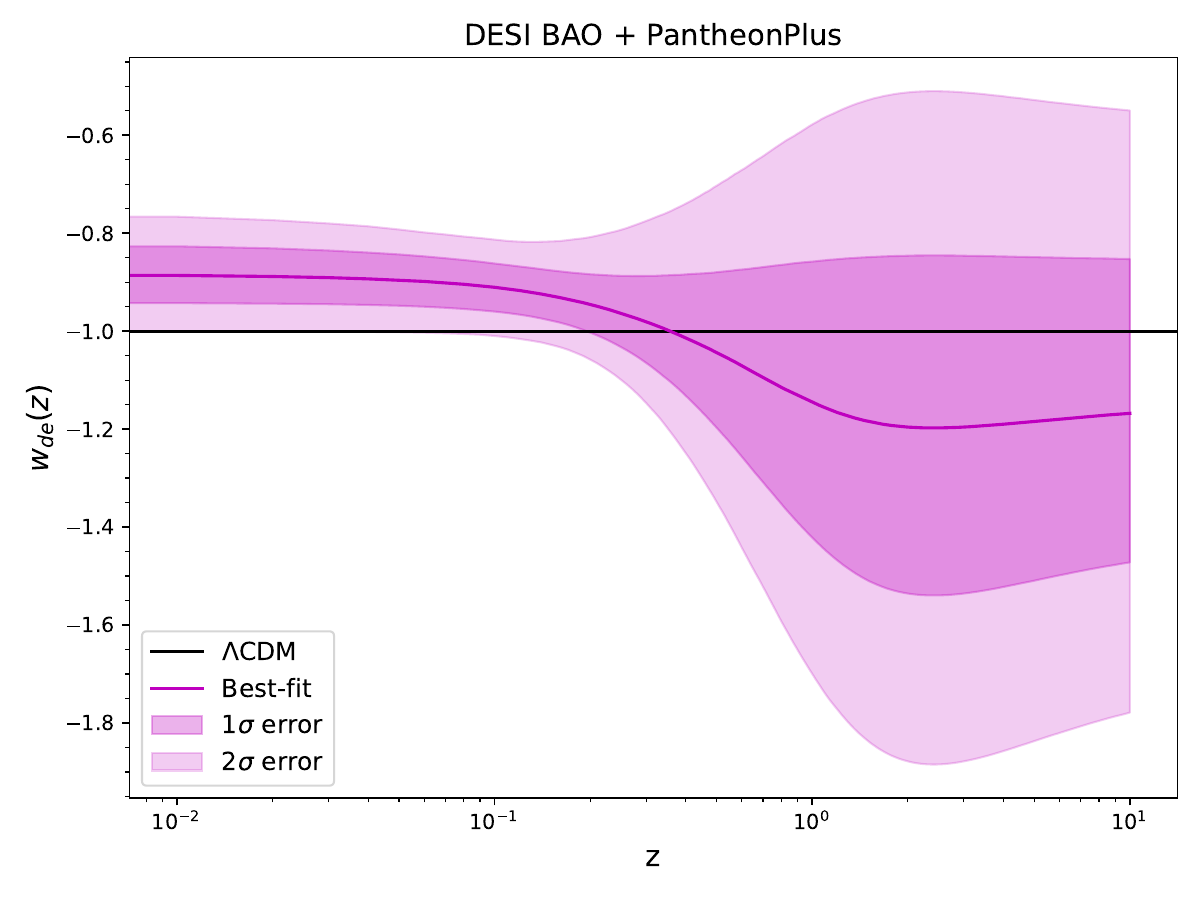}\includegraphics[width=5cm]{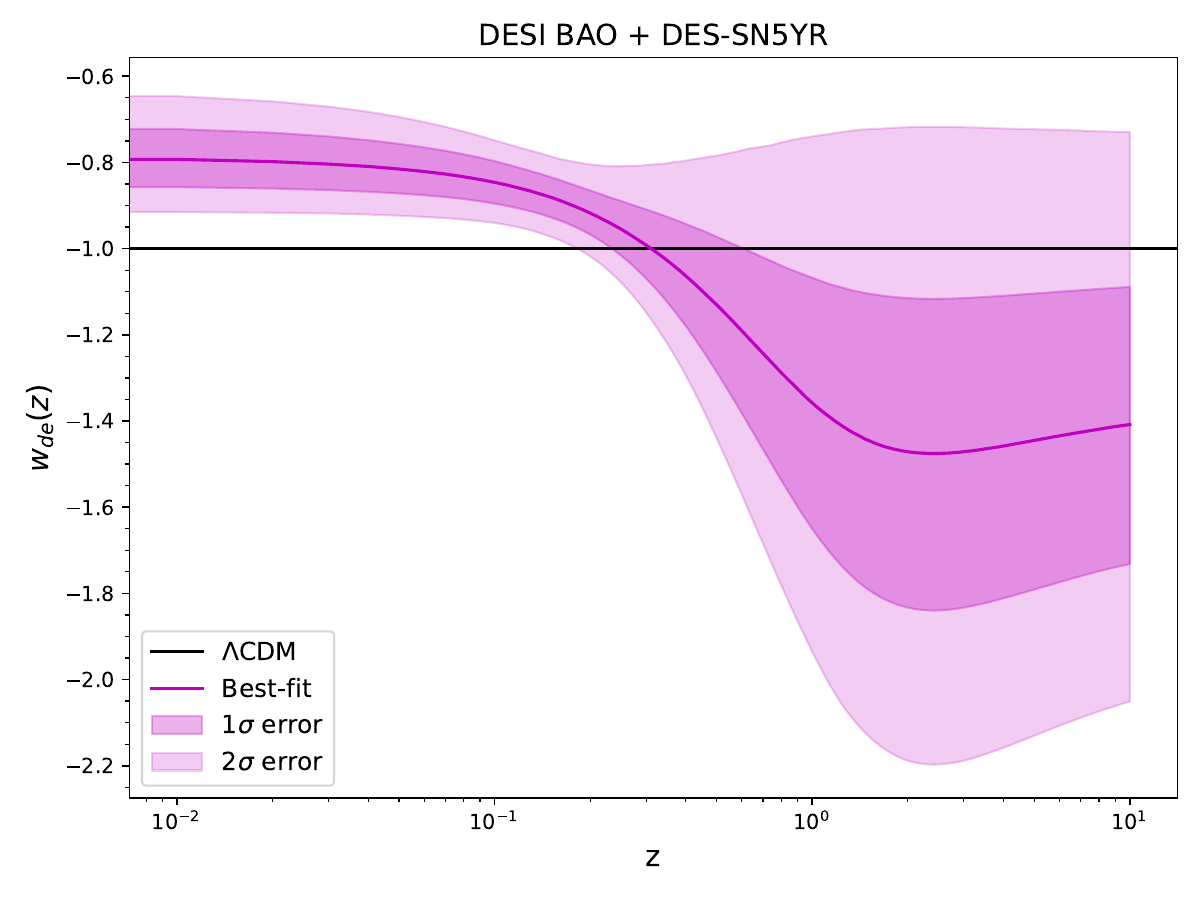}
\includegraphics[width=5cm]{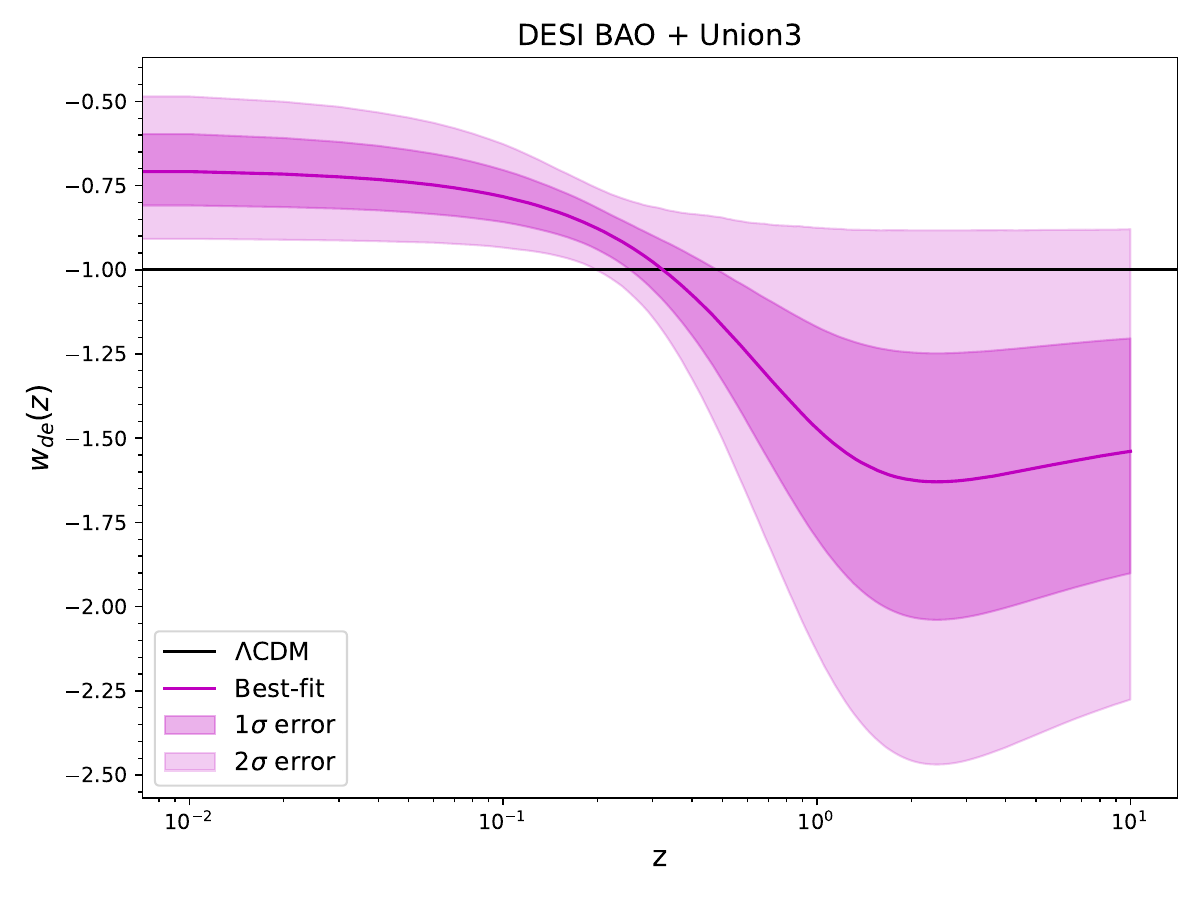}
   \includegraphics[width=5cm]{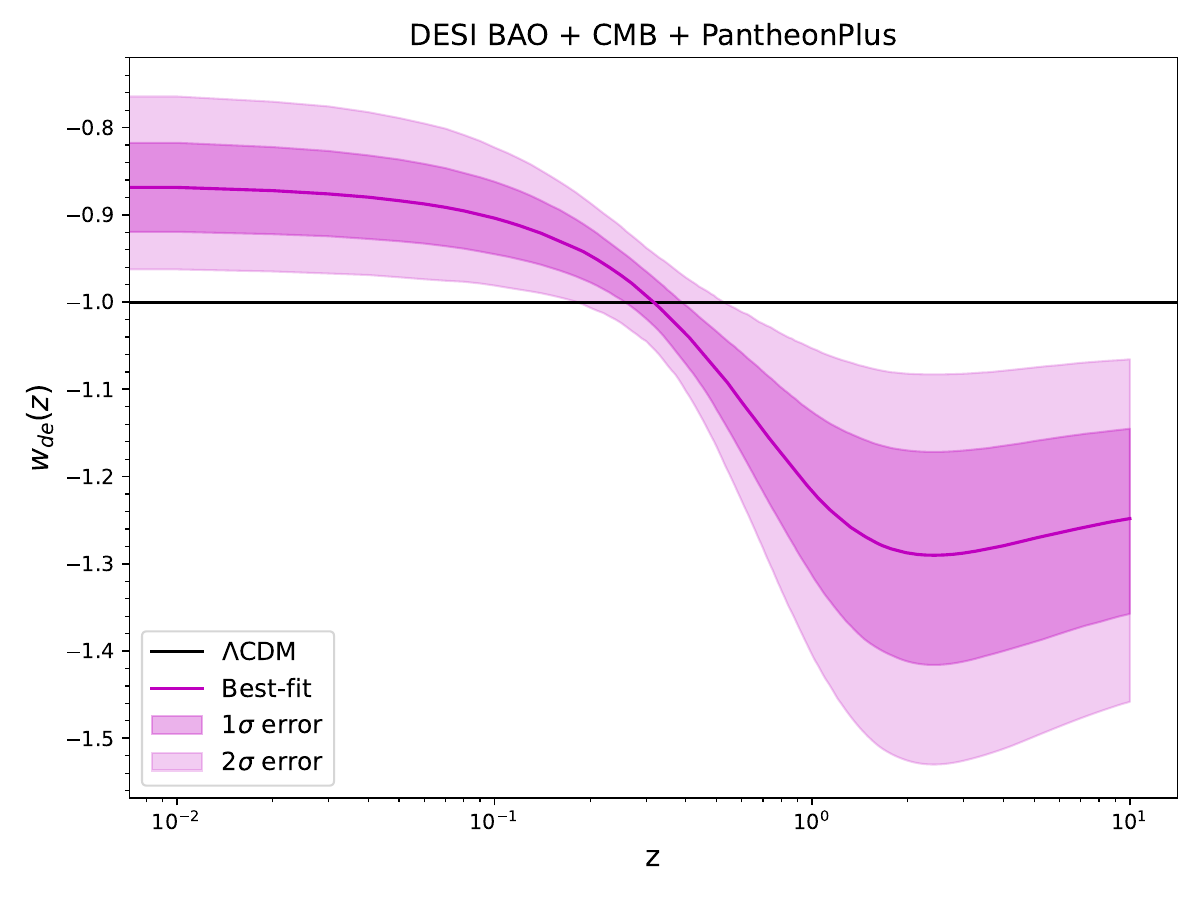}\includegraphics[width=5cm]{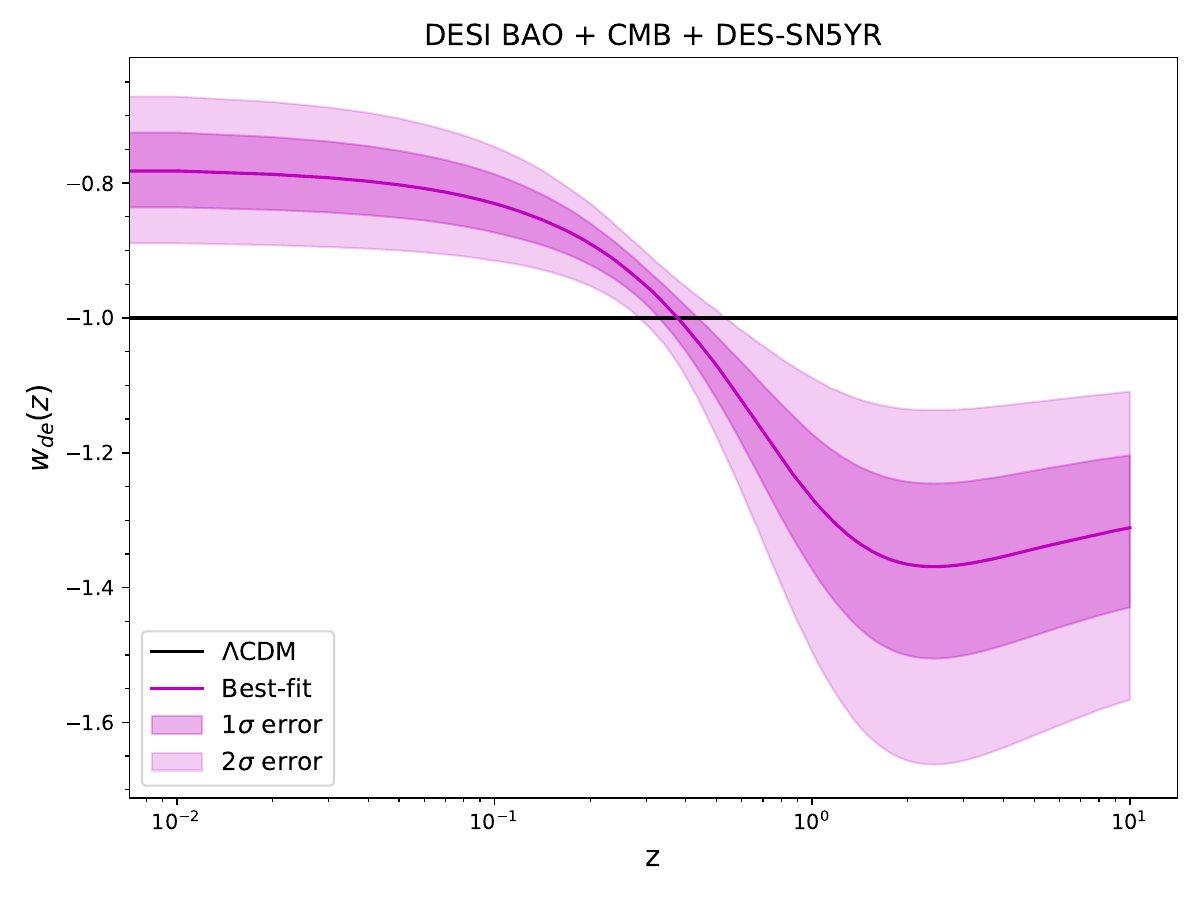}\includegraphics[width=5cm]{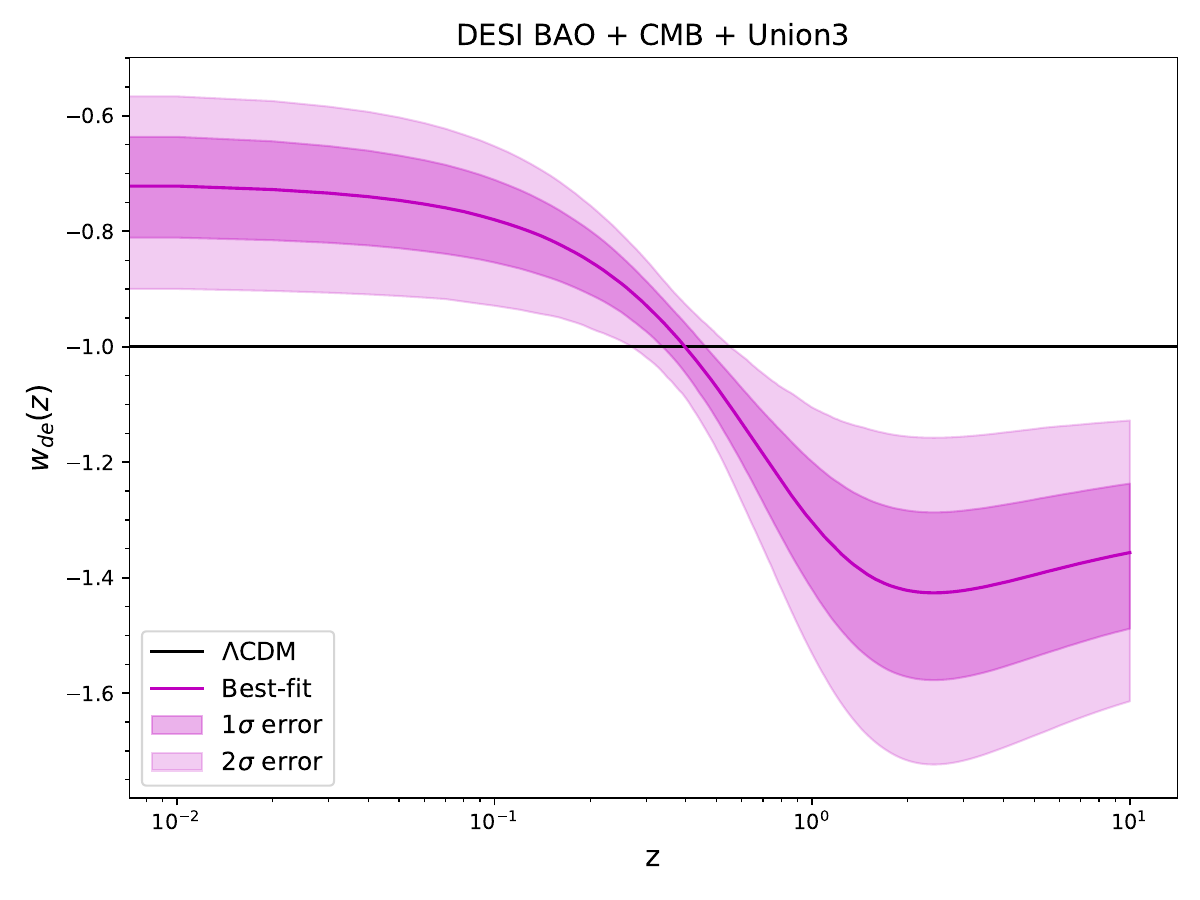}
	\caption{Evolution of $w_{de}(z)$ in BA parametrization using the various combinations of the observational data.}
	\label{fig:w_BA}
\end{figure*}

\begin{figure*} 
	\centering
    \includegraphics[width=5cm]{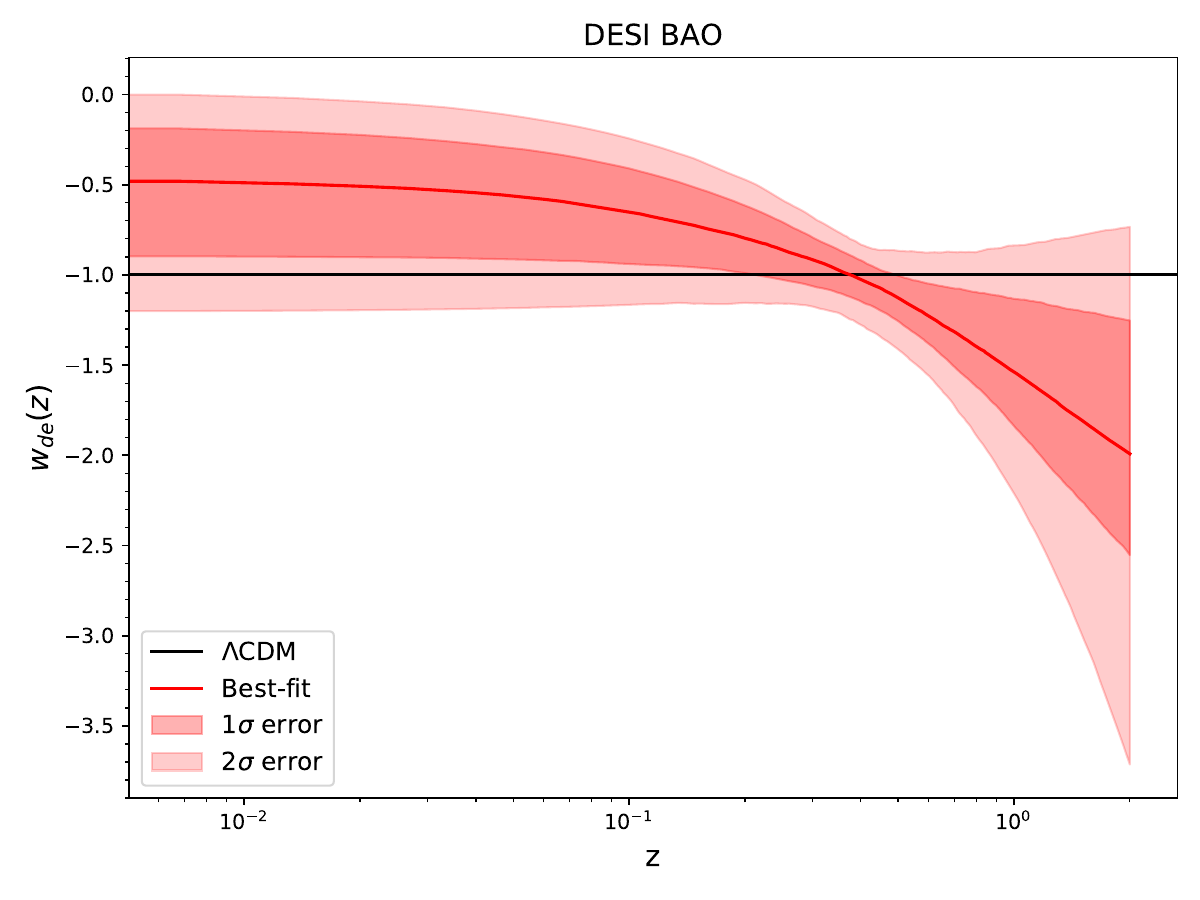}\includegraphics[width=5cm]{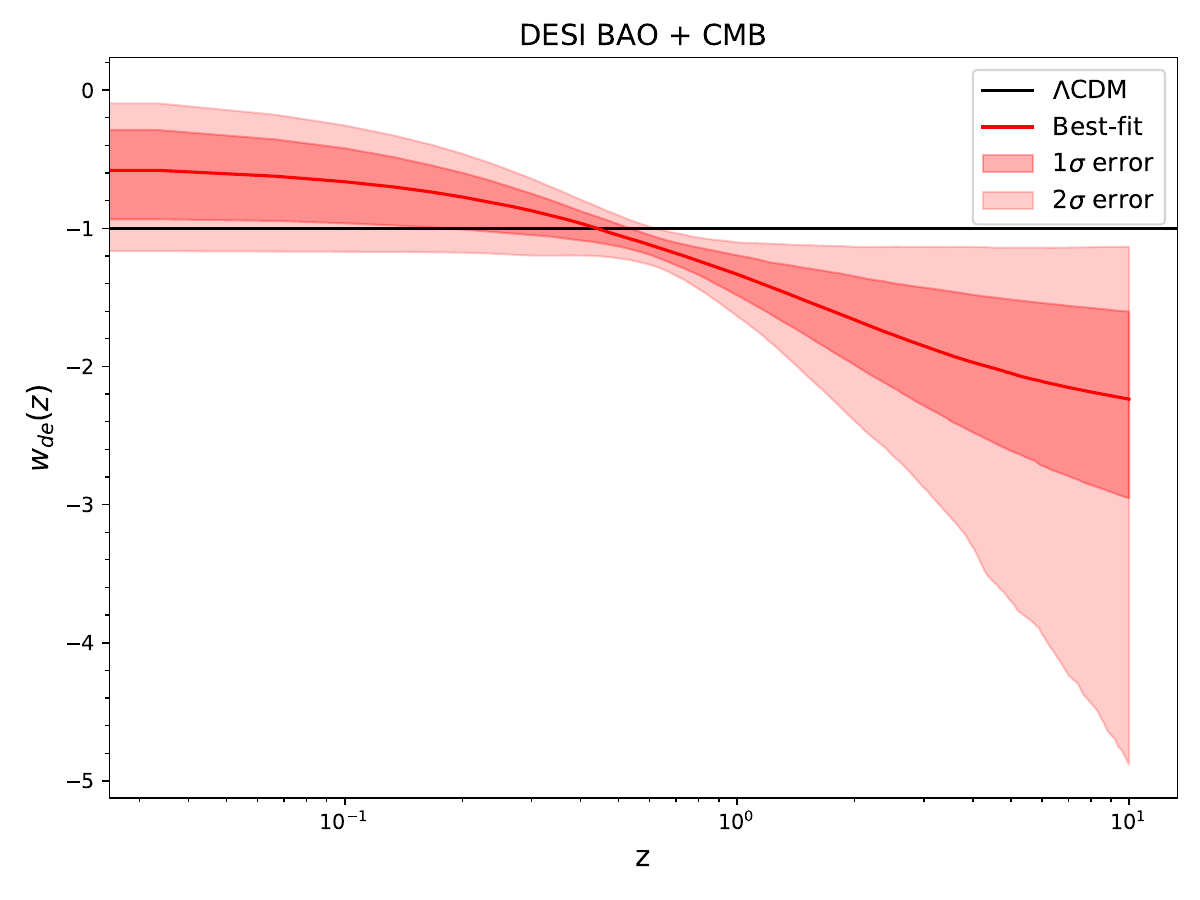}
\includegraphics[width=5cm]{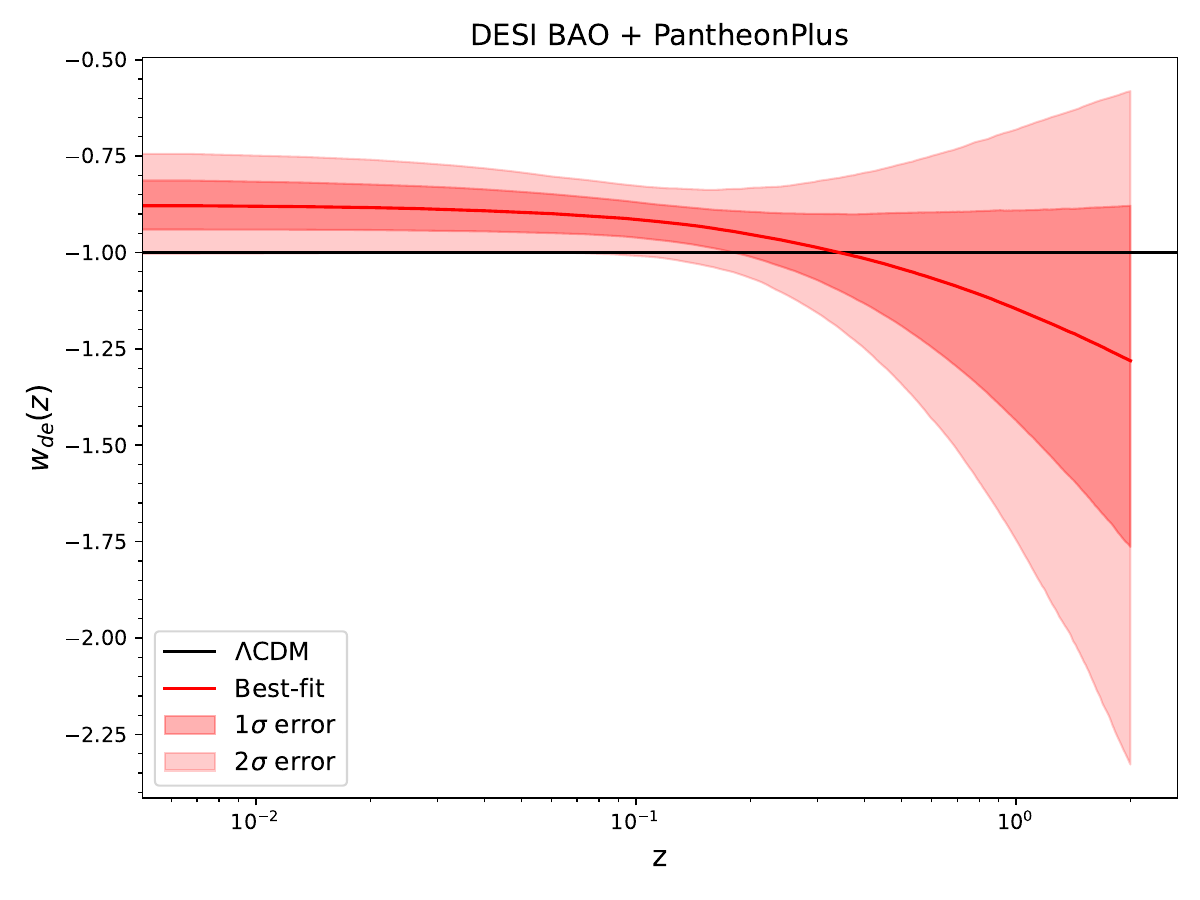}\includegraphics[width=5cm]{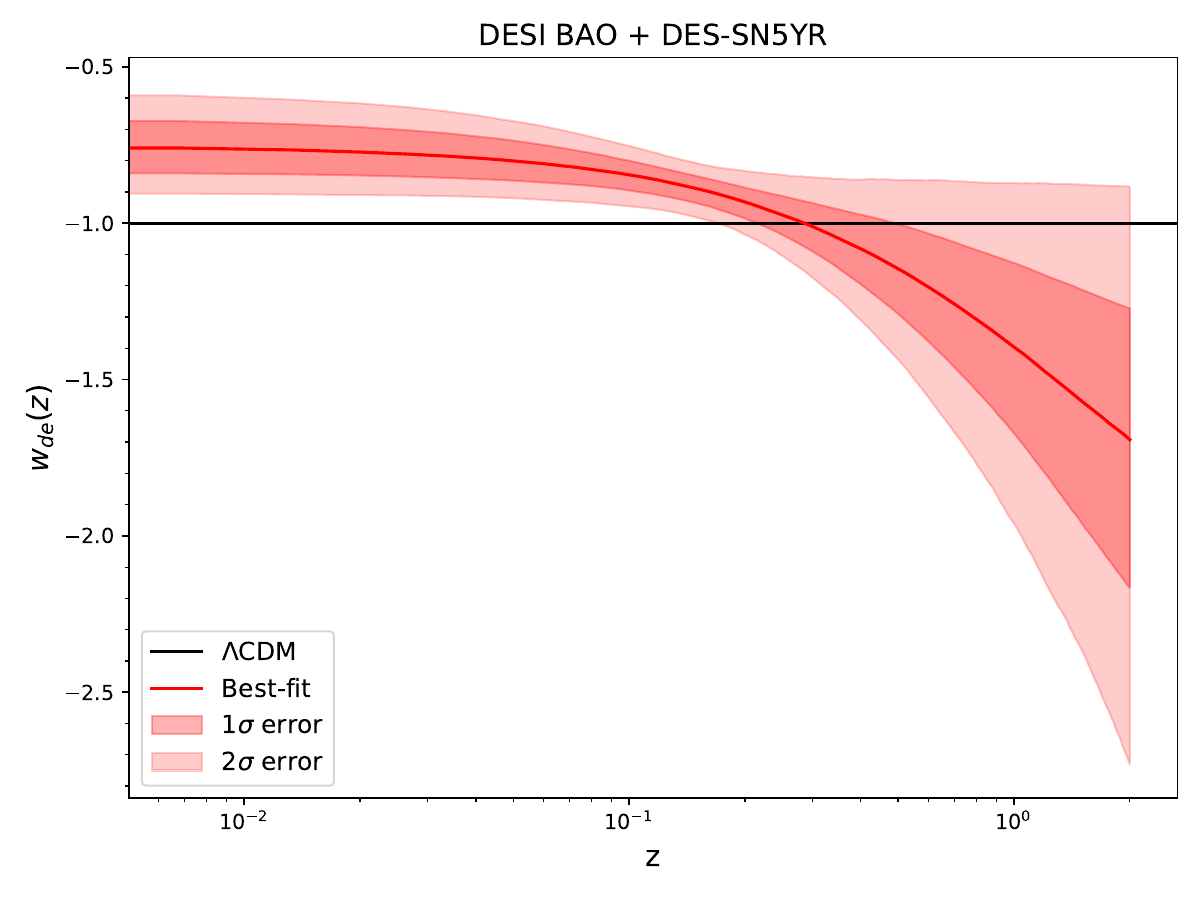}
\includegraphics[width=5cm]{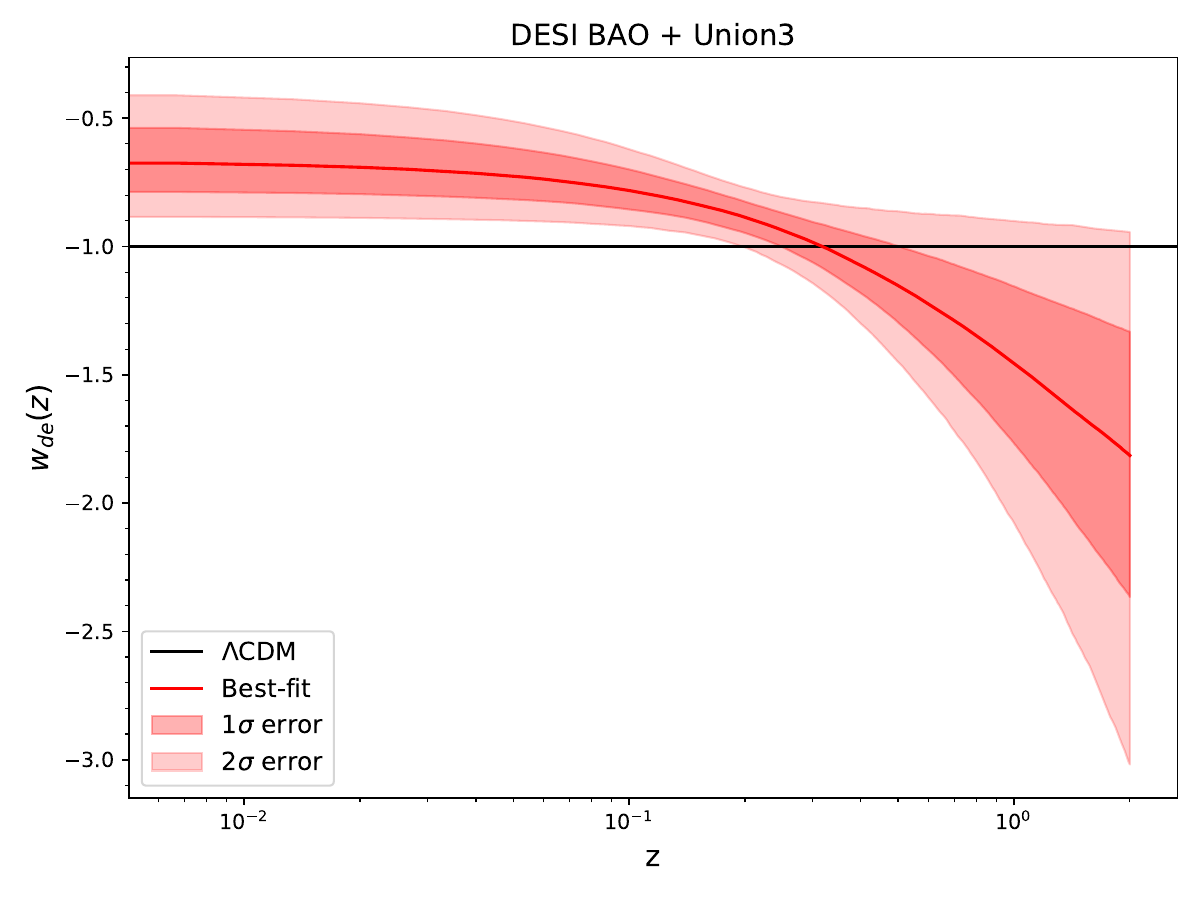}
   \includegraphics[width=5cm]{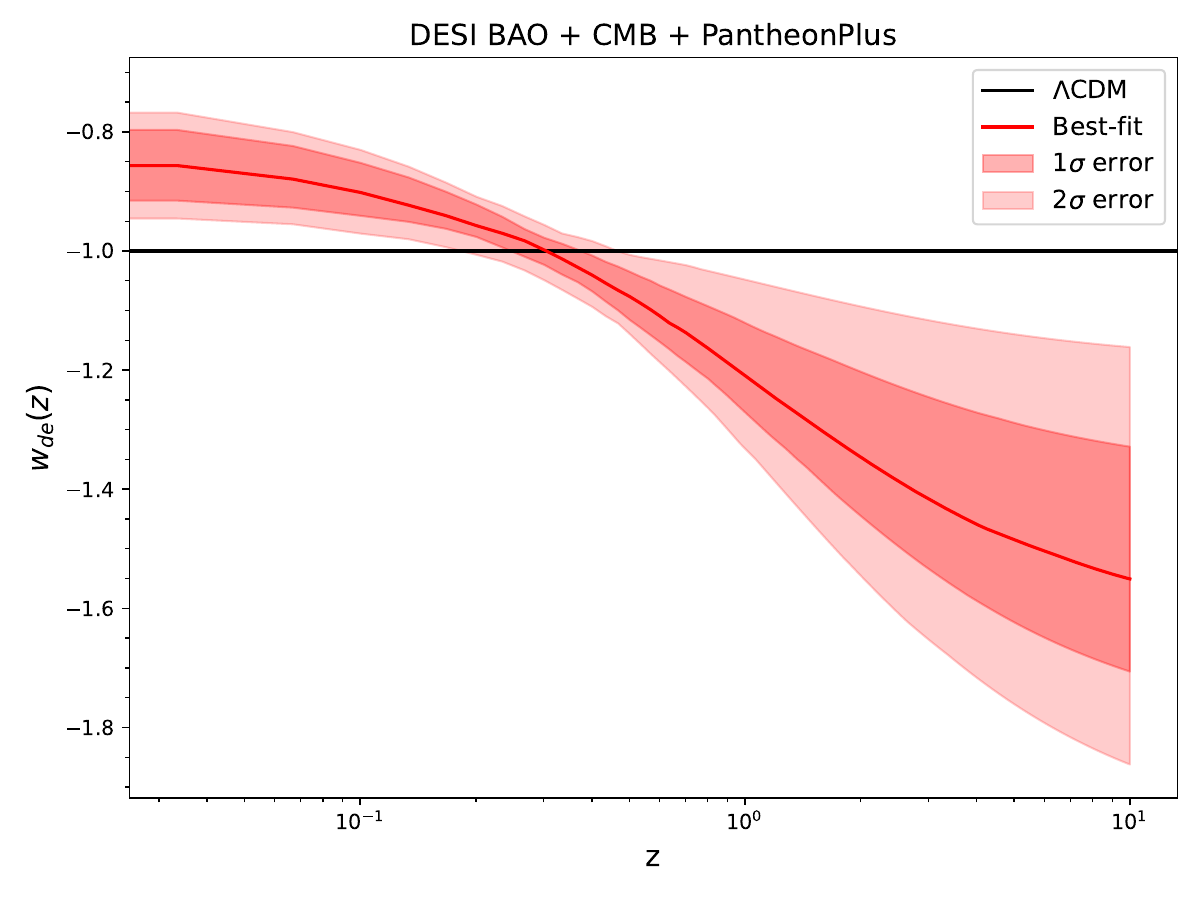}\includegraphics[width=5cm]{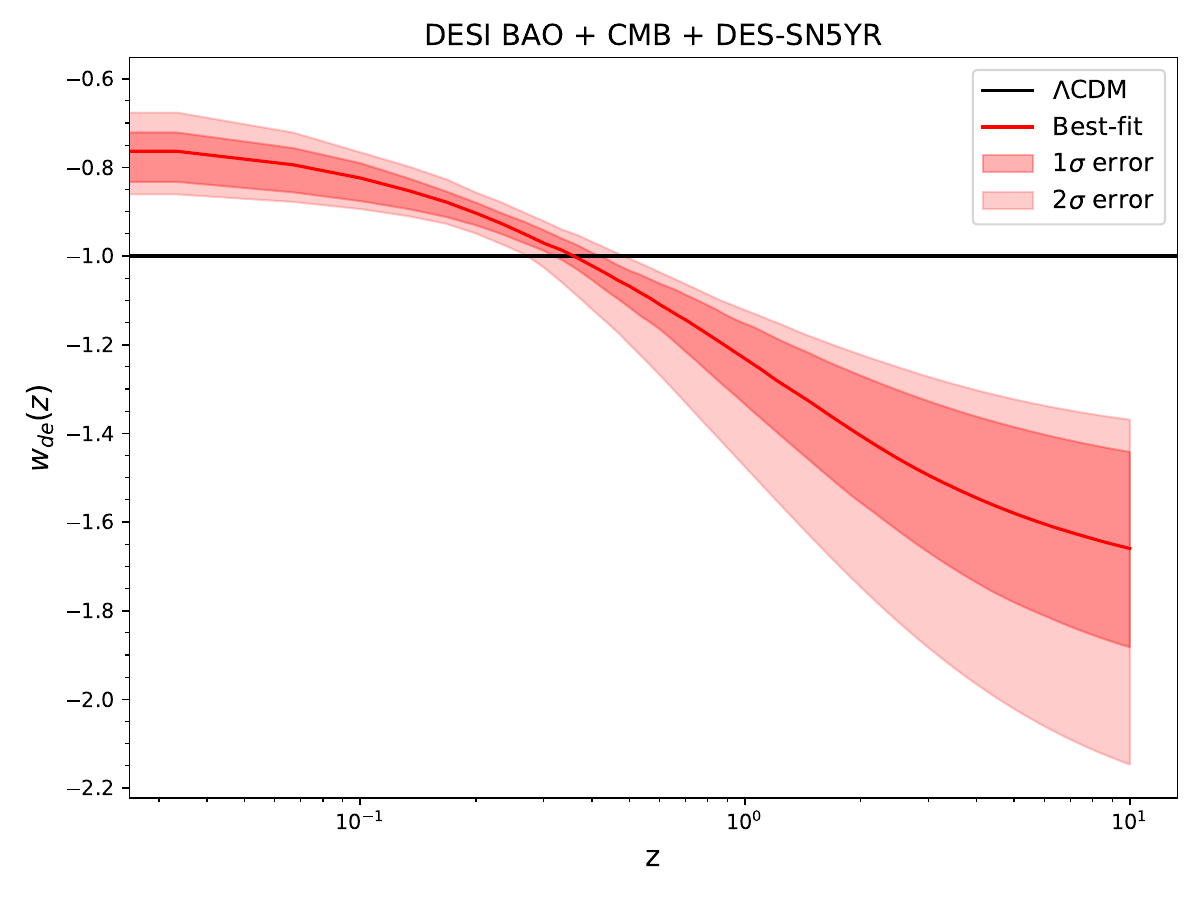}\includegraphics[width=5cm]{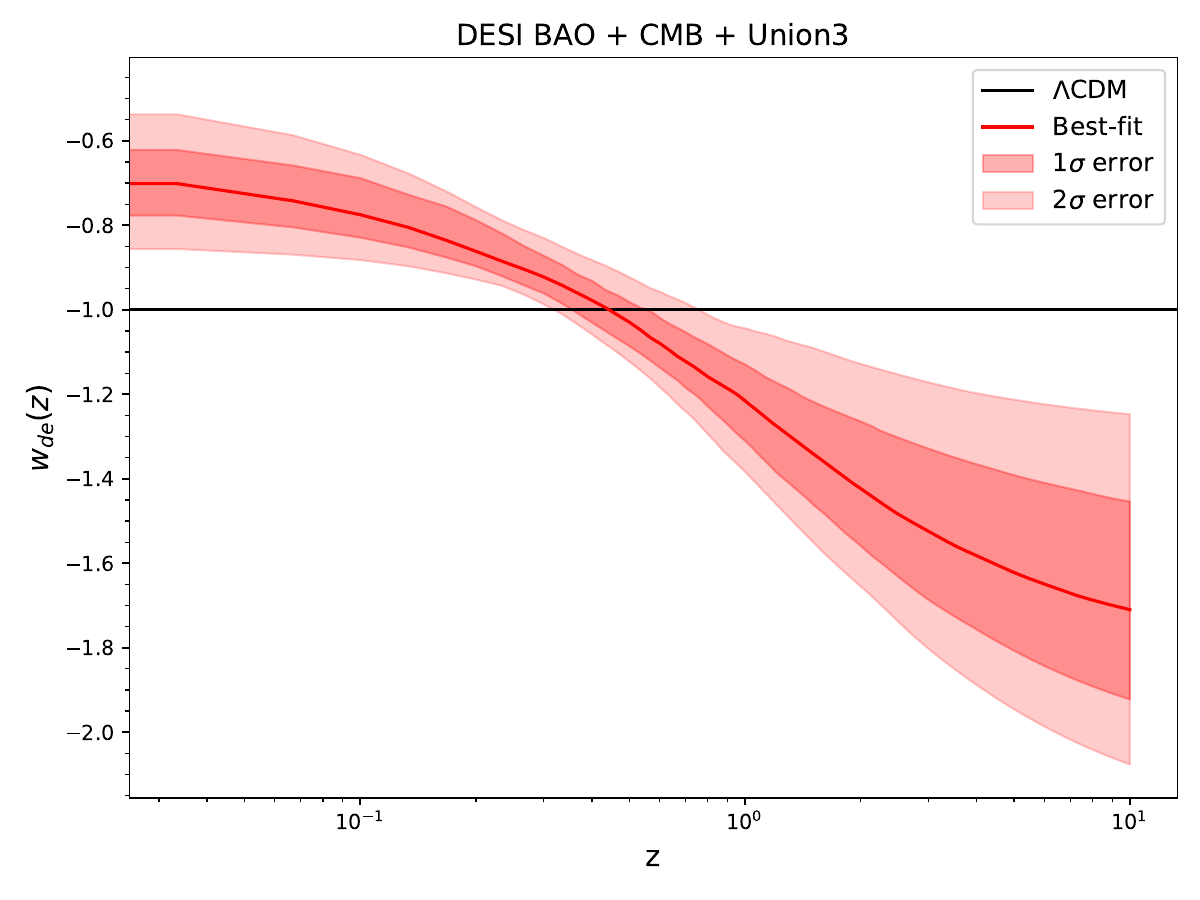}
	\caption{Evolution of $w_{de}(z)$ in Padé parametrization using the various combinations of the observational data.}
	\label{fig:w_PadéI}
\end{figure*}

\begin{figure*} 
	\centering
    \includegraphics[width=5cm]{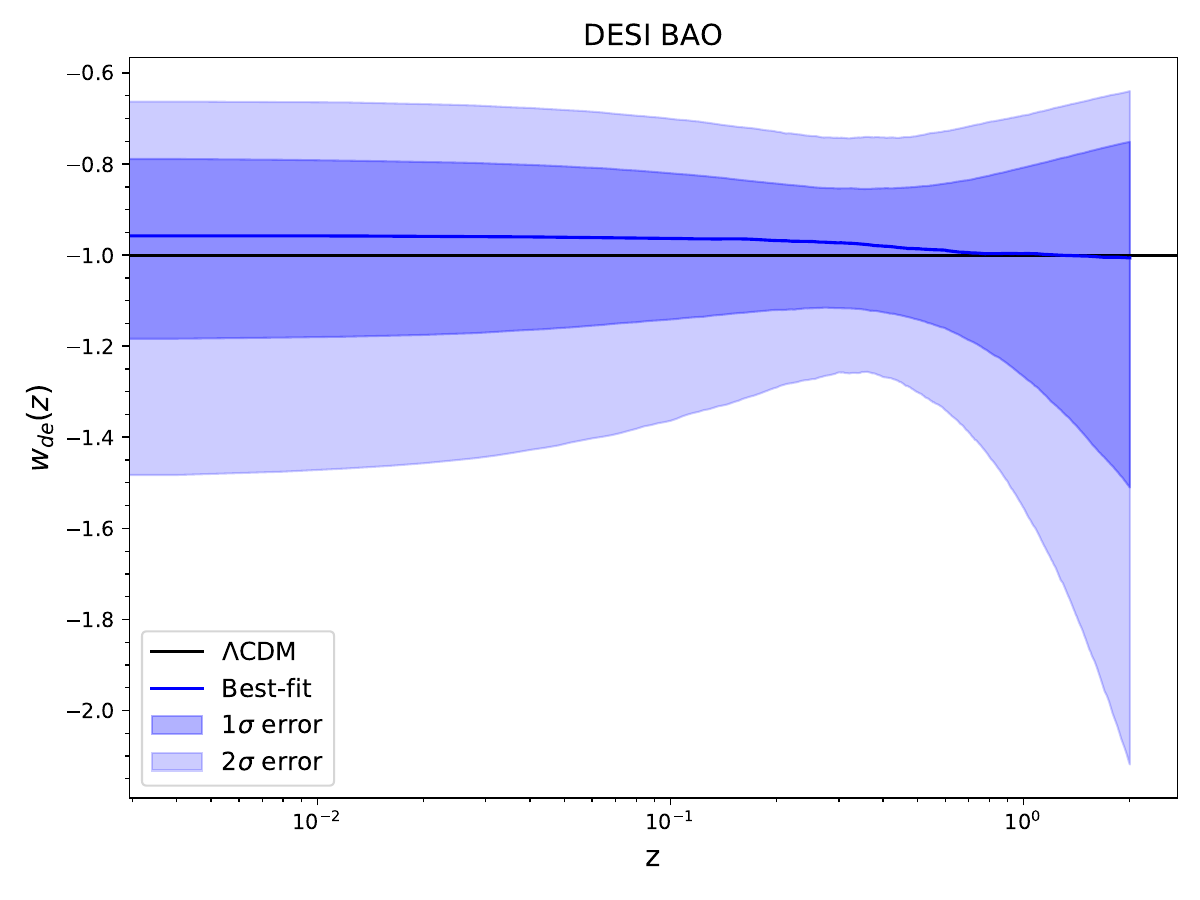}\includegraphics[width=5cm]{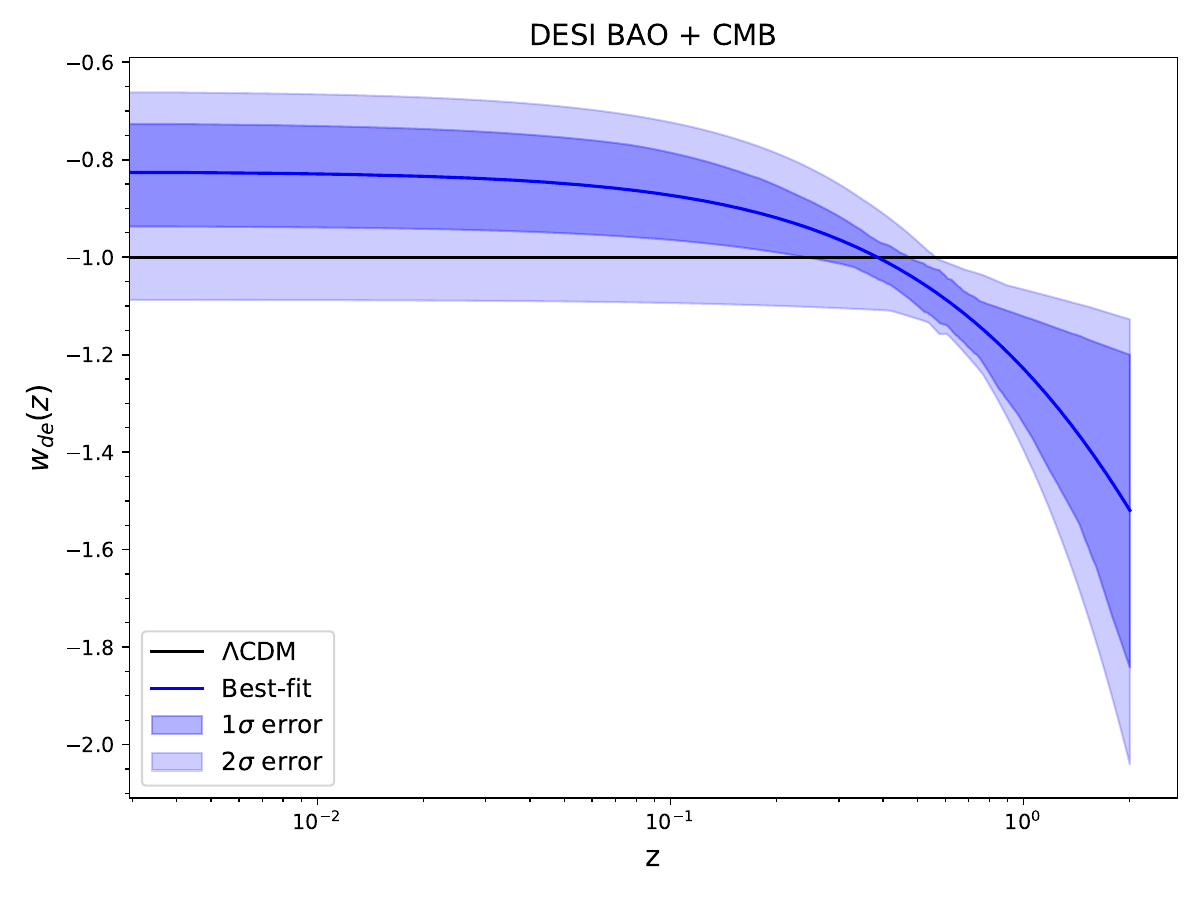}
\includegraphics[width=5cm]{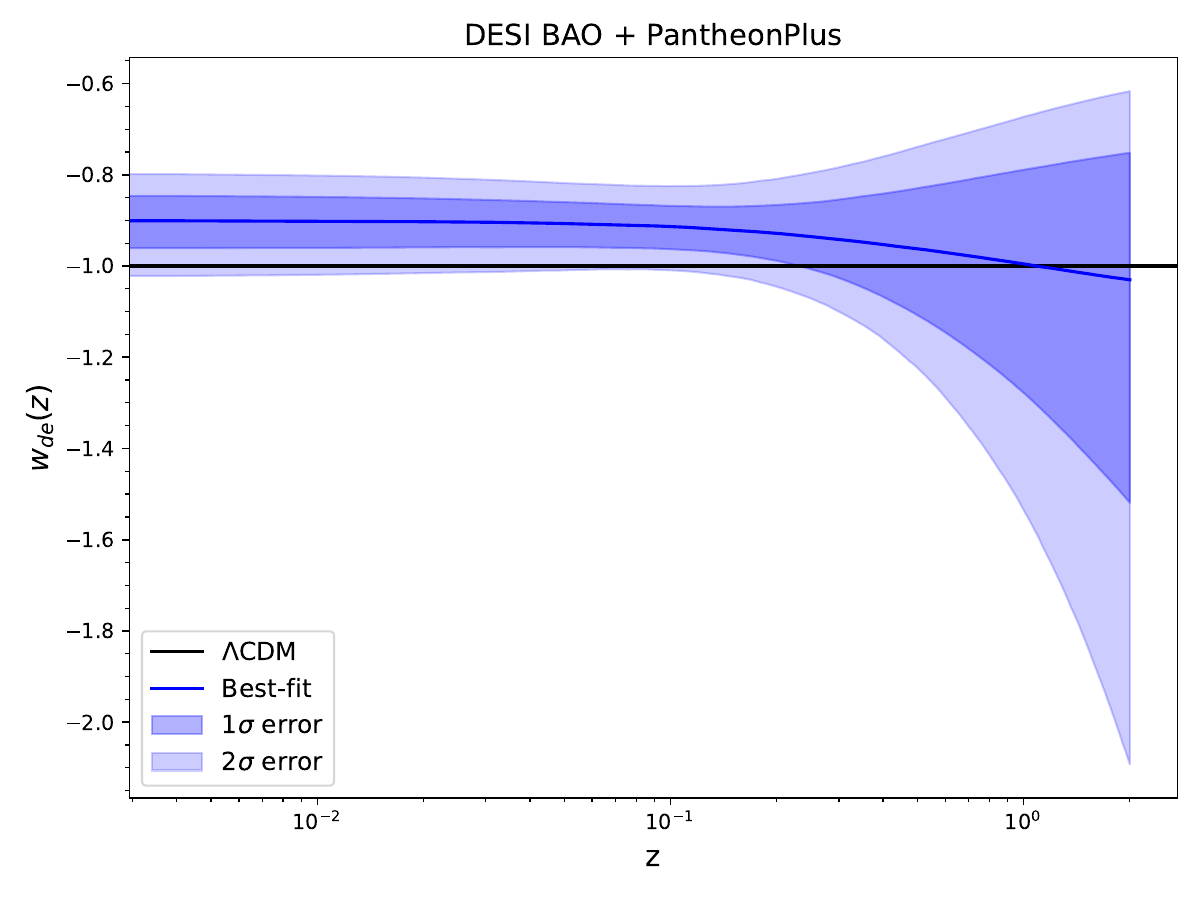}\includegraphics[width=5cm]{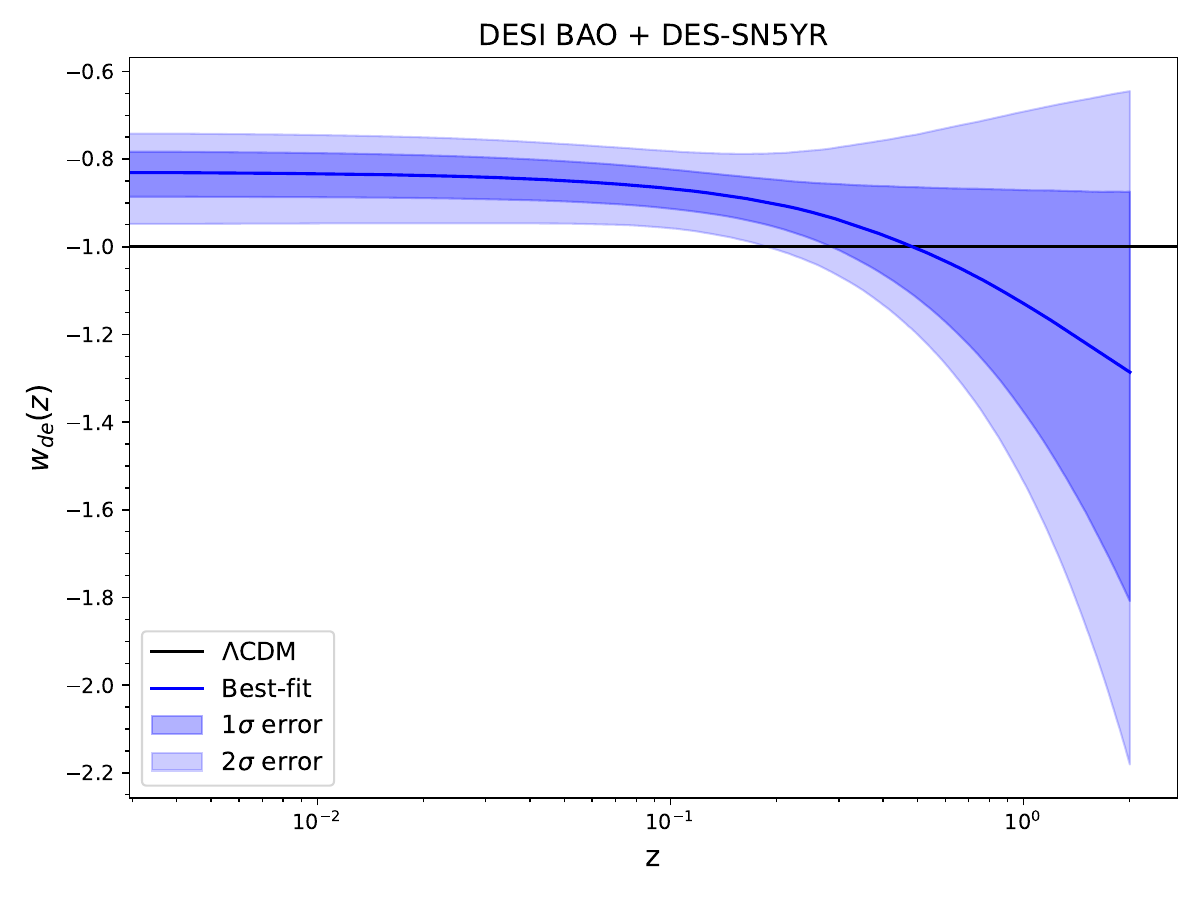}
\includegraphics[width=5cm]{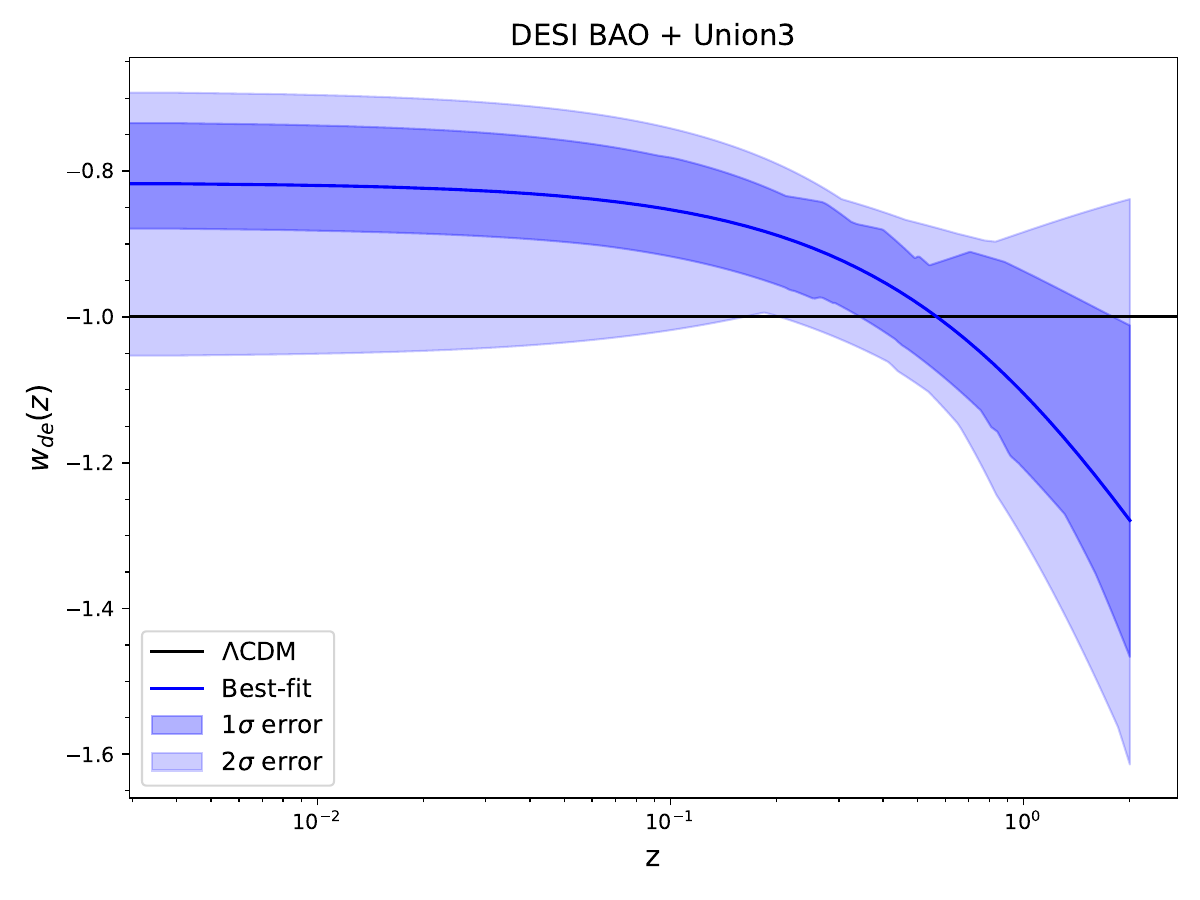}
   \includegraphics[width=5cm]{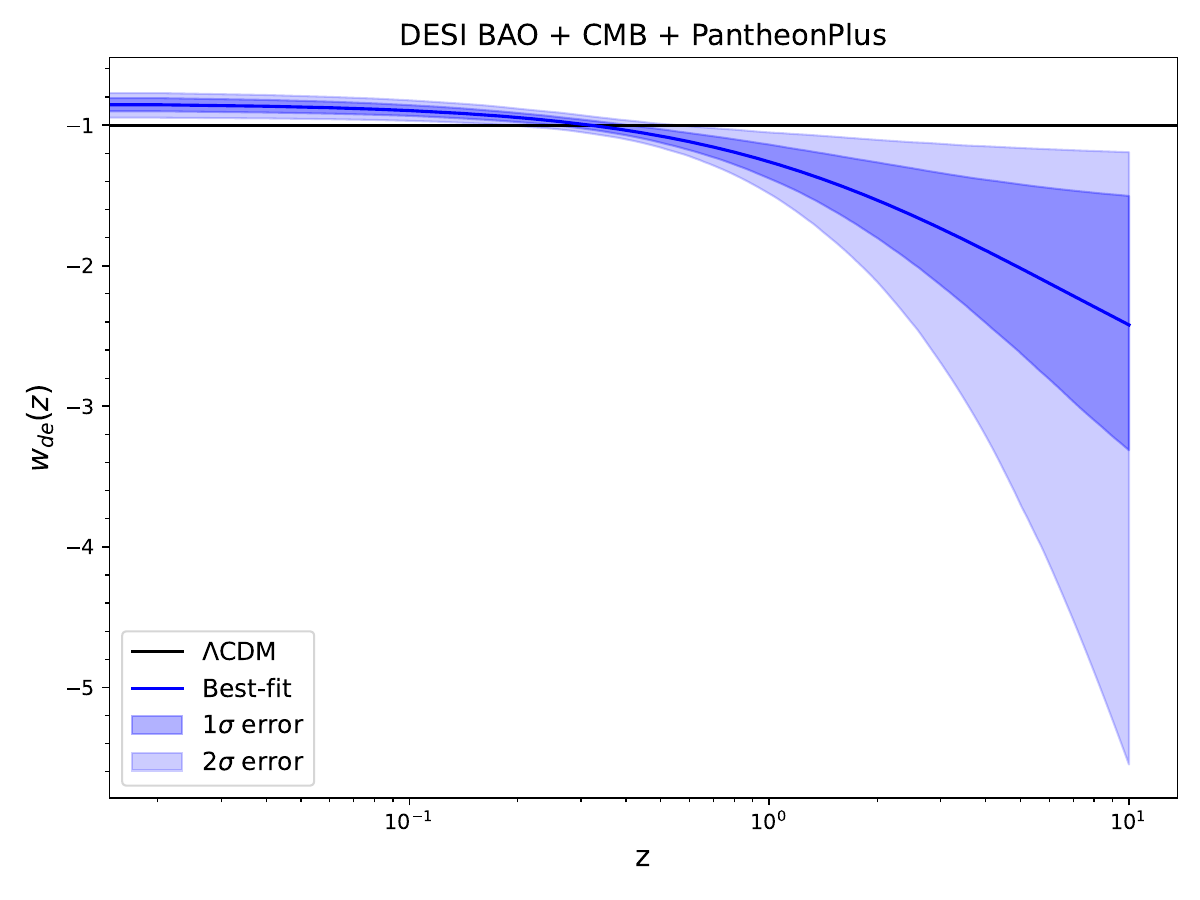}\includegraphics[width=5cm]{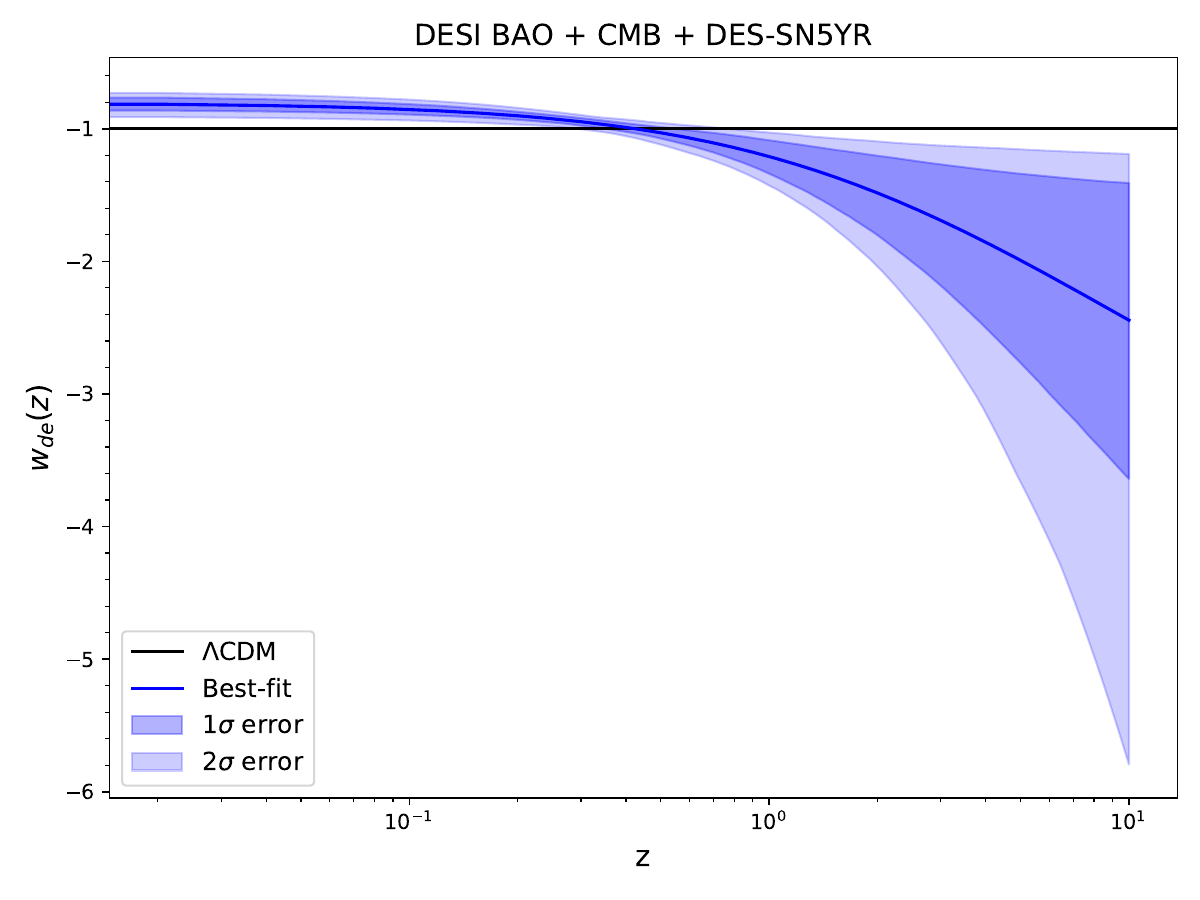}\includegraphics[width=5cm]{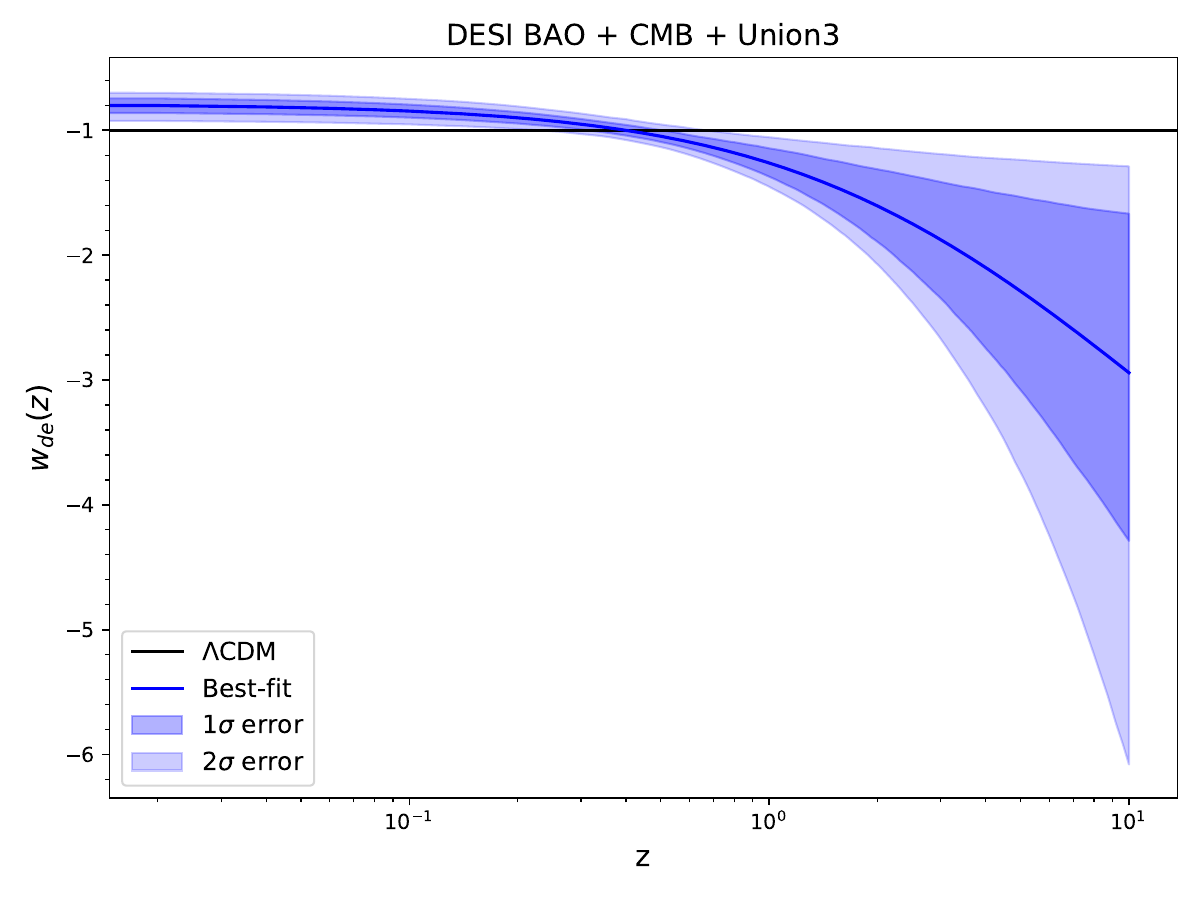}
	\caption{Evolution of $w_{de}(z)$ in Simplified Padé parametrization using the various combinations of the observational data.}
	\label{fig:w_SimPadé}
\end{figure*}

In Figures (\ref{fig:w_CPL}, \ref{fig:w_BA}, \ref{fig:w_PadéI}, \& \ref{fig:w_SimPadé}), we present the redshift evolution of the EoS parameter, $w_{de}$, for the CPL and BA models (using constraints on $w_0$ and $w_a$) and the Padé/simplified Padé models (using constraints on $w_0$, $w_1$ and $w_2$), as reported in Tables (\ref{Tab:CPL} - \ref{Tab:sim_Padé}) for various dataset combinations.
Across all DE parametrizations, we observe the evolving behavior of DE, transitioning from the phantom regime at higher redshifts to the quintessence regime at the present time. 
However, for DESI BAO alone, DESI BAO + CMB, and DESI BAO + various SNIa samples (PantheonPlus, Union3, and DES-SN5YR), this evolution is not significant due to the large uncertainties in $w_{de}$. This behavior aligns with our conclusion that statistically, there is no significant evidence for preferring the evolving DE scenario over the standard $\Lambda$CDM cosmology when using observational datasets: DESI BAO alone, DESI BAO + SNIa (various PantheonPlus, Union3, and DES-SN5YR compilations), and DESI BAO + CMB.
In the cases of DESI BAO + CMB + different SNIa compilations, we observe a more pronounced evolutionary behavior of $w_{de}$ from the phantom regime at higher redshifts to the quintessence phase at the present time for all DE parametrizations. This evolutionary behavior supports our conclusion that there is significant statistical evidence for the possibility of an evolving DE scenario over the standard flat-$\Lambda$CDM cosmology when utilizing all DESI BAO, CMB, and SNIa (Union3 and DESI-SN5YR compilations). Note that the evolutionary behavior of DE from the early phantom regime to the late quintessence regime for the DESI BAO + CMB + PantheonPlus combination corresponds to a moderate deviation ($2\sigma-3\sigma$) from the standard flat-$\Lambda$CDM cosmology.

\end{document}